\documentclass[%
 reprint,
superscriptaddress,
nofootinbib,
 amsmath,amssymb,
 aps,
]{revtex4-1}

\usepackage[dvipsnames]{xcolor}
\usepackage{tikz}
\usepackage{graphicx}
\usepackage{dcolumn}
\usepackage{bm}
\usepackage{hyperref}
\usepackage{enumitem}
\usepackage{lineno}
\usepackage{xcolor,colortbl}
\usepackage{mathtools}

\usepackage{multirow}
\newcommand\Tstrut{\rule{0pt}{3.3ex}}         
\newcommand\Bstrut{\rule[-2ex]{0pt}{0pt}}   

\hypersetup{
  colorlinks=true,
  citecolor=blue,
  urlcolor=purple
}
\usepackage{aas_macros}

\newcommand{\B}[1]{\ensuremath{\boldsymbol{#1}}}

\def\fun#1#2{\lower3.6pt\vbox{\baselineskip0pt\lineskip.9pt
        \ialign{$\mathsurround=0pt#1\hfill##\hfil$\crcr#2\crcr\sim\crcr}}}

\newcommand{\beq}{\begin{equation}}
\newcommand{\eeq}{\end{equation}}
\newcommand{\beqa}{\begin{eqnarray}}
\newcommand{\eeqa}{\end{eqnarray}}

\begin{document}


\title{Testing one-loop galaxy bias: cosmological constraints from the power spectrum}

\author{Andrea Pezzotta}
\email{pezzotta@mpe.mpg.de}
\affiliation{%
  Max-Planck-Institut f\"{u}r extraterrestrische Physik, Postfach 1312, Giessenbachstr., 85741 Garching, Germany
}%
\affiliation{%
  Institute of Space Sciences (ICE, CSIC), Campus UAB, Carrer de Can Magrans, s/n, 08193 Barcelona, Spain
}%
\affiliation{%
  Institut d’Estudis Espacials de Catalunya (IEEC), 08034 Barcelona, Spain
}%

\author{Martin Crocce}
\affiliation{%
  Institute of Space Sciences (ICE, CSIC), Campus UAB, Carrer de Can Magrans, s/n, 08193 Barcelona, Spain
}%
\affiliation{%
  Institut d’Estudis Espacials de Catalunya (IEEC), 08034 Barcelona, Spain
}%

\author{Alexander Eggemeier}
\affiliation{%
  Institute for Computational Cosmology, Department of Physics, Durham University, South Road, Durham DH1 3LE,
  United Kingdom
}%

\author{Ariel G. S\'{a}nchez}
\affiliation{%
  Max-Planck-Institut f\"{u}r extraterrestrische Physik, Postfach 1312, Giessenbachstr., 85741 Garching, Germany
}%

\author{Rom\'an Scoccimarro}
\affiliation{
  Center for Cosmology and Particle Physics, Department of Physics, New York University, NY 10003, New York, USA
}%

\date{\today}

\begin{abstract} 

We investigate the impact of different assumptions in the modeling of one-loop galaxy bias on the recovery of cosmological parameters, as a follow up of the analysis done in the first paper of the series at fixed cosmology. We use three different synthetic galaxy samples whose clustering properties match the ones of the CMASS and LOWZ catalogues of BOSS and the SDSS Main Galaxy Sample. We investigate the relevance of allowing for either short range non-locality or scale-dependent stochasticity by fitting the real-space galaxy auto power spectrum or the combination of galaxy-galaxy and galaxy-matter power spectrum. From a comparison among the goodness-of-fit ($\chi^2$), unbiasedness of cosmological parameters (FoB), and figure-of-merit (FoM), we find that a four-parameter model (linear, quadratic, cubic non-local bias, and constant shot-noise) with fixed quadratic tidal bias provides a robust modelling choice for the auto power spectrum of the three samples, up to $k_{\rm max}=0.3\,h\,\mathrm{Mpc}^{-1}$ and for an effective volume of $6\,h^{-3}\,\mathrm{Gpc}^3$. Instead, a joint analysis of the two observables fails at larger scales, and a model extension with either higher derivatives or scale-dependent shot-noise is necessary to reach a similar $k_{\rm max}$, with the latter providing the most stable results. These findings are obtained with three, either hybrid or perturbative, prescriptions for the matter power spectrum, \texttt{RESPRESSO}, gRPT and EFT. In all cases, the inclusion of scale-dependent shot-noise increases the range of validity of the model in terms of FoB and $\chi^2$. Interestingly, these model extensions with additional free parameters do not necessarily lead to an increase in the maximally achievable FoM for the cosmological parameters $\left(h,\,\Omega_ch^2,\,A_s\right)$, which are generally consistent to those of the simpler model at smaller $k_{\rm max}$.

\end{abstract}

\pacs{Valid PACS appear here}
\maketitle


\section{Introduction}
\label{sec:introduction}

Over the past decades galaxy redshift surveys have provided a wealth of information on the large-scale distribution of galaxies across the Universe. Clustering measurements of two-point statistics -- the galaxy power spectrum $P(k)$ and the two-point correlation function $\xi(s)$ -- from large data samples can indeed provide precise measurements about the underlying cosmological model \cite{DavPee8304, MadEfsSut9001, Tegmark2004, Cole2005}.

The inference of cosmological parameters from large-scale structure is made intrinsically more difficult by the realisation that galaxies are a biased tracer of the total matter density field \cite{HauPee7311, Kai8409, DavEfsFre8505, BarBonKai8605, TegBlaStr04, ZehZheWei05, Desjacques:2018}. This translates into the necessity of having an accurate description of the relationship between the galaxy and matter density fields, a phenomenon which is commonly referred to as \textit{galaxy bias}. Even though the latter can be partially understood in a phenomenological way, e.g. using results from N-body simulations, the complexity of dealing with tracers featuring different morphological properties makes desirable to develop an analytic formulation that is based on a more theoretical background. From this point of view, perturbative approaches stand as a natural way of describing galaxy bias in a physically motivated way.

The main idea behind this formulation is that the galaxy overdensity $\delta_{\rm g}$ can be expressed in terms of a series of operators involving spatial derivatives of the gravitational and velocity potentials. At leading order, this relation is captured by a single multiplicative factor, i.e. $\delta_{\rm g}=b_1\,\delta$, where $\delta$ is the matter density contrast and $b_1$ is a multiplicative factor called \textit{linear bias} \cite{Kai8409}. Higher-order contributions become progressively more important on mildly non-linear scales, as expected from a spherically-symmetric gravitational collapse \cite{MoWhi9609, MoJinWhi97}, in a way that the expression for $\delta_{\rm g}$ can be expanded to include higher powers of $\delta$, i.e. $\delta_{\rm g}=\sum_n\,b_n/n!\,\delta^n$ \cite{Kai8409, FriGaz94}. At the same time it has been shown that anisotropies in the process of gravitational collapse are responsible for the generation of non-negligible tidal effects, which also contribute to the local distribution of galaxies \cite{Chan:2012, Baldauf:2012}. This finding followed the realization that the local-in-matter-density bias model had limitations in providing a proper description of the clustering of dark matter halos \cite{ManGaz1107, RotPor1107} and was leading to incompatible constraints on the quadratic bias $b_2$ from measurements of the power spectrum and bispectrum \cite{Pollack:2012, PolSmiPor1405}.

The most important aspect of the previously described model is that galaxy bias is treated as a spatially local quantity. However, it is well known that the formation of halos and galaxies is triggered by the gravitational collapse of matter from a spatially finite region, and therefore the local assumption is bound to fail when approaching scales that roughly correspond to the Lagrangian radius $R$ of the host halos. In terms of $\delta_{\rm g}$, this effect can be taken into account by considering not only its dependency on density and tidal fields, but also on functionals of $\delta$ \cite{Desjacques:2018}. This is equivalent to introducing higher derivatives of the matter density field, that at leading order provide a contribution to $\delta_g$ of the form $R^2\,\nabla^2\delta$ \cite{McDonald:2009, Des0811, DesCroSco1011}. A further ingredient to the galaxy-matter relation is represented by stochastic terms, that on large scales behave as an additional contribution to Poissonian shot-noise \cite{Scherrer:1998}. Stochasticity is the direct result of small-scale perturbations, which are not correlated over long distances under the assumption of Gaussian initial conditions \cite{Dekel:1999, TarSod9909, Mat9911}. At higher wave modes the halo-exclusion effect \cite{MoWhi9609, SheLem99, SmiScoShe0703} imprints a scale dependence on the stochastic contributions, whose strength is controlled by the Lagrangian radius \cite{BalSelSmi1310}, similarly to higher derivatives. At next-to-leading order, this contribution scales as $k^2$, and might become relevant even for clustering analyses based on two-point statistics, as shown in \cite{EggScoCro2011}.

Galaxy bias models based on the one-loop perturbative expansion have been used to extract cosmological constraints from big data collaborations, such as BOSS,  using clustering measurements both in configuration \cite{SanScoCro1701} and in Fourier space \cite{GilPerVer1702, BeuSeoSai1704, GriSanSal1705}. More recently, the same data have been re-analysed in a number of works \cite{IvaSimZal1909, AmiGleKok1909, TroSanAsg2001} with novel techniques that nevertheless assume the same biasing scheme for the galaxy-matter relationship. The majority of these analyses assume a different ansatz in terms of the degrees of freedom of the model, by fixing some of the free bias parameters to some physically motivated relations. For example \cite{SanScoCro1701}\cite{GriSanSal1705} and \cite{TroSanAsg2001} fix the cubic non-local bias parameter to the \textit{Local Lagrangian} (LL) relation, while \cite{BeuSeoSai1704} also fix the quadratic non-local parameter. On the other side, \cite{IvaSimZal1909} set the cubic term to zero and leave the quadratic tidal bias completely free.

This paper is part of a set of works \cite{EggScoCro2011}\cite{EggScoSmi2021} in which we explore the impact of different bias modeling choices using a set of three performance metrics, namely the goodness-of-fit, the unbiasedness of sampling parameters, and the merit of the model. To do that, we make use of a set of three different simulated galaxy samples with an effective volume of $6\,h^{-3}\,\mathrm{Gpc}^3$, whose clustering properties and number densities reproduce the ones of three real catalogues, the CMASS and LOWZ samples of BOSS \cite{Eisenstein2011, Dawson2012, reid2015}, and the Main Galaxy Sample of SDSS \cite{Strauss2002}. In order to exclusively concentrate on the modeling of one-loop galaxy bias, we carry out this analysis in real-space removing the impact of redshift-space distortions, which would require an additional modeling layer. In \cite{EggScoCro2011} we performed a fixed cosmology analysis by using the measured non-linear matter power spectrum. In that case, we used the linear bias parameter as a proxy for the goodness of our model. Here, we additionally model the impact of matter non-linear evolution and sample over cosmological parameters to determine the impact of one-loop galaxy bias on the recovery of such parameters. In the first part of the paper we use the hybrid perturbative-simulated approach, \texttt{RESPRESSO}, while comparing results for different PT-based dark matter models in a later section.

Our paper is organised as follows. In Section \ref{sec:theory} we summarise the main ingredients of our galaxy bias model including a description of all the terms contributing at one-loop in perturbation theory. This includes also a short review of the three non-linear matter predictions we employ in this work. In Section \ref{sec:data} we describe the simulated galaxy samples we use to test one-loop galaxy bias, along with fitting procedure, parameter priors and the performance metrics we introduced above. Main results from fit of the auto galaxy power spectrum and the combination between the latter and the galaxy-matter cross power spectrum are given in Section \ref{sec:results}. We finally draw our conclusions in Section \ref{sec:conclusions}.

\section{One-loop perturbation theory for biased tracers}
\label{sec:theory}

The theory describing the evolution of the clustering of biased tracers on mildly non-linear scales is a well established topic (for a review of galaxy bias see \cite{Desjacques:2018}) that can be naturally described in the framework of perturbation theory. This section stands as an overview of the most important results obtained at one-loop in standard perturbation theory (SPT) and derived approaches. Notice that for sake of readability we omit all the dependences on redshift $z$.

\subsection{Galaxy bias expansion}
\label{sec:galaxy-bias}

The general perturbative expansion of galaxy bias can be interpreted as a sum of different operators that are functions of the gravitational potential $\Phi$ and velocity potential $\Phi_{\rm v}$. Focusing on one-loop contributions, the relationship between biased tracers and the underlying dark matter density field can be described considering terms up to third order in the matter perturbation $\delta$. Following the notation of \cite{EggScoSmi1906} we can write
\begin{equation}
\label{eq:galaxy_bias}
\begin{split}
\delta_{\rm g}(\B{x}) = \;\, & \overline{b}_1\,\delta(\B{x})+\frac{\overline{b}_2}{2}\,\delta^2(\B{x})+\overline{\gamma}_2\,\mathcal{G}_2(\Phi_{\rm v}|\,\B{x})\\ 
& +\overline{\gamma}_{21}\,\mathcal{G}_2(\varphi_1,\varphi_2|\,\B{x})+\overline{\beta}_1\,\nabla^2\delta(\B{x})+\ldots\,,
\end{split}
\end{equation}
where the first two terms on the right-hand side are part of the standard local expansion in powers of $\delta$. 

In the previous equation $\mathcal{G}_2$ is a Galilean invariant operator, representing the tidal stress tensor generated by the velocity potential $\Phi_{\rm v}$
, and it is given by
\begin{equation}
\label{eq:galileon_G2_LO}
\mathcal{G}_2(\Phi_{\rm v})=(\nabla_{ij}\Phi_{\rm v})^2-(\nabla^2\Phi_{\rm v})^2.
\end{equation}
In Fourier space \footnote{We adopt the usual convention for the Fourier transform,
  $\delta(\B{x}) = \int_{\B{k}} \exp{(-i\B{k}\cdot\B{x})}\,\delta(\B{k})$, and use the short-hand notation for the 3d integrals,
  $\int_{\B{k}_1,\ldots,\B{k}_n} \equiv \int \text{d}^3k_1/(2\pi)^3 \cdots \text{d}^3k_n/(2\pi)^3$.} this translates into
\begin{equation}
\label{eq:galileon_G2_LO_Fourier}
\mathcal{G}_2(\B{k})=\!\!\int_{\B{q}}\!\left[\frac{\B{q}\cdot(\B{k}-\B{q})}{q^2\,|\B{k}-\B{q}|^2}-1\right]\theta(\B{q})\,\theta(\B{k}-\B{q})\,,
\end{equation}
where $\theta$ is the divergence of the matter velocity field, such that $\nabla^2\Phi_{\rm v} \equiv \theta$.

Differently from the first two terms of Equation (\ref{eq:galaxy_bias}), terms involving $\mathcal{G}_2$ incorporate non-local (in matter density $\delta$) contributions that spontaneously arise from the non-linear evolution of the matter density field. The second of these terms is obtained by expressing the non-linear velocity potential up to second order (i.e. $\Phi_{\rm v}=\Phi_{\rm v}^{(1)}+\Phi_{\rm v}^{(2)}, \,\varphi_1=-\Phi_{\rm v}^{(1)},\,\varphi_2=-\Phi_{\rm v}^{(2)}$), and keeping the next-to-leading order correction  \cite{ChaSheSco1711}, leading to 
\begin{equation}
\label{eq:galileon_G2_NLO}
\mathcal{G}_2(\varphi_1,\varphi_2)=\nabla_{ij}\varphi_1\nabla_{ij}\varphi_2-\nabla^2\varphi_1\nabla^2\varphi_2,
\end{equation}
where $\nabla^2\varphi_1=-\theta$ is the linear velocity divergence field, and $\nabla^2\varphi_2=-\mathcal{G}_2(\varphi_1)$ is the next-to-leading order. The net result of adding this higher-order correction is that $\delta_{\rm g}$ collects contributions up to third order in the matter perturbation $\delta$. 

In addition to the standard expansion, any biased tracer also comes with a physical scale which regulates the importance of higher-derivative operators \cite{BarBonKai8605,Mat9911,McDonald:2009,DesCroSco1011,MusShe1206}, whose leading order scales as $\nabla^2\delta$. This scale quantifies the size of the region in which galaxy formation occurs, and it is therefore influenced by any short-range gravitational effect and baryonic corrections. For halos, this scale is close to the Lagrangian radius \cite{McDonald:2009, LazSch1911}, while it differs for other kinds of tracers, such as galaxies and quasars, depending on their type.

An important aspect of the previously described biasing scheme is the renormalization of the parameters on which it is based. As a matter of fact, the \textit{bare} bias parameters listed in Equation (\ref{eq:galaxy_bias}) are sensitive to the UV cutoff scale used to make one-loop integrals convergent. In particular, the current basis carries a dependence on the variance of the matter density field $\sigma^2=\mathcal{h}\delta^2(\B{x})\mathcal{i}$, and more generally to any higher-order correlator $\mathcal{h}\delta^{(n)}\mathcal{i}$. This dependence is completely non-physical and can be reabsorbed by means of an adequate renormalization of the bias parameters \cite{McDonald:2006,Assassi:2014}. The new basis (for convention this is denoted without the overscript hat) can be identified using the peak-background split formalism \cite{SchJeoDes1307}, and it is constructed in a way that its components quantify the response of the cosmic mean abundance of tracers to a change in the background density, with no dependence on the one-loop cutoff scale. 

An alternative approach to renormalization can be obtained expanding $\delta_{\rm g}$ in terms of the galaxy \textit{multi-point propagators}, a topic which was firstly described in \cite{EggScoSmi1906}. Consistently with the naming convention adopted in the context of renormalized perturbation theory \cite{CroSco0603a}, these quantities are defined as the ensemble averaged derivatives of $\delta_{\rm g}$ with respect to $\delta_{\rm L}$,
\begin{equation}
\begin{split}
    \left<\frac{\partial\delta_{\rm g}(\B{k})}{\partial\delta_{\rm L}(\B{k}_1)\,\cdots\,\partial\delta_{\rm L}(\B{k}_n)}\right>
    \equiv \;\, &(2\pi)^3\,\Gamma_{\rm g}^{(n)}(\B{k}_1,\ldots,\B{k}_n) \\ &\times\,\delta_{\rm D}(\B{k} - \B{k}_{1 \cdots n})\,,
  \end{split}
\end{equation}
and, being observables themselves, they are already normalized by construction. For this reason, the multi-point propagators can be identified with the scale-dependent bias parameters in a way that
\begin{equation}
\delta_{\rm g} = \Gamma_{\rm g}^{(1)} \otimes {\cal H}_1 + \Gamma_{\rm g}^{(2)} \otimes {\cal H}_2 + \ldots\,,
\end{equation}
where the ${\cal H}_n$ are the Wiener-Hermite functionals \cite{EggScoSmi1906, Bernardeau:2002}, and the operator $\otimes$ is defined as
\begin{equation}
  \begin{split}
    \left[\Gamma_{\rm g}^{(n)} \otimes {\cal H}_n\right](\B{k}) \equiv \;\, &(2\pi)^3 \int_{\B{k}_1,\ldots,\B{k}_n}
    \delta_{\rm D}(\B{k}-\B{k}_{1 \cdots n}) \\ &\times\,\Gamma_{\rm g}^{(n)}(\B{k}_1,\ldots,\B{k}_n)\,{\cal
      H}_n(\B{k}_1,\ldots,\B{k}_n) \,,
  \end{split}
\end{equation}

Using this approach, we can define our observables in a fully consistent framework. In this paper we are interested in the galaxy auto power spectrum $P_{\rm gg}$ and galaxy-matter cross power spectrum $P_{\rm gm}$, which are defined as the auto and cross-correlation of the two density fields $\delta_{\rm g}$ and $\delta_{\rm m}$, as
\begin{align}
\label{eq:Pgg_Pgm_def}
\mathcal{h}\delta_{\rm g}(\B{k})\,\delta_{\rm g}(\B{k}')\mathcal{i}&\equiv(2\pi)^3 P_{\rm gg}(k)\,\delta_{\rm D}(\B{k}+\B{k}')\,,\\
\mathcal{h}\delta_{\rm g}(\B{k})\,\delta_{\rm m}(\B{k}')\mathcal{i}&\equiv(2\pi)^3 P_{\rm gm}(k)\,\delta_{\rm D}(\B{k}+\B{k}')\,.
\end{align}
Substituting Equation (\ref{eq:galaxy_bias}) in the previous set, we get to the full expressions for the galaxy auto and cross power spectra at one-loop,
\begin{align}
\begin{split}
P_{\rm gg}(k)=  &\;\, b_1^2 \,P_{\rm mm}(k) + b_1b_2\,P_{b_1b_2}(k) + b_1\gamma_2\, P_{b_1\gamma_2}(k) \\ &+ b_2^2\,P_{b_2b_2}(k) + b_2\gamma_2\, P_{b_2\gamma_2}(k) + \gamma_2^2\,P_{\gamma_2\gamma_2}(k)\\& + b_1\gamma_{21}\, P_{b_1\gamma_{21}}(k) -2b_1\beta_1 \,k^2 P_{\rm L}(k)\,, 
\label{eq:Pgg_1-loop}
\end{split}
\\[2ex]
\begin{split}
P_{\rm gm}(k)= &\;\, b_1 \,P_{\rm mm}(k) + \frac{1}{2}\bigg[b_2\,P_{b_1b_2}(k)+\gamma_2\,P_{b_1\gamma_2}(k)\\&+\gamma_{21}\,P_{b_1\gamma_{21}}(k)\bigg]-\beta_1\,k^2P_{\rm L}(k)\,, 
\label{eq:Pgm_1-loop}
\end{split}
\end{align}
where $P_{\rm L}$ is the linear matter power spectrum, $P_{\rm mm}$ is the non-linear matter power spectrum, and the one-loop bias corrections read
\begin{flalign}
\label{eq:Pbibj}
\begin{split}
P_{b_1b_2}(k) = &\;\, 2\!\!\int_{\B{q}}\! F_2(\B{k}-\B{q}, \B{q})P_{\rm L}(|\B{k}-\B{q}|)P_{\rm L}(q)\,,
\end{split}
\\[2ex]
\begin{split}
P_{b_1\gamma_2}(k) = &\;\,P_{b_1\gamma_2}^{\rm mc}(k) + P_{b_1\gamma_2}^{\rm prop}(k) \\
                                    = &\;\,4\!\!\int_{\B{q}}\! F_2(\B{k}-\B{q},\B{q})S(\B{k}-\B{q},\B{q})P_{\rm L}(|\B{k}-\B{q}|)P_{\rm L}(q)\\
                                    & + 8P_{\rm L}(k)\!\!\int_{\B{q}}\! G_2(\B{k},\B{q}) S(\B{k}-\B{q},\B{q})P_{\rm L}(q)\,,
\end{split}
\\[2ex]
\begin{split}
P_{b_2b_2}(k) = &\;\, \frac{1}{2}\int_{\B{q}} \!\left[P_{\rm L}(|\B{k}-\B{q}|) P_{\rm L}(q)-P_{\rm L}^2(q)\right]\,,\label{eq:Pb2b2}
\end{split}
\\[2ex]
\begin{split}
P_{b_2\gamma_2}(k) = &\;\, 2\!\!\int_{\B{q}}\! S(\B{k}-\B{q},\B{q})P_{\rm L}(|\B{k}-\B{q}|)P_{\rm L}(q)\,,
\end{split}
\\[2ex]
\begin{split}
P_{\gamma_2\gamma_2}(k) = &\;\, 2\!\!\int_{\B{q}}\! S^2(\B{k}-\B{q},\B{q})P_{\rm L}(|\B{k}-\B{q}|)P_{\rm L}(q)\,,
\end{split}
\\[2ex]
\begin{split}
P_{b_1\gamma_{21}}(k) = &\;\, 4P_{\rm L}(k)\!\!\int_{\B{q}}\! S(\B{k}-\B{q},\B{q})S(\B{k},\B{q})P_{\rm L}(q)\,.
\end{split}
\end{flalign}
Here $F_2$ and $G_2$ are the symmetrised second-order mode-coupling kernels \cite{Bernardeau:2002}, and $S(\B{k}_1, \B{k}_2)=\left(\hat{\B{k}}_1\cdot\hat{\B{k}}_2\right)^2-1$ is the Fourier transform of the kernel describing $\mathcal{G}_2(\Phi_{\rm v})$, as shown in Equation (\ref{eq:galileon_G2_LO_Fourier}). 

The only two propagator-like contributions are perfectly degenerate with each other, and follow the relation 
\begin{equation}
\label{eq:b1g2_b1g21_degeneracy}
P_{b_1\gamma_{21}}(k)=-\frac{7}{6}P_{b_1\gamma_2}^{\rm prop}(k)\,.
\end{equation}
For this reason, in a real analysis, it is common practice to either neglect one of the two tidal field related parameters (e.g. \cite{IvaSimZal1909}) or to assume perfectly local-in-matter-density relations (see Section \ref{sec:co-evolution}) to express one of them in terms of lower order local bias parameters (e.g. \cite{SanScoCro1701,GriSanSal1604}).

Since the $P_{b_2b_2}$ contribution does not asymptote to 0 in the large-scale limit, we renormalise it as in Equation (\ref{eq:Pb2b2}), and absorb the constant low-$k$ amplitude as an additional contribution to the shot noise error, that will be discussed in Section \ref{sec:stochasticity}.

\subsection{Matter modeling}
\label{sec:matter}

In this section we discuss the modeling options for the non-linear matter power spectrum $P_{\rm mm}$ in Equations (\ref{eq:Pgg_1-loop}) and (\ref{eq:Pgm_1-loop}). As a matter of fact an accurate modeling of $P_{\rm mm}$ is essential, as any systematic effect in the description of the matter density field in the range of scales we are considering might lead to invalid interpretations of the galaxy-matter bias relationship.

Differently from \cite{EggScoCro2011}, where one-loop bias was investigated at fixed cosmology and adopting the measured matter power spectrum as reference, here the main goal is to assess the level of accuracy of our model in terms of cosmological parameters. For this reason, we have to explicitly assume a model for $P_{\rm mm}$. In the rest of this section we provide a description of the three models we are going to test. In particular we use 1) a refined RPT-derived model, based on the preservation of Galilean invariance (dubbed gRPT), 2) an EFT-like approach based on BAO damping and \textcolor{red}{a} non-trivial stress tensor, and 3) a mixed approach, \texttt{RESPRESSO}, based on accurate N-body simulations and a perturbative expansion around the fiducial cosmology.

\subsubsection{Standard perturbation theory}

The basic assumption of SPT is that dark matter behaves as a perfect pressureless fluid on large enough scales, where matter is not subject to shell crossing as in multi-streaming regions. Under this assumption, and after having expanded the matter density contrast in a Taylor series, i.e. $\delta=\delta^{(1)}+\delta^{(2)}+\delta^{(3)}+\ldots$, we find solutions at every order in perturbation as \cite{Bernardeau:2002}
\begin{equation}
\label{eq:delta_expansion}
\begin{split}
\delta^{(n)}(\B{k})=\int_{\B{q}_1\ldots\B{q}_n} & \delta_{\rm D}(\B{k}-\B{q}_1-\ldots-\B{q}_n)\,F_n(\B{q}_1,\ldots,\B{q}_n) \\ & \times\delta_{\rm L}(\B{q}_1)\ldots\delta_{\rm L}(\B{q}_n)\,,
\end{split}
\end{equation}
where $F_n$ is the $n$-th order symmetrized kernel describing the non-linear mode coupling between fluctuations at different wavelengths.

Moving to two-point statistics, the expansion for the matter power spectrum can be written as
\begin{equation}
\label{eq:power_expansion}
P_{\rm mm}(k)=P_{\rm L}(k)+P^{1\mbox{-}\mathrm{loop}}(k)+\ldots\,,
\end{equation}
where at one-loop the only non-vanishing contributions are
\begin{equation}
\label{eq:power_1-loop}
\begin{split}
P^{1\mbox{-}\mathrm{loop}}(k)=&\:\,P_{22}(k)+P_{13}(k)\\
=&\:\,2\!\!\int_{\B{q}}\! F_2(\B{k}-\B{q},\B{q})P_{\rm L}(\B{k}-\B{q})P_{\rm L}(q)\\
&+ 6\,P_{\rm L}(k)\!\!\int_{\B{q}}\!F_3(\B{q},-\B{q}, \B{k})P_{\rm L}(q)\,.
\end{split}
\end{equation}
It is now well established that a SPT approach like the one described above lead to significant residuals if compared to the output of numerical simulations, even including higher-order corrections \cite{BlaGarKon1309}. The source of this inaccuracy can be mostly identified in two separate effects, whose description is the subject of the next two sections.

\subsubsection{BAO damping from large-scale ``infrared" modes}
\label{sec:BAO}

One of the most acknowledged deviations between one-loop SPT predictions  and the matter power spectrum measured from numerical simulations is the shape of the BAO features. Since the characteristic scale of the BAO peak is much larger than the scale at which non-linear contributions become important, one may expect a standard perturbative approach to provide accurate predictions on that scale. However, large-scale bulk motions produce a non-negligible effect on the amplitude of the power spectrum at the BAO scale, the most significant of which is a smearing of the BAO signal due to the large-scale relative displacement field \cite{CroSco0603b,EisSeoWhi0708,EisSeoSir0708,MatPie08,CroSco0801}. 

These corrections to the matter power spectrum were firstly resummed in the context of RPT \cite{CroSco0603b}, and at leading order the net effect is to apply a damping factor to the BAO wiggles. Practically, we can express the matter power spectrum as the sum of a smooth ($P_{\rm nw}$) and wiggly ($P_{\rm w}$) term \cite{SeoSieEis0810},
\begin{equation}
\label{eq:sum_smooth_wiggly}
P(k)=P_{\rm nw}(k)+P_{\rm w}(k).
\end{equation}
The smooth-wiggle split can be realised adopting several different recipes, and throughout this paper we follow the approach described in \cite{VlaSelChu1603} and \cite{OsaNisBer1903}, where the smooth component is defined as a rescaling of the featureless spectrum firstly defined in \cite{EisHu1901} to account for broadband difference with the linear power spectrum.

At leading order, the damping factor can be calculated assuming the Zel'dovich approximation, and subsequently applied to the wiggly component, so that
\begin{equation}
\label{eq:IR_LO}
P_{\rm LO}(k)=P_{\rm nw}(k)+e^{-k^2\Sigma^2}P_{\rm w}(k),
\end{equation}
where
\begin{equation}
\label{eq:IR_sigma}
\Sigma^2=\!\int_0^{k_{\rm S}}\!P_{\rm nw}(q)\left[1-j_0\left(\frac{q}{k_\mathrm{BAO}}\right)+2j_2\left(\frac{q}{k_\mathrm{BAO}}\right)\right]\frac{dq}{6\pi^2}\,,
\end{equation}
is the relative displacement field two-point function at the BAO scale \cite{EisSeoWhi0708}. Here $j_n$ is the $n$-th order spherical Bessel function, $k_\mathrm{BAO}=\pi/l_\mathrm{BAO}$ is the wavemode corresponding to the reference BAO scale $l_\mathrm{BAO}=110\,h^{-1}\,\mathrm{Mpc}$, and $k_{\rm S}$ is the UV limit of integration. To properly account for the resummation of IR modes at any given scale $k$, one should integrate over all modes $q<k$, in a way that $k_{\rm S}=k_{\rm S}(k)$. However, pushing the integration to significantly large values of $k_{\rm S}$ would result in the breaking of the range of validity of the pertubative IR expansion. At the same time, it can be shown that the integrand of Equation (\ref{eq:IR_sigma}) gives significant contributions only up to $k_{\rm S}\sim0.2\,h\,\mathrm{Mpc}^{-1}$ and therefore we restrict the integration to the range $[0,\,0.2]$ when computing the value of $\Sigma$ \cite{BlaGarIva1607,IvaSimZal1909}.

At next-to-leading order, the IR-resummed matter power spectrum can be written as \cite{BlaGarIva1607}
\begin{equation}
\label{eq:IR_NLO}
\begin{split}
P_{\rm NLO}(k)=&\,P_{\rm nw}(k)+\left(1+k^2\Sigma^2\right)e^{-k^2\Sigma^2}P_{\rm w}(k)\\&+\left(P_{\rm nw}^{1\mbox{-}\mathrm{loop}}(k)+e^{-k^2\Sigma^2}P_{\rm w}^{1\mbox{-}\mathrm{loop}}(k)\right),
\end{split}
\end{equation}
where $P_{\rm nw}^{1\mbox{-}\mathrm{loop}}$ is the one-loop matter correction defined in Equation (\ref{eq:power_1-loop}) but evaluated using the smooth component $P_{\rm nw}$ rather than the full linear power spectrum $P_{\rm L}$, and $P_{\rm w}^{1\mbox{-}\mathrm{loop}}=P^{1\mbox{-}\mathrm{loop}}-P_{\rm nw}^{1\mbox{-}\mathrm{loop}}$.

\subsubsection{Small-scale corrections: non-trivial stress tensor}
\label{sec:EFT}

Along with the resummation of infrared modes, we also have to consider the impact of small-scale physics on long wavelength fluctuations. This happens because the original assumption of a perfectly pressureless fluid is bound to fail on non-linear scales, where 
dark matter experiences shell-crossing in multistreaming regions \cite{Bernardeau:2002}. Moreover, the effect of baryonic physics, such as galaxy formation, cooling and feedback, also contributes in generating a baryonic pressure that impacts the clustering of dark matter on larger scales.

The net effect of UV scales on dark matter clustering is to generate a non-zero stress tensor \cite{PueSco0908} whose leading contribution to the matter power spectrum is to add a counter-term of the form \cite{PueSco0908, CarHerSen1206, BauNicSen1207}
\begin{equation}
\label{eq:counter-term}
P_{\rm ctr}(k)=-2\,c_1\,k^2P_{\rm L}(k).
\end{equation}
Here $c_1$ can be treated as an effective speed of sound, that reflects the influence of short wavelength perturbations, and in particular of the complex physics beyond galaxy formation. Given the poor knowledge about these types of processes, a standard assumption is to treat $c_1$ as a free parameter (see e.g. \cite{CarForGre1310, BalMerZal1512}) and marginalise over it to obtain the posterior distribution of the parameters of interest. 

By inspection of the individual terms contributing to one-loop formulas in Equations (\ref{eq:Pgg_1-loop}) and (\ref{eq:Pgm_1-loop}), we can notice that the counter-term defined in Equation (\ref{eq:counter-term}) is completely degenerate with the leading order higher-derivative contribution defined in Section \ref{sec:galaxy-bias}, as they both scale as $k^2P_{\rm L}(k)$. In principle we may remove this degeneracy when jointly fitting the auto and cross galaxy power spectra, as the non-vanishing stress tensor affects only the one-loop matter power spectrum $P_{\rm mm}$. This results in different imprints on $P_{\rm gg}$ and $P_{\rm gm}$,
\begin{equation}
\label{eq:beta_c_degeneracy}
\begin{split}
&P_{\rm gg}(k)\supset -2b_1\left(b_1c_1+\beta_1\right)k^2P_{\rm L}(k)\,,\\
&P_{\rm gm}(k)\supset -\left(2b_1c_1+\beta_1\right)k^2P_{\rm L}(k)\,.
\end{split}
\end{equation}
In practice, the sensitivity to UV modes might affect in different ways $P_{\rm gg}$ and $P_{\rm gm}$, leading to inconsistent values of $c_1$ between the two observables. For this reason in the rest of the paper we will employ two independent free parameters $\beta_P$ and $\beta_P^\times$ characterising the $k^2P_{\rm L}(k)$ contributions coming from $P_{\rm gg}$ and $P_{\rm gm}$, respectively.

\subsubsection{Modeling of the non-linear matter power spectrum}
\label{sec:matter_models}

In this section we briefly summarise the three different prescriptions we adopt to model the non-linear matter power spectrum $P_{\rm mm}$ throughout the rest of the paper. 

The first of such models is based on a perturbative-simulated mixed approach, which revolves around a fiducial high-resolution measurement of the non-linear matter power spectrum from N-body simulations, and a two-loop perturbative expansion in the cosmological parameter space for the \textit{response function} \cite{BerTarNis1401, NisBerTar1611}. The latter quantifies the variation of the non-linear power spectrum at scale $k$ induced by a variation of the linear power spectrum at scale $q$, namely
\begin{equation}
\label{eq:response_function}
K(k,q)\equiv q\,\frac{\partial P_{\rm mm}(k)}{\partial P_{\rm L}(q)}.
\end{equation}
Under the assumption of having a reliable measurement of $P_{\rm mm}$ for a fiducial cosmology $\B{\theta}_\mathrm{fid}$, it is possible to predict the same observable at a generic position $\B{\theta}$ as
\begin{equation}
\label{eq:model_respresso}
\begin{split}
P_{\rm mm}(k\,|\,\B{\theta}) = &\,P_{\rm mm}(k\,|\,\B{\theta}_\mathrm{fid})\int d(\log q) \,K(k,q)\\
&\times\left[P_{\rm L}(q\,|\,\B{\theta})-P_{\rm L}(q\,|\,\B{\theta}_\mathrm{fid})\right]\,.
\end{split}
\end{equation}
The range of validity of this mixed approach becomes progressively less accurate for cosmologies that are far way from $\B{\theta}_\mathrm{fid}$, but this issue can be overcome by employing a multi-step reconstruction starting from the fiducial cosmology.

The \texttt{RESPRESSO} public package \cite{NisBerTar1712} makes use of this approach, starting from a fiducial measurements of $P_{\rm mm}$ from a set of high-resolution N-body simulations with the Planck 2015 cosmology \cite{Planck1609}. 

The second model we consider is based on the Effective Field Theory of Large Scale Structure \cite{CarHerSen1206, AngFasSen1503}, and it is close to what was recently used in the full shape analysis of the BOSS DR12 galaxy power spectrum \cite{IvaSimZal1909}. At one-loop in real-space, this model is based on SPT results, but it also accounts for the effect of IR and UV modes on the evolution of the matter power spectrum, as described in the previous two sections. Namely, we can write a simple expression for the one-loop matter power spectrum, that reads
\begin{equation}
\label{eq:model_eft}
P_{\rm mm}(k) = P_{\rm NLO}(k)+P_{\rm ctr}(k)\,,
\end{equation}
where $P_{\rm NLO}(k)$ and $P_{\rm ctr}(k)$ are defined in Equations (\ref{eq:IR_NLO}) and (\ref{eq:counter-term}), respectively.

The third model we consider is based on a particular flavour of Renormalised Perturbation Theory (RPT) \cite{CroSco0603a, CroSco0603b}. In this kind of approach, the non-linear matter power spectrum is typically separated into a component that evolves the initial density contrast independently at each wavelength, called \textit{propagator} $G(k)$, and a mode-coupling term that accounts for the mixing of scales due to non-linear evolution, so that we can write
\begin{equation}
\label{eq:model_rpt}
P(k)=G^2(k)P_{\rm L}(k)+P^{\rm MC}(k)\,.
\end{equation}
In a RPT-based approach (e.g. \cite{BerCroSco0811, TarNisBer1304}), the propagator is resummed while keeping the mode-coupling term at a fixed order, leading to a breaking of the Galilean invariance (GI) of equal-time correlators \cite{ScoFri9607}. This translates into an unphysical damping of the broadband power, which becomes mostly significant in the UV regime. The RPT flavour we consider here, known as gRPT, effectively attempts to resum the mode-coupling term in a way that is consistent with the resummation of the propagator (see equation 17 of \cite{EggScoCro2011} for the explicit formula). At the same time, this approach naturally incorporates IR resummation consistently with what is described in Section \ref{sec:EFT}. We notice that, although even in this case the impact of the non-zero stress tensor should be taken into account with the addition of an effective speed of sound, the broadband predicted by gRPT is slightly suppressed with respect to SPT predictions. This would in principle lead to an even smaller UV counter-term, and for this reason we fix $c_1=0$ when modeling $P_{\rm mm}$ with gRPT.

Being partially calibrated on numerical measurements, \texttt{RESPRESSO} provides much more accurate results on non-linear scales, as it intrinsically incorporates higher-order corrections with respect to the previous two models, which include only one-loop contributions to the matter density field. For this reason, in the first part of this paper we fix the description of dark matter non-linear evolution following this approach, and check the impact of using different matter models only in Section \ref{sec:results_models}. This choice is also motivated by the analysis performed in \cite{EggScoCro2011}, that showed how \texttt{RESPRESSO} is the model that most closely reproduces the performances of the \textit{true} matter power spectrum measured from simulations.

\subsection{Stochasticity}
\label{sec:stochasticity}

One ingredient that is still missing from Equation (\ref{eq:Pgg_1-loop}) is the stochastic contribution to the galaxy power spectrum. As a matter of fact, the previous relations are completely deterministic, and assume a one-to-one correspondence between the distribution of galaxies and the combined effect of the matter density and tidal fields. However, galaxy formation is determined not only by large-scales perturbations, but also by short wavelength modes. Under the assumption of Gaussian initial conditions, these modes are completely uncorrelated from large-scales fluctuations, and give birth to an additional stochastic field $\varepsilon_{\rm g}$ which is dependent on the local distribution of matter.

In practice, we can define the stochastic contribution $P_{\varepsilon_{\rm g}\varepsilon_{\rm g}}$ to the power spectrum as \cite{Desjacques:2018}
\begin{equation}
\mathcal{h}\varepsilon_{\rm g}(\B{k})\,\varepsilon_{\rm g}(\B{k}')\mathcal{i} = (2\pi)^3 P_{\varepsilon_{\rm g}\varepsilon_{\rm g}}(k)\,\delta_{\rm D}(\B{k}+\B{k}'),
\end{equation} 
and add this contribution to the one-loop galaxy power spectrum in Equation (\ref{eq:Pgg_1-loop}). As for the modeling of $P_{\varepsilon_{\rm g}\varepsilon_{\rm g}}$, we notice that the relation between the distribution of galaxies and the high-$k$ modes is not exactly local, as it depends on the distribution of matter within a small finite region, similarly to the higher-derivatives of the galaxy density field. For this reason we can write
\begin{equation}
\label{eq:P_egeg}
P_{\varepsilon_{\rm g}\varepsilon_{\rm g}}(k)=\frac{1}{\overline{n}}\,\left(1+N_0+N_2\,k^2+\ldots\right)\,,
\end{equation} 
where $\overline{n}$ is the mean number density of galaxies in the considered volume. The constant contribution $N_0$ represents deviations from purely Poissonian shot-noise ($1/\overline{n}$), while higher-order corrections, of which the leading term scales as $k^2$, are generated to account for the short-range non-locality described above.

$N_0$ is expected from the halo exclusion effect \cite{SheLem99, SmiScoShe0703, BalSelSmi1310} for which two different dark matter halos cannot overlap (the same principle is what drives the effective matter pressure in multi-streaming regions). This implies a deviation from Poissonian shot-noise that can be either positive (super-Poisson) or negative (sub-Poisson), depending on the considered tracer. Sub-Poissonian shot-noise is more expected for central galaxies of massive halos, while super-Poissonian values are more typical of galaxy populations with high satellite fractions \cite{BalSelSmi1310, CasMoShe0207}.

Notice that assuming Gaussian initial conditions all cross-correlators of the form $\mathcal{h}\varepsilon_{\rm g}\,\delta\mathcal{i}$ are null by construction, and therefore Equation (\ref{eq:P_egeg}) is the only stochastic contribution to the galaxy auto power spectrum $P_{\rm gg}$. In reality, non-linear gravitational evolution introduces a degree of correlation between long and short wavelengths, so that $\mathcal{h}\varepsilon_{\rm g}\,\delta\mathcal{i}\neq0$ at later times, but all of these contributions are subdominant in the case of the power spectrum, and can thus be neglected.

As anticipated in Section \ref{sec:galaxy-bias}, the quadratic term $P_{b_2b_2}$ needs to be renormalized in order to provide a null contribution in the low-$k$ limit. For this reason, we subtract from Equation (\ref{eq:Pb2b2}) its large-scale asymptote \cite{McDonald:2006, McDonald:2009}, defined by
\begin{equation}
\label{eq:P_noise}
P_{b_2b_2}^\mathrm{noise}=\frac{b_2}{2}\int_{\B{q}} P_{\rm L}^2(q)\,,
\end{equation}
and reabsorb it into the constant shot-noise parameter $N_0$.

Along with $P_{\rm gg}$ it can be shown that also $P_{\rm gm}$ requires an additional stochastic component. In this case, stochasticity is not sourced by $\varepsilon_{\rm g}$, as once again all correlators of the form $\mathcal{h}\varepsilon_{\rm g}\,\delta\mathcal{i}$ vanish, but rather by the matter density field itself via a new stochastic field $\varepsilon_{\rm m}$. The reason of this is that dark matter ceases to behave as an ideal pressureless fluid on small-scales, where the dynamics of gravitational collapse is subject to shell crossing. This translates into an effective pressure exerted by the matter density field, whose contribution to the galaxy-matter cross power spectrum scales as $k^2$ in the low-$k$ limit \cite{AboMirPaj1605}. For this reason, it follows \cite{Desjacques:2018}
\begin{equation}
\mathcal{h}\varepsilon_{\rm g}(\B{k})\,\varepsilon_{\rm m}(\B{k}')\mathcal{i} = (2\pi)^3 P_{\varepsilon_{\rm g}\varepsilon_{\rm m}}(k)\,\delta_{\rm D}(\B{k}+\B{k}')\,,
\end{equation}
where
\begin{equation}
\label{eq:P_egem}
P_{\varepsilon_{\rm g}\varepsilon_{\rm m}}(k)=\frac{1}{\overline{n}}\,(N_2^\times k^2+\ldots)\,.
\end{equation}

\subsection{Co-evolution relations}
\label{sec:co-evolution}

Although the previously described galaxy bias expansion is complete at one-loop, the large number of free bias parameters makes its applicability to real datasets difficult, particularly when using the information contained only in the two-point statistics (using additional constraints from e.g. the galaxy bispectrum can break some of the degeneracies between parameters). For this reason it is common practice to adopt empirical relations among bias parameters in order to reduce the degrees of freedom of our model.

The most natural expressions are the so-called \textit{local Lagrangian} relations \cite{CatLucMat9807, Catelan:2000, McDonald:2009, Matsubara:2011}. The latter are based on the assumption that galaxies (or more generally any biased tracer of the matter density field) are formed instantaneously at an infinite past time, that galaxy formation is driven exclusively by the local matter density field, and that the number of tracers is conserved after their formation. Under these assumptions, it is possible to describe the subsequent evolution of the galaxy density field under the effect of gravity, which leads to the appearance of higher-order non-local operators even in the presence of a purely local relation at the time of formation.

It is possible to relax the Lagrangian local-in-matter-density assumption, while still requiring the conservation of the total number of tracers. In this way, we can express the Eulerian non-local parameters $\gamma_2$ and $\gamma_{21}$ as a function of the corresponding Lagrangian counterparts, 
\begin{align}
\gamma_2 &=-\frac{2}{7}(b_1-1)+\gamma_{2,\mathcal{L}}\,, \label{eq:g2_coev}\\
\gamma_{21} &=-\frac{2}{21}(b_1-1)+\frac{6}{7}\gamma_2+\gamma_{21,\mathcal{L}}\,, \label{eq:g21_coev}
\end{align}
where the subscript $\mathcal{L}$ indicates Lagrangian quantities, and the remaining terms in the right-hand sides are the result of gravitational evolution \cite{Chan:2012, Baldauf:2012, EggScoSmi1906}.

Although local Lagrangian relations (i.e. $\gamma_{2,\mathcal{L}}=\gamma_{21,\mathcal{L}}=0$) have been proven to be much more accurate than simply neglecting non-local terms, recent measurements of non-local bias parameters from a wide range of halo samples showed a slight deviation with respect to observations \cite{LazSch1712, AbiBal1807}. An alternative approach for the quadratic tidal parameter is based on the excursion set theory \cite{SheChaSco1304}, for which $\gamma_2$ can be predicted as a function of $b_1$ using a quadratic fit,
\begin{equation}
\label{eq:g2_ex-set}
\gamma_{2,\,\rm{ex}}(b_1)= 0.524-0.547\,b_1+0.046\,b_1^2\,.
\end{equation}
As shown in Figure 1 of \cite{EggScoCro2011}, this relation is more accurate than the local Lagrangian approximation for tracers with $b_1\gtrsim1.3$. Therefore this relation should apply much better to our datasets, which show a linear bias consistently larger than this value (see Table \ref{tab:tracers}). For this reason, we fix $\gamma_2$ to Equation (\ref{eq:g2_ex-set}) throughout the rest of the paper.

\section{Data and methodology}
\label{sec:data}

\subsection{Simulated galaxy samples}
\label{sec:catalogs}

The robustness of our biasing scheme can be assessed only by validating the model over a wide range of tracers, featuring different host halo masses, galaxy bias, and redshifts. For this purpose, we make use of a set of three different synthetic galaxy samples, whose main properties are summarised in Table \ref{tab:tracers}. 
 
\begin{table*}
 \renewcommand{\arraystretch}{1.5}
  \centering
  \caption{Main properties of the three synthetic galaxy samples used in this work. The table shows the label we use to identify each sample, the N-body run on which it is based, the total number of independent realizations, redshift, galaxy number density, effective volume scaling ratio (see Section \ref{sec:measurements_P_cov}), linear galaxy bias and large-scale deviation from Poisson shot-noise (in units of $1/\overline{n}$). In order to obtain fiducial values for the last two columns, we adopt the same strategy described in \cite{EggScoCro2011} that makes use of the large-scale limit of the quantities $P_{\rm gg}$, $P_{\rm gm}$ and $P_{\rm mm}$.}
  
  \begin{ruledtabular}
    \begin{tabular}{cccccccc}
      Identifier & Simulation & $N_{\rm R}$ & $z$ &  $\overline{n}$ $\left[\left(h/\mathrm{Mpc}\right)^3\right]$ & $\eta$ & $b_1$ 
      & $N_0$ \Tstrut\Bstrut\\ \hline
      MGS & \textsc{LasDamas} Carmen & $40$ & $0.132$ & $1.1 \times 10^{-3}$ & $7.04$ & $1.414 \pm 0.003$ & $-0.16 \pm 0.07$ \Tstrut\\
      LOWZ & \textsc{LasDamas} Oriana & $40$ &$0.342$ & $9.4 \times 10^{-5}$ & $1$ & $2.235 \pm 0.012$ & $-0.176 \pm 0.018$ \\ 
      CMASS & \textsc{Minerva} & $100$ &$0.57$ &$4.0 \times 10^{-4}$ & $2.47$ & $2.022 \pm 0.003$ & $-0.29 \pm 0.02$ \Bstrut\\ 
      
    \end{tabular}
  \end{ruledtabular}
  \label{tab:tracers}
\end{table*}

\begin{table}
\renewcommand{\arraystretch}{1.5}
  \centering
  \caption{Cosmological parameters of the dark matter simulations used to generate the galaxy samples analysed in this work. Columns show the matter, dark energy and baryon density, Hubble constant, scalar index, and rms density fluctuations within a sphere of radius $8\,h^{-1}\,\mathrm{Mpc}$.}
  \begin{ruledtabular}
    \begin{tabular}{ccccccc}
      Simulation & $\Omega_m$ & $\Omega_{\Lambda}$ & $\Omega_b$ & $h$ & $n_s$ & $\sigma_8$ \Tstrut\Bstrut\\
      \hline
      \textsc{LasDamas} & 0.25 & 0.75 & 0.04 & 0.7 & 1.0 & 0.8 \Tstrut \\
      \textsc{Minerva} & 0.285 & 0.715 & 0.046 & 0.695 & 0.9632 & 0.828 \Bstrut\\
    \end{tabular}
  \end{ruledtabular}
  \label{tab:cosmology}
\end{table} 
 
The three catalogues were generated by populating dark matter halos with galaxies using Halo Occupation Distribution (HOD) prescriptions. Since the HOD parameters were calibrated to obtain number densities consistent with the ones of pre-existing real observations, for sake of easiness we label our simulated catalogues with the name of the corresponding data samples. Nevertheless, we remind the reader that here we make use of the full comoving snapshot volume, without considering any selection effect, as the goal is to investigate the range of validity of 1-loop galaxy bias, leaving aside the impact of observational systematics.

The CMASS sample is based on the \textsc{Minerva} simulations \cite{GriSanSal1604}, which consist in a set of 100 realizations of a $(1500\,h^{-1}\,\mathrm{Mpc})^3$ cubic box with periodic boundary conditions. The simulations were run using the public \textsc{Gadget} code \cite{Springel2001, Springel2005}, that regulated the motion of $1000^3$ dark matter particles within the aforementioned volume. Initial conditions were set up using the linear power spectrum obtained from CAMB \cite{Lewis2002} and displacing particles according to second-order Lagrangian Perturbation Theory (2LPT) \cite{CroPueSco0611}. The comoving snapshot has a redshift of $z=0.57$ and it is meant to reproduce the properties of the BOSS CMASS galaxy sample \cite{AlaAtaBai1709}.

The LOWZ $(z=0.342)$ and MGS $(z=0.132)$ samples are based on the Oriana and Carmen boxes of the \textsc{LasDamas} N-body simulations \cite{McBBerSco0901, SinBerMcB1807}. These are a set of 40 independent realizations with periodic boundary conditions, with a volume of $(2400\,h^{-1}\,\mathrm{Mpc})^3$ and $(1000\,h^{-1}\,\mathrm{Mpc})^3$, and a mass resolution of $4.6 \times 10^{11}\,h^{-1}\,M_{\odot}$ and $4.9 \times 10^{10}\,h^{-1}\,M_{\odot}$, respectively. Initial conditions were also set up using 2LPT, but in this case the initial power spectrum is computed using CMBFast \cite{SelZal96}. The HOD parameters of these synthetic catalogues were selected to reproduce the properties of the BOSS LOWZ and SDSS Main Galaxy Sample (MGS) at $M_r<-21$.

The complete set of cosmological parameters for \textsc{Minerva} and \textsc{LasDamas} is summarised in Table \ref{tab:cosmology}.

\subsection{Measurements of power spectra and their covariances}
\label{sec:measurements_P_cov}

We make use of the estimator described in \cite{SefCroSco1512}, based on fourth order particle assignment scheme and interlacing optimization to reduce the effect of large modes aliasing, to measure the galaxy auto power spectrum $P_{\rm gg}$ and the galaxy-matter cross power spectrum $P_{\rm gm}$ for all the tracers described in Table \ref{tab:tracers}. We adopt a linear $k$ binning from $k_{\rm min}=k_{f}$, where $k_{f}=2\pi/L$ is the fundamental frequency of the box of size $L$, to $k_{\rm max}=k_{\rm Nyq}$, where $k_{\rm Nyq}=\pi N_{\rm grid}/L$ is the Nyquist frequency corresponding to a FFT grid of size $N_{\rm grid}$. We use a linear binning with step $\Delta_k=k_f$ for CMASS and LOWZ, and $\Delta_k=2k_f$ for MGS. Since we model the stochastic contribution to the galaxy auto power spectrum in terms of deviations from Poisson shot-noise, we correct our raw measurements of $P_{\rm gg}$ by subtracting the constant factor $P^\mathrm{noise} = 1/\overline{n}$.

Our final data vector consists in the ensemble average over the number of independent realizations $N_{\rm R}$ (100 for CMASS, 40 for both LOWZ and MGS). First, we define the averaged observable 
\begin{equation}
  \overline{X}_i = \frac{1}{N_{\rm R}} \sum_{n=1}^{N_{\rm R}} X^{(n)}_i 
\end{equation}
where $X\in\{P_{\rm gg},\,P_{\rm gm}\}$, $i$ is the index running through the $k$ binning, and the superscript $n$ refers to the $n$-th realization of the considered observable. From this definition we estimate the auto- and cross- covariance matrices as
\begin{equation}
  C_{X \times Y,ij} = \frac{1}{N_{\rm R}}\sum_{n=1}^{N_{\rm R}} \left(X_i^{(n)}-\overline{X}_i\right)\,\left(Y_j^{(n)}-\overline{Y}_j\right)\,,
\end{equation}
where once again $X,\,Y\in\{P_{\rm gg},\,P_{\rm gm}\}$ . 

Similarly to what was done in \cite{EggScoCro2011} we retain only the diagonal entries of the previously defined covariance matrices. This approximation is justified given the significant low number density of the synthetic samples we consider (actual values are listed in Table \ref{tab:tracers}). These numbers translate in a significant shot-noise contribution to the power spectrum, whose main effect is to boost the diagonal entries of the covariance matrix, leading to sub-dominant off-diagonal terms. Therefore we approximate our covariance matrices with a block-diagonal shape, and consider only the auto- and cross-correlation at the same wavelength (i.e. $C_{ij}=0$ for $k_i\neq k_j$). Additionally, in order to reduce noise due to the limited number of independent realizations, we compare the raw covariance matrix to Gaussian predictions \cite{GriSanSal1604} for each $k$ bin, and retain the maximum of the two values.

The error budgets of our galaxy samples are subsequently rescaled in order to match the same \emph{effective} volume \cite{FelKaiPea9405, Teg9711}, defined as 
\begin{equation}
  V_{\mathrm{eff}}(k^\star) = \left[\frac{\overline{n}\,P_{\rm gg}(k^\star)}{1 + \overline{n}\,P_{\rm gg}(k^\star)}\right]^2\,V\,,
\end{equation}
where $\overline{n}$ and $V$ are the mean number density and volume of the considered sample, respectively, and we choose the reference scale to evaluate the effective volume as $k^\star = 0.1\,h\,\mathrm{Mpc}^{-1}$. In this way the constraining power and the signal-to-noise ratio of the three galaxy catalogues are artificially set to roughly match the same amplitude. We choose to rescale the covariance of both CMASS and MGS to match the effective volume of the LOWZ sample $\left(\approx 6\,(\mathrm{Gpc}/h)^3\right)$. Table \ref{tab:tracers}
 shows the rescaling factor for the three samples, that is simply defined as the sample-to-LOWZ ratio between the respective effective volumes.

\subsection{Fitting procedure and prior choices}
\label{sec:likelihood}
 
We make use of the large number of independent realizations for each synthetic galaxy sample to define the overall likelihood function as the product of the individual $N_{\rm R}$ likelihoods \cite{EggScoCro2011, OddSefPor2003}. In this way, under the assumption that the individual likelihoods are well described by a multivariate Gaussian distribution, we can define the overall likelihood function as
\begin{equation}
  \label{eq:data.likelihood}
  \begin{split}
    -2 \log{{\cal L}_{\mathrm{tot}}} &= -\frac{2}{N_{\rm R}} \sum_{n=1}^{N_{\rm R}} \log{{\cal L}_{(n)}} =
    \frac{1}{N_{\rm R}}\sum_{n=1}^{N_{\rm R}} \chi_{(n)}^2 \\ &\hspace{-1.5em}= \frac{1}{N_{\rm R}}\sum_{n=1}^{N_{\rm R}}
    \sum_{i,j=1}^{N_{\rm b}} \left(X_i^{(n)} - \mu_i\right)\,C_{X,ij}^{-1}\,\left(X_j^{(n)} -
      \mu_j\right)\,,
  \end{split}
\end{equation}
where $X_i^{(n)}$ is the measurement from the $n$-th realization of the $i$-th $k$ bin, $\mu_i$ is the corresponding model prediction, and $C_{ij}$ is the rescaled covariance matrix as described in Sec.~\ref{sec:measurements_P_cov}. With this definition, our likelihood ends up coinciding with the likelihood of the mean of the data with covariance $C_{ij}$. In practice, there is a constant factor between the two definitions, that depends on the number of independent realizations $N_{\rm R}$ and the number of data points $N_{\rm b}$. This factor is taken into account when deriving the goodness of fit for each of the tested configurations.

The inference of model parameters is carried out through a least-$\chi^2$ analysis based on a standard Metropolis-Hastings MCMC algorithm. The likelihood incorporates a complete recipe for one-loop galaxy clustering and an interface to \textsc{CAMB}. For each tested configuration of the MCMC, we first run some preliminary chains that are needed to obtain a robust estimate of the parameter covariance, which is essential for the good convergence of the Markov chain in a highly dimensional parameter space. We iterate this process twice, each time specifying the parameter covariance obtained at the previous step, before running the final set of chains. These are terminated as soon as the chains satisfy a standard Gelman-Rubin convergence criterium, i.e. as soon as the between-chain and within-chain variances (we run 8 independent chains for each case, changing the initial random seed) are in agreement within $R<1.1$ \cite{GelRub9201}. After removing the burn-in, these chains are combined into a single object, which we pass to the software package \texttt{getdist} \cite{Lew2019} to extract the parameter posteriors.

\begin{table}
 \renewcommand*{\arraystretch}{1.4}
  \centering
  \caption{Adopted priors for the full list of cosmological and nuisance parameters of the fits. We impose a uniform prior ($\mathrm{U}$) for all the sampling parameters. The scalar amplitude $A_{\rm s}$ is not sampled over when fitting only $P_{\rm gg}$ but it is when adding the additional constraint from $P_{\rm gm}$. In all cases the second-order tidal bias $\gamma_2$ is fixed to the excursion set relation defined in Equation (\ref{eq:g2_ex-set}). }
  \bigskip
  \begin{ruledtabular}
  \setlength{\tabcolsep}{15pt}
    \begin{tabular}{c|c|c}
    \multicolumn{3}{c}{\cellcolor{blue!25}COSMOLOGY}\\
    \hline
	 $h$ & \multicolumn{2}{c}{$\mathrm{U}[0.5,\,1]$}\\
	  $\Omega_{\rm c} h^2$ & \multicolumn{2}{c}{$\mathrm{U}[0.08,\,0.16]$}\\
	  \multirow{2}{*}{$10^9 A_{\rm s}$} & $P_{\rm gg}:$& $P_{\rm gg}+P_{\rm gm}:$\\
	  & Fixed to fiducial & $\mathrm{U}[0.7,\,3.3]$\\
	  \hline
	  \multicolumn{3}{c}{\cellcolor{blue!25}BIAS}\\
	  \hline
	  $b_1 \sigma_8$ & \multicolumn{2}{c}{$\mathrm{U}[0.25,\,4]$}\\
	  $b_2 \sigma_8^2$ & \multicolumn{2}{c}{$\mathrm{U}[-10,\,10]$}\\
	  $\gamma_2$ &  \multicolumn{2}{c}{Fixed to $\gamma_{2,ex}(b_1)$}\\
	  $\gamma_{21} \sigma_8^3$ & \multicolumn{2}{c}{$\mathrm{U}[-8,\,8]$}\\
	  $\beta_P\,[k_\mathrm{HD}^{-2}]$ & \multicolumn{2}{c}{$\mathrm{U}[-10,\,10]$}\\
	  $\beta_P^\times\,[k_\mathrm{HD}^{-2}]$ & \multicolumn{2}{c}{$\mathrm{U}[-10,\,10]$}\\
	  \hline
	  \multicolumn{3}{c}{\cellcolor{blue!25}SHOT-NOISE}\\
	  \hline
	  $N_0\,[\overline{n}^{-1}]$ & \multicolumn{2}{c}{$\mathrm{U}[-1,\,0.5]$}\\
	  $N_2\,[k_\mathrm{HD}^{-2}\,\overline{n}^{-1}]$ & \multicolumn{2}{c}{$\mathrm{U}[-10,\,10]$}\\
	  $N_2^\times\,[k_\mathrm{HD}^{-2}\,\overline{n}^{-1}]$ & \multicolumn{2}{c}{$\mathrm{U}[-10,\,10]$}\\
	  \hline
	  \multicolumn{3}{c}{\cellcolor{blue!25}COUNTER-TERMS}\\
	  \hline
	  $c_1$ & \multicolumn{2}{c}{$\mathrm{U}[0,\,10]$}\\     
    \end{tabular}
  \end{ruledtabular}
  \label{tab:priors}
\end{table}

The models described in Section \ref{sec:theory} embed a large number of free parameters. In this analysis we vary both cosmological and nuisance terms (the latter include bias, shot-noise and matter counter-term). Since we are constraining two-point statistics only via galaxy clustering, we fix both the baryon density parameter $\Omega_{\rm b} h^2$ and the primordial spectral index $n_{\rm s}$ to their fiducial values (listed in Table \ref{tab:cosmology}), and vary the CDM density parameter $\Omega_{\rm c} h^2$ and the Hubble constant $h$. We add the primordial scalar amplitude $A_{\rm s}$, but only when fitting the combination of $P_{\rm gg}$ and $P_{\rm gm}$, since the fit of the auto galaxy power spectrum alone cannot break the strong degeneracy between $A_{\rm s}$ and $b_1$. On the contrary, while both $P_{\rm gg}$ and $P_{\rm gm}$ have the same dependency on $A_{\rm s}$, they scale differently with the linear bias, i.e. $P_{\rm gg}\propto b_1^2$ and $P_{\rm gm}\propto b_1$ in the linear limit. We choose a flat prior for the three cosmological parameters that is large enough to fully contain their posterior distribution up to $2\sigma$ even for the least constraining configuration.

The same choice of uniform priors is chosen for the nuisance parameters. We sample the combination of the bias parameters with the corresponding power of $\sigma_8$ in order to avoid strong degeneracies between the one-loop bias expansion and the intrinsic non-linearity of the matter power spectrum $P_{mm}$, that at first order is well captured by $\sigma_8$. As anticipated in Section \ref{sec:co-evolution}, we fix $\gamma_2$ to Equation (\ref{eq:g2_ex-set}), and vary all other bias parameters freely. We quote higher-derivatives parameters with respect to an arbitrary fiducial scale $k_\mathrm{HD}=0.4\,h\,\mathrm{Mpc}^{-1}$, while all the shot-noise parameters can be naturally expressed in units of $1/\overline{n}$. Since terms involving $N_2$ and $N_2^\times$ both carry a $k^2$-dependency, we also express the latter in units of $k_\mathrm{HD}$.

The full list of priors for the model parameters is shown in Table \ref{tab:priors}.

\subsection{Rescaling of the input linear power spectrum}
\label{sec:input_power_spectrum}

\begin{figure}
  \centering
  \includegraphics[width=.495\textwidth]{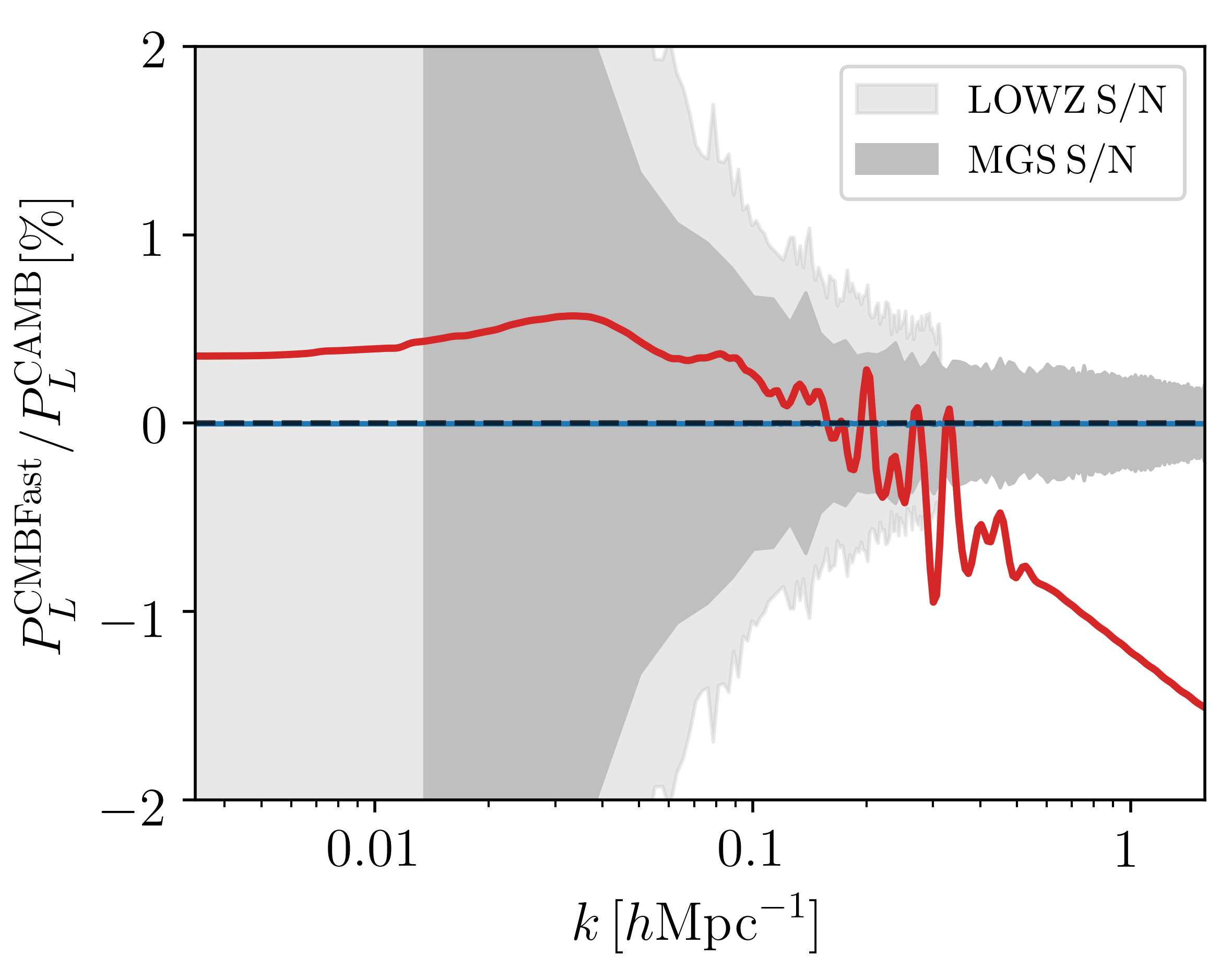}
  \caption{Ratio between the CMBFast and CAMB linear prediction at the reference cosmology of the \textsc{LasDamas} simulations. The two shaded grey bands mark the intrinsic $1\mbox{-}$sigma error of the LOWZ and MGS galaxy power spectrum $P_{\rm gg}$.}
  \label{fig:CMBFast_to_CAMB_ratio}
\end{figure}

As discussed in Section \ref{sec:data}, \textsc{Minerva} and \textsc{LasDamas} adopt different Boltzmann solvers to obtain the power spectra for the initial particle displacements. In particular, the \textsc{Minerva} runs assume initial conditions based on CAMB, while the \textsc{LasDamas} simulations make use of the power spectrum computed with \textsc{CMBFast}. Figure \ref{fig:CMBFast_to_CAMB_ratio} shows the fractional deviation between the output of the two codes at the reference cosmology of the \textsc{LasDamas} runs and at the redshift of the MGS sample. This difference is compared to LOWZ and MGS intrinsic signal-to-noise ratio, represented by the two shaded areas. The two spectra deviate at the level of $\sim0.5\%$ in the range of scales that are relevant to this analysis, and this might represent an issue when modeling the galaxy-galaxy and galaxy-matter power spectra. For this reason, when fitting LOWZ and MGS, we employ a rescaling scheme based on the renormalization of the input power spectrum of \textsc{CAMB} by the ratio shown in Figure \ref{fig:CMBFast_to_CAMB_ratio}. This approximation is valid as long as the sampled cosmology is not far from the fiducial one, and this becomes a completely legitimate assumption when hitting exactly the true cosmology. However, we claim that even for cosmologies that are slightly off with respect to the fiducial one, this rescaling scheme is already better than directly using the \textsc{CAMB} prediction with no further correction.

In order to perform this rescaling, we first apply the units transformation $h\,\mathrm{Mpc}^{-1}\rightarrow \mathrm{Mpc}^{-1}$. This is meant to assure that the power spectrum ratio is obtained at the same set of scales, avoiding the dependency on the Hubble parameter $h$ \cite{Sanchez2012}. At each step of the Markov chain, the current CAMB prediction is firstly transformed into $\mathrm{Mpc}^{-1}$ units, rescaled using the approach described above, and finally transformed back into $h\,\mathrm{Mpc}^{-1}$ units.

\subsection{Performance metrics}
\label{sec:metrics}

The validation of one-loop galaxy bias as described in Section \ref{sec:theory} is mostly based on its range of validity. In order to select the maximum mode at which our model stops providing good performances, we run multiple MCMC chains varying $k_{\rm max}$ in the interval $(0.1,\,0.3)\,h\,\mathrm{Mpc}^{-1}$ with a step of $0.025\,h\,\mathrm{Mpc}^{-1}$, leading to 9 different fitting ranges for each configuration. We stop at $k_{\rm max}=0.3\,h\,\mathrm{Mpc}^{-1}$ as pushing the model to even smaller scales would inevitably require accounting for two-loop contributions in the description of galaxy bias and non-linear matter evolution. Clearly, exploiting the small-scale information contained in the non-linear regime naturally provides a better constraining power for the model parameters, but at the cost of introducing theory systematics in their posterior distribution. As described in \cite{EggScoCro2011} (see \cite{OsaNisBer1903} for a similar approach), we make use of a set of three different performance metrics, whose goal is to provide a quantitative way to compare the posteriors extracted from our Markov chains, and to determine when a particular model no longer gives an accurate description of our datasets.

\subsubsection{Figure of bias}
\label{sec:fob}

The first quantity we are interested in evaluating is the ability of the model to provide unbiased measurements of its free parameters. In particular, in this work we focus mostly on the systematic error of the cosmological parameters: the Hubble constant $h$, the cold dark matter density $\Omega_{\rm c} h^2$, and, when considering the joint fit between $P_{\rm gg}$ and $P_{\rm gm}$, the primordial scalar amplitude $A_{\rm s}$.

We define the \textit{Figure of Bias} (FoB) of the considered model for a given parameter set \mbox{\boldmath$\theta$} as
\begin{equation}
 \mathrm{FoB}(\B{\theta}) \equiv \left[\left(\,\overline{\B{\theta}}-\B{\theta}_\mathrm{fid}\right)^\intercal {\bold S}^{-1} \left(\,\overline{\B{\theta}}-\B{\theta}_\mathrm{fid}\right)\right]^{1/2}, 
  \label{eq:figure_of_bias}
\end{equation}
where $\overline{\B{\theta}}$ represents the mean of the posterior distribution, $\B{\theta}_\mathrm{fid}$ represents the fiducial position, and ${\bold S}$ is the parameter covariance expressed in matrix form. In this way, we are simply quantifying the relative separation of the measured parameter from its true value in terms of the variance of the posterior distribution. If the FoB is evaluated only for one parameter, then the standard $68\% -95\%$ percentile thresholds correspond to a FoB of $1 -2$, respectively. On the contrary, when more than one parameter is considered to compute the FoB, the two percentiles can be calculated by direct integration of a multivariate Gaussian distribution with the corresponding dimensionality. For the case we consider $n=2\, (n=3)$ it follows that the $68\%-95\%$ percentiles correspond to a FoB of $1.52-2.49\, (1.88-2.83)$, respectively.

\subsubsection{Goodness-of-fit}
\label{sec:goodness-if-fit}

Along with finding unbiased constraints on the parameters of interest, we also ask our model to provide a good description of the observables we use in the fit. We quantify the \textit{goodness-of-fit} in terms of the standard $\chi^2$ extracted from the Markov chains (Equation (\ref{eq:data.likelihood})), but after rescaling it to account for the additional factor $\eta$ introduced in the covariance of the data (see Section \ref{sec:measurements_P_cov}). With this rescaling, we can compare the recovered $\chi^2$ to the $68\%$ and $95\%$ percentiles of a $\chi^2$ distribution with $N_\mathrm{dof}$ degrees of freedom, where
\begin{equation}
N_\mathrm{dof} = N_{\rm R} \times N_{\rm b} - N_{\rm p}\,.
\label{eq:numbers_degrees_of_freedom}
\end{equation}   
Here $N_{\rm R}$ is the number of independent realizations for each galaxy sample, $N_{\rm b}$ is the total number of $k$ bins included in the fit, and $N_{\rm p}$ is the number of free model parameters.

\subsubsection{Figure of merit}
\label{sec:fom}

The final metric we employ is based on the merit of the considered model with respect to the cosmological parameters that are varied in the fit. We define the \textit{Figure of Merit} (FoM) of a given model for a subset of parameters $\B{\theta}$ as the inverse of the determinant of the parameter covariance matrix corresponding to the subset $\B{\theta}$, i.e.
\begin{equation}
\mathrm{FoM}(\B{\theta}) \equiv  \frac{1}{\sqrt{\mathrm{det}\left[\,\overline{\bold S}(\B{\theta})\right]}}\,.
\end{equation}
Here $\overline{\bold S}(\B{\theta})\equiv {\bold S}(\B{\theta})/\B{\theta}^2_\mathrm{fid}$ is the block corresponding to the parameter subset $\B{\theta}$ rescaled by their fiducial values. With this last normalization we are explicitly asking for a \textit{relative} FoM rather than an absolute quantity, that would be more difficult to compare to the one of other parameters.

\section{Testing one-loop galaxy bias} 
\label{sec:results}

\begin{table}
\renewcommand{\arraystretch}{1.5}
  \centering
  \caption{List of all the considered combinations between the modeling of $P_{\rm mm}$ and the extensions to one-loop galaxy bias, as described in Section \ref{sec:theory}. For each model, we denote with a check mark all the free parameters of the $P_{\rm gg}$ fit, while we use a double check mark for the free parameters that are exclusive to the joint $P_{\rm gg}+P_{\rm gm}$ fit.}
  \begin{ruledtabular}
    \begin{tabular}{c|c|c|c|c|c|c|c|c|}
      $P_{mm}$ & \multicolumn{3}{c|}{\texttt{RESPRESSO}}& \multicolumn{3}{c|}{gRPT} & \multicolumn{2}{c|}{EFT}\\
      \hline
      Bias model & STD & k2N & HD & STD & k2N & HD & STD & k2N\\
      \hline
      $h$ & \checkmark & \checkmark & \checkmark & \checkmark & \checkmark & \checkmark & \checkmark & \checkmark\\
      $\Omega_{\rm c} h^2$ & \checkmark & \checkmark & \checkmark & \checkmark & \checkmark & \checkmark & \checkmark & \checkmark\\
      $A_{\rm s}$ & \checkmark\checkmark & \checkmark\checkmark & \checkmark\checkmark & \checkmark\checkmark & \checkmark\checkmark & \checkmark\checkmark & \checkmark\checkmark & \checkmark\checkmark\\
      $b_1\sigma_8$ & \checkmark & \checkmark & \checkmark & \checkmark & \checkmark & \checkmark & \checkmark & \checkmark\\
      $b_2\sigma_8^2$ & \checkmark & \checkmark & \checkmark & \checkmark & \checkmark & \checkmark & \checkmark & \checkmark\\
      $\gamma_{21}\sigma_8^3$ & \checkmark & \checkmark & \checkmark & \checkmark & \checkmark & \checkmark & \checkmark & \checkmark\\
	  $\beta_P$ &  &  & \checkmark &  &  & \checkmark &  & \\
      $\beta_P^\times$ &  &  & \checkmark\checkmark &  &  & \checkmark\checkmark &  & \\
      $N_0$ & \checkmark & \checkmark & \checkmark & \checkmark & \checkmark & \checkmark & \checkmark & \checkmark\\
      $N_2$ &  & \checkmark &  &  & \checkmark &  &  & \checkmark \\
      $N_2^\times$ &  & \checkmark\checkmark &  &  & \checkmark\checkmark &  &  & \checkmark\checkmark \\
      $c_1$ &  &  &  &  &  &  & \checkmark & \checkmark \\
      
    \end{tabular}
  \end{ruledtabular}
  \label{tab:models}
\end{table} 

In this section we analyse parameter posteriors extracted from several Markov chains, where we vary both the maximum fitting scale $k_{\rm max}$ and the final expression we adopt to model one-loop galaxy power spectra. The complete list of model configurations is summarised in Table \ref{tab:models}, along with the total number of  free parameters in each case. Based on results from \cite{EggScoCro2011}, we decide to employ \texttt{RESPRESSO} to model $P_{\rm mm}$ in this section, and to leave the comparison between different matter models to Section \ref{sec:results_models}. In this way we can focus exclusively on assessing the validity of one-loop bias in real-space leaving aside both the impact of non-linear matter evolution and redshift-space distortions (that will be the topic of a future paper). 

For each configuration, we compute the three performance metrics defined in Section \ref{sec:metrics} and show their trends as a function of $k_{\rm max}$. At the same time, we perform sanity checks on the linear bias $b_1\sigma_8$ and the large-scale shot-noise parameter $N_0$, for which we have fiducial values to compare with (see Table \ref{tab:tracers}). As anticipated in Section \ref{sec:co-evolution}, throughout the rest of the paper we fix the second-order non-local bias parameter $\gamma_2$ to the excursion-set relation (Equation (\ref{eq:g2_ex-set})), in order to break the strong degeneracy between the former and $\gamma_{21}$. This choice is also based on results from \cite{EggScoCro2011}, where it was shown how this choice led to stable and overall accurate measurement of the linear bias.

In the next two sections we analyse the posterior derived from fits of the auto galaxy power spectrum $P_{\rm gg}$ alone, and from the combination of the auto and cross galaxy power spectra, $P_{\rm gg}+P_{\rm gm}$. In the latter case, the additional information contained in $P_{\rm gm}$ allows to break the degeneracy between the linear bias and the amplitude of primordial fluctuations, allowing us to also sample over $A_{\rm s}$.

In both cases, we adopt a criterion for defining the range of validity of a given model, which is based on a combination of FoB and goodness-of-fit. Following \cite{EggScoCro2011}, we define the model-breaking statistics as
\begin{equation}
\sigma_\mathrm{MB}(k) \equiv \frac{\mathrm{FoB}(k)}{\mathrm{FoB}_{68\%}}+\frac{\chi^2(k)-1}{\chi^2_{95\%}(k)-1},
\label{eq:model_breaking}
\end{equation}
where the subscript percentages correspond to the percentiles of the corresponding distribution (FoB and $\chi^2$). We say that a given model breaks down at a scale $k^\dag$ when the model-breaking statistics assumes a critical value $\sigma_\mathrm{crit}$ at $k^\dag$, i.e.
\begin{equation}
\sigma_\mathrm{MB}(k^\dag) = \sigma_\mathrm{crit}\,.
\end{equation}
We arbitrary choose the critical threshold $\sigma_\mathrm{crit}=1.5$. In this way we would accept models with a maximum FoB one and a half times larger than the value corresponding to the $1$-sigma of the FoB distribution, but only under a perfect recovery of the shape of the input dataset. Practically, this case is unrealistic at high $k_{\rm max}$, as the $\chi^2$ progressively deviates from the number of degrees of freedom because of the weakening of the model. For this reason $\sigma_\mathrm{MB}$ receives individual contributions from the FoB and the goodness-of-fit.

\subsection{Validity of one-loop galaxy bias for the auto power spectrum}
\label{sec:validity-auto-power}

\begin{figure*}
  \centering
  \includegraphics[width=\textwidth]{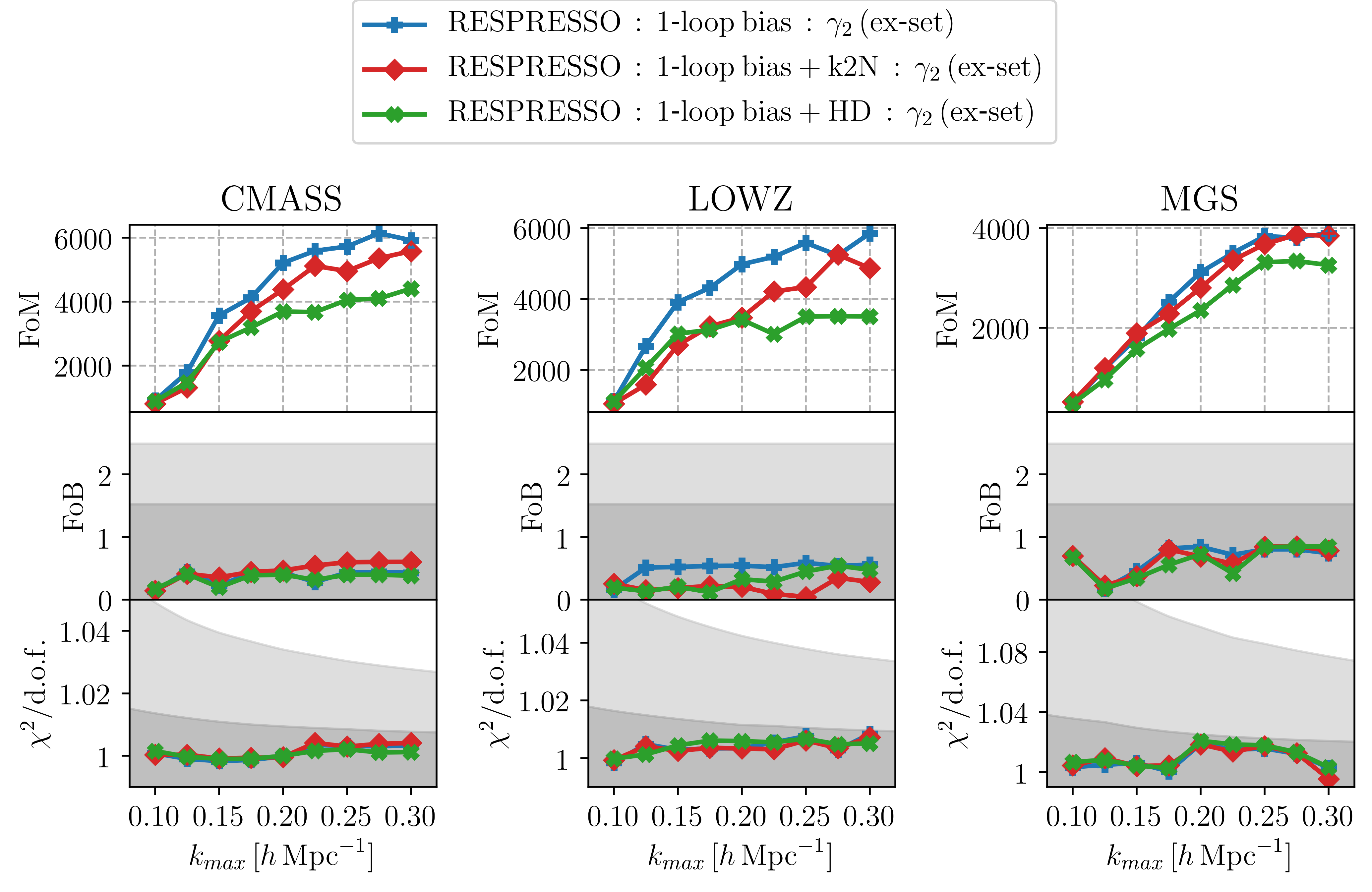}
  \caption{FoM, FoB and goodness-of-fit extracted from chains corresponding to the fits of the galaxy auto power spectrum $P_{\rm gg}$. Performance metrics refer to the combinations of $h$ and $\Omega_{\rm c} h^2$, which are the only two free cosmological parameters of the model. The standard (STD), $k^2$-dependent shot-noise (k2N), and higher derivatives (HD) models are colour-coded as shown in the legend. In all three cases, the non-linear matter power spectrum is modelled with \texttt{RESPRESSO}, and the second-order tidal bias $\gamma_2$ is fixed to Equation (\ref{eq:g2_ex-set}). The two shaded grey regions mark the $68$th and $95$th percentiles of the FoB and $\chi^2$ distribution.}
  \label{fig:fcf_Pgg_respresso}
\end{figure*}

In this section we analyse results from the fits to the auto power spectrum $P_{\rm gg}$. We fix the description of the non-linear matter power spectrum and test different extensions of one-loop galaxy bias. The standard (STD) model is based on two free cosmological parameters $(h,\,\Omega_{\rm c} h^2)$ plus four free nuisance parameters $(b_1\sigma_8,\,b_2\sigma_8^2,\,\gamma_{21}\sigma_8^3,\,N_0)$. The $k^2$-dependent shot-noise (k2N) and higher-derivatives (HD) extensions have one additional free parameter each, i.e. $N_2$ and $\beta_P$, respectively.   

In Figure \ref{fig:fcf_Pgg_respresso} we show the three performance metrics extracted from the three different $P_{\rm gg}$ measurements (CMASS, LOWZ and MGS), plotted against the maximum scale $k_{\rm max}$ included in the fit. In this case we show the FoB and FoM corresponding to the combination $h,\,\Omega_{\rm c} h^2)$, i.e. to all the cosmological parameters that are varied in the model. The two grey shaded areas in the panels corresponding to FoB and goodness-of-fit mark the $68$th and $95$th percentiles of the corresponding quantity. As described in Section \ref{sec:fob}, the $68\%-95\%$ FoB limits do not correspond to values of FoB $=1$ and FoB $=2$, but rather have to be computed integrating a multivariate Gaussian distribution with $n=2$ number of dimensions. 

For the three galaxy samples, we find that the STD model (blue line) alone is enough to provide a good description of the dataset in terms of goodness-of-fit, as the reduced $\chi^2$ constantly stays within the $68\%$ (dark grey shaded area) of the corresponding $\chi^2$ distribution. At the same time, the combined constraint on the $(h,\,\Omega_{\rm c} h^2)$ pair is unbiased, showing a multivariate posterior distribution that is able to capture the fiducial position at typically $\sim0.5\,\sigma$. Adding either an additional $k^2$-dependent shot-noise (red line) or the leading higher derivatives correction (green line) does not significantly alter the two metrics, while reducing the amplitude of the corresponding FoM ($\sim10\%$ when adding the stochastic contribution $N_2$, $\sim20-30\%$ when adding the higher derivatives parameter $\beta_P$). Even for the STD model, for which we expect the model to fail earlier than the other two extensions) the model-breaking scale $k^\dag$ is higher than the maximum scale we probe $(k_{\rm max}=0.3\,h\,\mathrm{Mpc}^{-1})$. However, we decide to stop at this scale without exploring smaller scales, even if the model-breaking criterion were most likely satisfied at higher $k_{\rm max}$. As a matter of fact it would be hard to separate the effective goodness of the model from the impact of two-loop contributions to galaxy bias.

\begin{figure*}
\includegraphics[width=\textwidth]{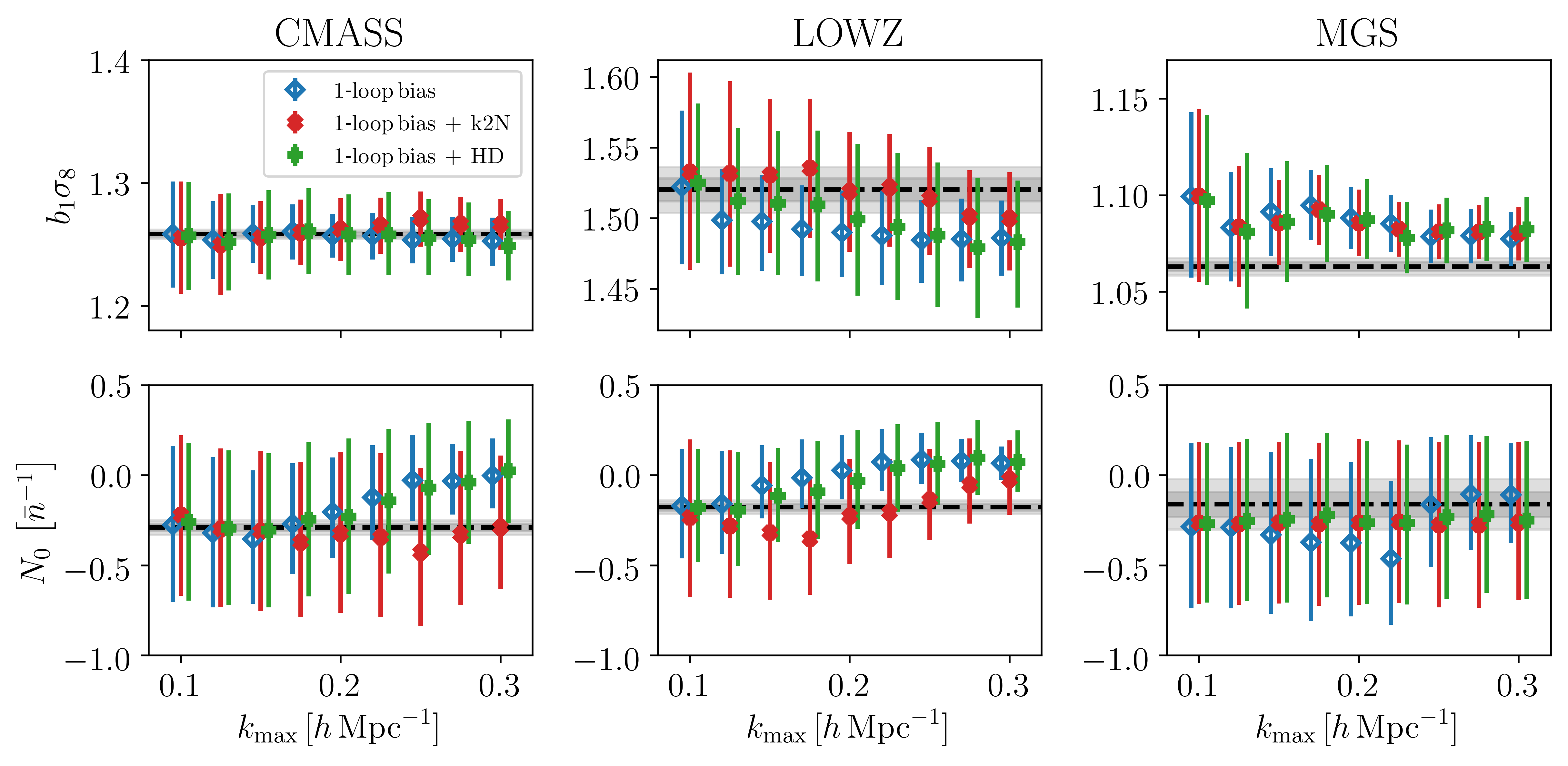}
\caption{Marginalised constraints on the linear bias $b_1\sigma_8$ (top) and the constant stochastic term $N_0$ (bottom) as a function of $k_\mathrm{max}$, for the fits of the galaxy auto power spectrum $P_{\rm gg}$. In all cases dark matter non-linear evolution is modelled using \texttt{RESPRESSO}. Fiducial values of the linear bias and constant shot-noise contribution are reproduced with black dashed lines. Uncertainties on the fiducial values are marked with shaded grey bands (marking $1$-and $2$-$\sigma$ confidence intervals). Fiducial measurements are listed in Table \ref{tab:tracers}.}
\label{fig:b1_N0_vs_kmax_auto}
\end{figure*}

We notice how, similarly to what was obtained in \cite{EggScoCro2011} when focusing on the linear galaxy bias, there is a tendency of the FoM to flatten above an approximate scale of $k_{\rm max}=0.25\,h\,\mathrm{Mpc}^{-1}$. This is an indication that pushing the non-linear model to increasingly higher wave modes does not necessarily imply a more stringent measurement of the cosmological parameters, as most of the additional constraining power is absorbed by nuisance parameters such as higher-order galaxy bias contributions.

In Figure \ref{fig:b1_N0_vs_kmax_auto} we show the dependence on $k_{\rm max}$ of the linear bias parameter $b_1\sigma_8$ and the constant stochastic contribution $N_0$, for which we have fiducial values obtained exploiting the large-scale limit of the measured galaxy and matter power spectra. Although the extension to either the k2N or the HD model enlarges the size of the errorbars, we notice how the former model is the only one capable of providing a simultaneous unbiased measurements of the two parameters, for the three galaxy samples we are considering. In particular the LOWZ sample shows a clear detection of $N_2$, that, if not accounted for, can lead to a significant $(>2$-$\sigma)$ systematic effect on $N_0$ above $k_{\rm max}\sim0.25\,h\,\mathrm{Mpc}^{-1}$, while partially recovering the true amplitude of $P_{\rm gg}$ with an underestimation of $b_1\sigma_8$. The only exception is represented by the MGS sample, whose linear bias is constantly overestimated by a $\sim2\%$ factor for all the three modeling assumptions, and that is consistent with the fiducial value only up to $2$-$\sigma$.

\subsection{Consistency between auto and cross power spectra}
\label{sec:consistency-ax}

\begin{figure*}
  \centering
  \includegraphics[width=\textwidth]{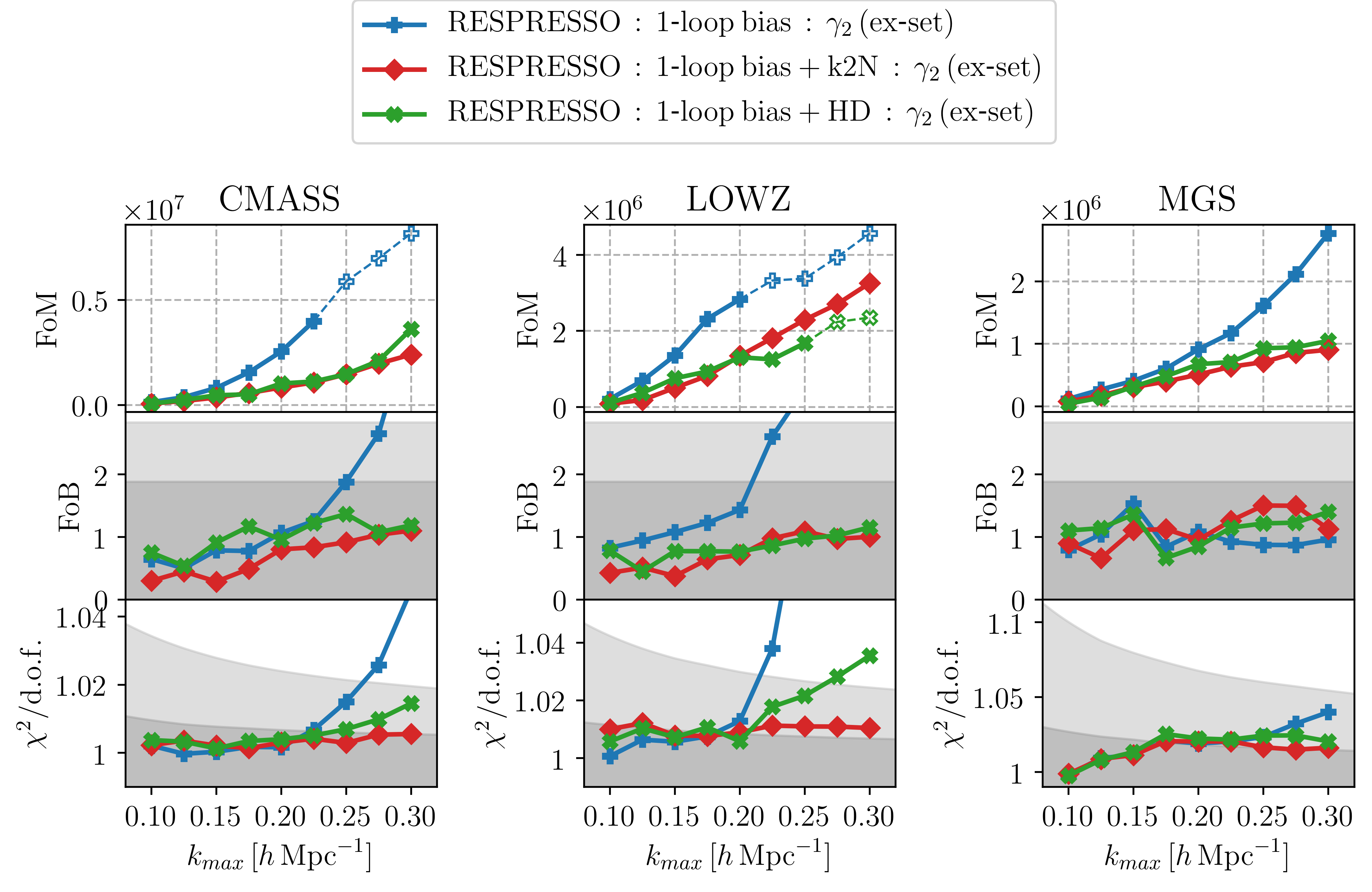}
  \caption{Same as Figure \ref{fig:fcf_Pgg_respresso} but for the combinations of the galaxy auto and cross power spectrum, $P_{\rm gg}$ and $P_{\rm gm}$. In this case, performance metrics (FoM and FoB) refer to the combinations of the three cosmological parameters $(h,\, \Omega_{\rm c} h^2,\,A_{\rm s})$. The standard (STD), $k^2$-dependent shot-noise (k2N), and higher derivatives (HD) models are colour-coded as shown in the legend. The solid-to-dashed transition mark the model-breaking scale $k^\dag$. In all three cases, the non-linear matter power spectrum is modelled with \texttt{RESPRESSO}. The two shaded grey regions mark the $68$th and $95$th percentiles of the FoB and $\chi^2$ distribution.}
  \label{fig:fcf_Pgg+Pgm_respresso}
\end{figure*}

In this section we focus on a much more stringent test, corresponding to the simultaneous fit of  both the galaxy auto power spectrum $P_{\rm gg}$ and the galaxy-matter cross power spectrum $P_{\rm gm}$. As a matter of fact, a good accuracy for this combined statistics might become crucial in analyses that aim to exploit the whole information contained in galaxy clustering and galaxy-galaxy weak lensing, as most of the upcoming large observational projects are going to do ($3\times2$-point analysis, see e.g. \cite{Abbott2018, heymans2020, Pandey2020}).

The different scaling of $P_{\rm gg}$ and $P_{\rm gm}$ on the linear bias $b_1$ leads to the breaking of the strong $b_1-A_{\rm s}$ degeneracy. Therefore, we  additionally sample the scalar amplitude $A_{\rm s}$, effectively extending by one the number of degrees of freedom of our model. Also in this case, we concentrate on the one-loop biasing scheme, fixing the description of the matter clustering to the output of \texttt{RESPRESSO}. We test the standard (STD) model against the $k^2$-dependent shot-noise (k2N) and higher derivatives (HD) templates, for which we introduce two more degrees of freedom (one free parameter for $P_{\rm gg}$ and $P_{\rm gm}$, each). In summary, the STD model is based on three free cosmological parameters $(h\,,\Omega_{\rm c} h^2,\,A_{\rm s}$) plus four free nuisance parameters $(b_1\sigma_8,\, b_2\sigma_8^2,\, \gamma_{21}\sigma_8^3\,,N_0)$. The k2N and HD models have two additional free parameters each, i.e. $(N_2,\, N_2^\times)$ and $(\beta_P,\,\beta_P^\times)$, respectively. Note that that the second-order tidal bias $\gamma_2$ is not a free parameter, but it is fixed to the excursion-set relation defined in Equation (\ref{eq:g2_ex-set}) for all the cases we consider.

In Figure \ref{fig:fcf_Pgg+Pgm_respresso} we show the three performance metrics in a similar way as we did for the $P_{\rm gg}$-only fits. In this case both FoM and FoB are computed from the combinations of the three cosmological parameters $(h,\,\Omega_{\rm c} h^2,\,A_{\rm s})$. We first notice how the use of the STD model is no longer sufficient to provide an accurate recovery of the cosmological parameters on the overall range of scales we are considering. Indeed, the FoB of both CMASS and LOWZ quickly increases above $2$-$\sigma$ at an approximate scale of $k_{\rm max}\sim0.23\,h\,\mathrm{Mpc}^{-1}$ and keeps getting worse at higher wave modes. The same trend is observed in panels referring to the goodness-of-fit, as the reduced $\chi^2$ starts to fall outside the $95\%$ confidence region at approximately the same scale cut. According to the model-breaking criterion, the range of validity of the STD model is limited to scales below $k\sim0.2\,h\,\mathrm{Mpc}^{-1}$ (this is reflected in the plot by the solid-to-dashed line transition).

Adding $k^2$-dependent stochasticity increases both the accuracy in the recovery of the cosmological parameters and the overall likelihood between data vectors and best-fitting model. Indeed, the k2N model shows a FoB which is constantly within $1$-$\sigma$ for the three galaxy samples. Moreover, the goodness-of-fit is significantly improved with respect to the STD model, with the $\chi^2$ measured at the highest $k_{\rm max}$ still being consistent within $1$-$\sigma$. The combination of FoB and goodness-of-fit can be observed via the model-breaking criterion on the FoM, for which the k2N model is the only one managing to be accepted up to $k_{\rm max}=0.3\,h\,\mathrm{Mpc}^{-1}$.

The HD model also provides a better recovery (compared to STD) of the cosmological parameters, within $1$-$\sigma$ from the fiducial position, but it fails earlier than the k2N model to provide the required goodness-of-fit for both CMASS and LOWZ, hinting for a stronger necessity of adding $k^2$-dependent stochasticity with respect to short-range non-localities.

\begin{figure*}
\includegraphics[width=\textwidth]{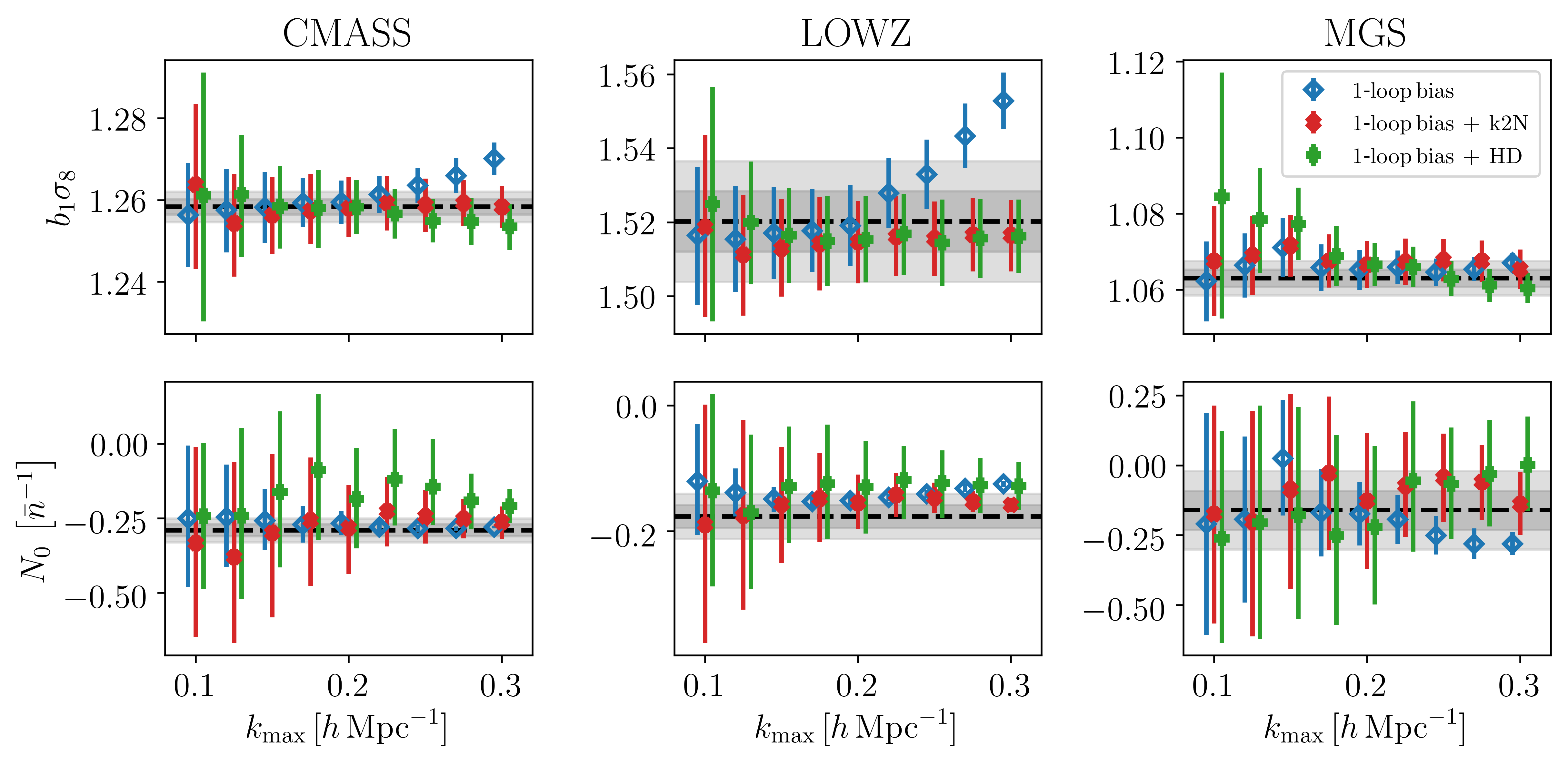}
\caption{Same as Figure \ref{fig:b1_N0_vs_kmax_auto} but for the combined fits of the galaxy auto power spectrum $P_{\rm gg}$ and the galaxy-matter cross power spectrum $P_{\rm gm}$. In all cases dark matter non-linear evolution is modelled using \texttt{RESPRESSO}. Fiducial values of the linear bias and constant shot-noise contribution are reproduced with black dashed lines. Uncertainties on the fiducial values are marked with shaded grey bands (marking $1\mbox{-}$ and $2\mbox{-}\sigma$ confidence interval).}
\label{fig:b1_N0_vs_kmax_cross}
\end{figure*}

In terms of FoM amplitude, the k2N and HD models behave in a mostly similar way, significantly reducing the statistical constraints of the cosmological parameters at any given $k_{\rm max}$ cut. However, we notice how according to the model-breaking criterion both models allow to push the analysis to higher $k_{\rm max}$ with respect to the STD case. In other words, a proper comparison of the merit of the three models should be done considering the individual scale $k^\dag$ after which the model breaks down. Considering the CMASS sample, the STD model provides a FoM consistent with the one obtained from the HD case, and $\sim30\%$ larger than the one of the k2N model. However, the latter two models have yet to reach the maximum scale $k^\dag$ corresponding to the breaking of the model. A simple exercise of extending the k2N model to slightly larger modes values shows that the latter turns out to be competitive against the STD model at a scale of $k_{\rm max}\sim0.35\,h\,\mathrm{Mpc}^{-1}$. As for LOWZ, the HD case is significantly less constraining than the STD model ($\sim50\%$), mostly because its range of validity does not run all the way up to $k_{\rm max}=0.30\,h\,\mathrm{Mpc}^{-1}$ but stops earlier. In contrast, the k2N model outperforms the standard model by a $\sim10\%$ factor, although also in this case the model breaking scale $k^\dag$ has yet to be reached.

A particular case is represented by MGS, that differently from the previous two samples shows a stable and accurate recovery of cosmological parameters all the way up to $k_{\rm max}=0.3\,h\,\mathrm{Mpc}^{-1}$ even with the STD model. The reason is most likely related to our choice of employing \texttt{RESPRESSO} to model the non-linear matter evolution. This ultimately provides an accurate description of $P_{\rm mm}$ even at the low redshift of MGS, leaving only the galaxy-matter relation to be modelled. Since MGS features the lowest large-scale bias among the samples we consider, it is likely that the one-loop standard expansion is enough to provide both a good $\chi^2$ and an accurate parameter recovery on the full range of scales up to $k_{\rm max}=0.3\,h\,\mathrm{Mpc}^{-1}$.

Differently from the fit of the galaxy auto-power spectrum, in this case the FoM for the three different galaxy samples do not show a flattening on the range of scales we are considering. Checking the independent contributions coming from the individual cosmological parameters, we notice how the FoM does actually flatten for both $h$ and $A_{\rm s}$, whereas it monotonically increases for $\Omega_{\rm c} h^2$. This is a hint showing that the matter density parameter still benefits of additional informations contained in the mildly non-linear regime, when combining the two galaxy power spectra.

In Figure \ref{fig:b1_N0_vs_kmax_cross} we show the dependence of the linear bias $b_1\sigma_8$ and the constant shot-noise contribution on $k_{\rm max}$, similarly to what we showed in Figure \ref{fig:b1_N0_vs_kmax_auto} for $P_{\rm gg}$ alone. The more stringent requirement of simultaneously fitting $P_{\rm gg}$ and $P_{\rm gm}$ makes the standard model fail in providing accurate measurements of both parameters. In particular, $b_1\sigma_8$ starts to be biased at approximately the same scale for which the model also provides incorrect measurements of the cosmological parameters. Adding either $k^2$-dependent shot noise terms or higher derivatives parameters help in recovering the true values of the bias, with a slight preference for the former. More generally the k2N model is the only one that is capable of simultaneously providing unbiased measurements of the cosmological parameters, the linear bias and the large-scale stochastic contribution to the galaxy power spectrum. This is also in agreement with the results obtained in \cite{EggScoCro2011} at fixed cosmology.

\subsection{Constraints on stochasticity and higher-derivative parameters}
\label{sec:noise-HD-constraints}

\begin{figure*}
\includegraphics[width=\textwidth]{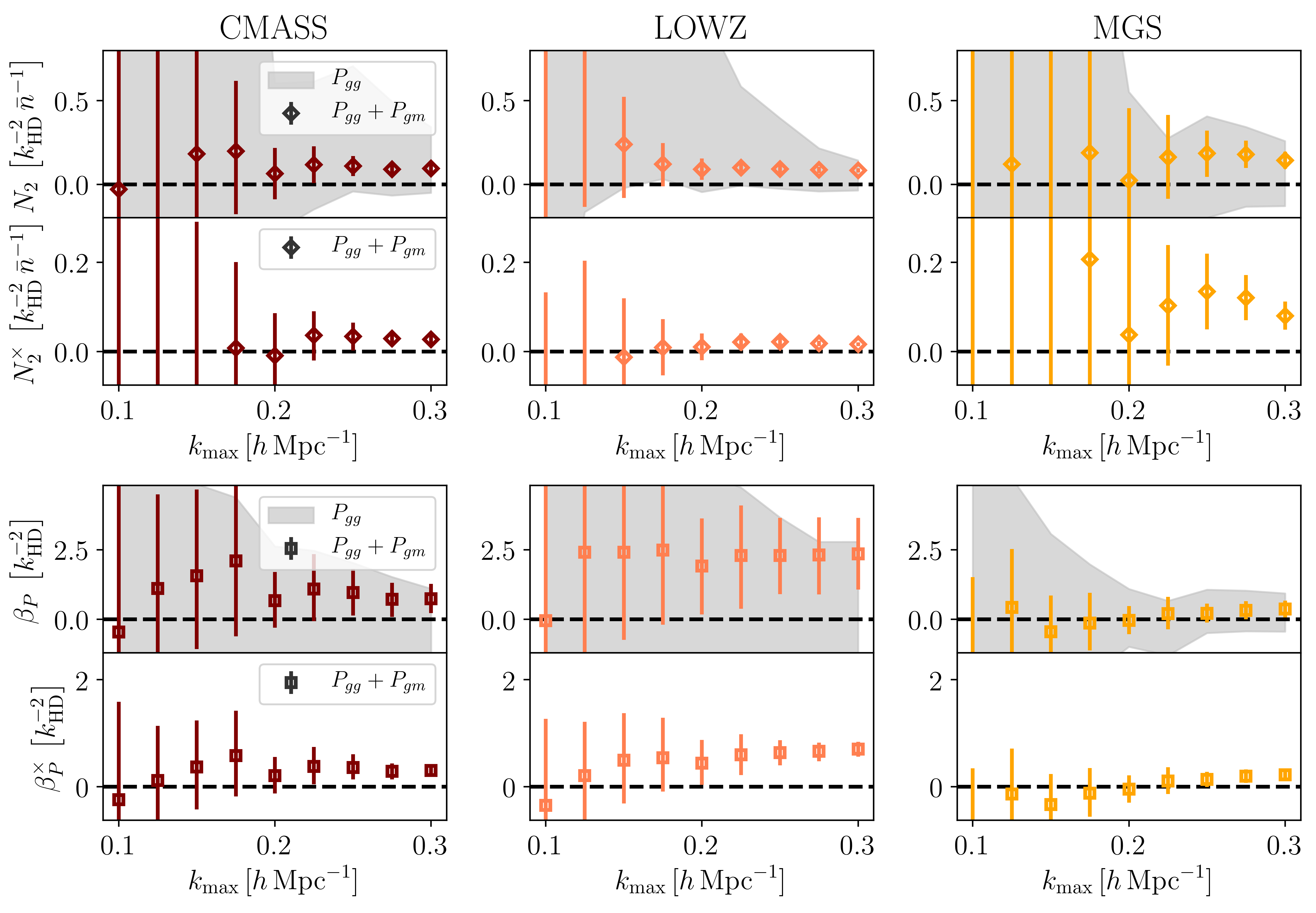}
\caption{Marginalised posterior of the $k^2$-dependent stochastic parameters $N_2$ and $N_2^\times$ (top), and the higher derivatives parameters $\beta_P$ and $\beta_P^\times$ (bottom), as a function of $k_\mathrm{max}$, for the combination of $P_{\rm gg}+P_{\rm gm}$. In all cases dark matter non-linear evolution is modelled using \texttt{RESPRESSO}. For consistency check we also show the $1$-$\sigma$ standard deviations of the parameters $N_2$ and $\beta_P$ obtained from the fit of $P_{\rm gg}$ only as a shaded grey area.}
\label{fig:N2_bP_vs_kmax}
\end{figure*}

After having focused on the marginalised posteriors of the cosmological parameters, linear bias and constant shot noise, here we try to put internal constraints on both the $k^2$-dependent stochastic parameters and the higher derivatives parameters. In particular, we want to check whether these terms can be detected with statistical significance under the assumption of using a reference effective volume like the one described in Section \ref{sec:data}.

Figure \ref{fig:N2_bP_vs_kmax} shows marginalised constraints for the two sets of parameters (extracted from the $P_{\rm gg}+P_{\rm gm}$ chains) as a function of $k_{\rm max}$.  For consistency check, in the panels referring to parameters that enter the modeling of the galaxy auto power spectrum, $N_2$ and $\beta_P$, we also show less stringent constraints extracted from the chains corresponding to the fit of $P_{\rm gg}$ (shaded grey band). 

The consistency of the $k^2$-dependent shot-noise model is additionally reinforced by the trend observed in the top panels of Figure \ref{fig:N2_bP_vs_kmax}. As a matter of fact, the relation between $N_2/N_2^\times$ and $k_{\rm max}$ can be well described by a straight line with zero slope, this fact intrinsically hinting at a correct behaviour for both parameters, as their profiles remain stable when adding constraints at higher wave modes. Moreover, there is a clear detection of the $k^2$-dependent shot-noise parameters, with a statistical significance that further reinforce the impact of adding $N_2$ and $N_2^\times$ in terms of goodness-of-fit. Indeed $N_2$ and $N_2^\times$ are detected at $k_{\rm max}=0.3\,h\,\mathrm{Mpc}^{-1}$ with a statistical significance of $3.5$-$\sigma$ and $2.5$-$\sigma$, respectively, for the CMASS sample. With LOWZ the detection becomes even larger, with numbers that are close to $6$-$\sigma$ and $4$-$\sigma$, respectively. The MGS sample is the one showing the least significant detection of both parameters, with a possible modification of the recovered value of $N_2^\times$ at $k_{\rm max}=0.3\,h\,\mathrm{Mpc}^{-1}$. However, we remind that the STD model performs surprisingly well on this dataset, and therefore the weaker detection of the two parameters is partially expected.

The higher derivatives parameters also show stable results as a function of $k_{\rm max}$, with the highest detection represented by $\beta_P^\times$ for LOWZ. However on these scales the HD model is already broken, as shown in Figure \ref{fig:fcf_Pgg+Pgm_respresso}. For the other cases, the typical significance of the detection is set to $\sim2$-$\sigma$ and $\sim4$-$\sigma$ for $\beta_P$ and $\beta_P^\times$, respectively.

Differently from \cite{EggScoCro2011}, we do not observe incompatible results between the marginalised posteriors of $N_2$ and $\beta_P$ from fits of $P_{\rm gg}$ and $P_{\rm gg}+P_{\rm gm}$. However in this case we significantly extended the dimensionality of the parameter space leading to weaker constraints on the model parameters, so that the deviation may be hidden by the larger statistical noise.

\subsection{Results for alternative matter models}
\label{sec:results_models}

\begin{figure*}
  \centering
  \includegraphics[width=\textwidth]{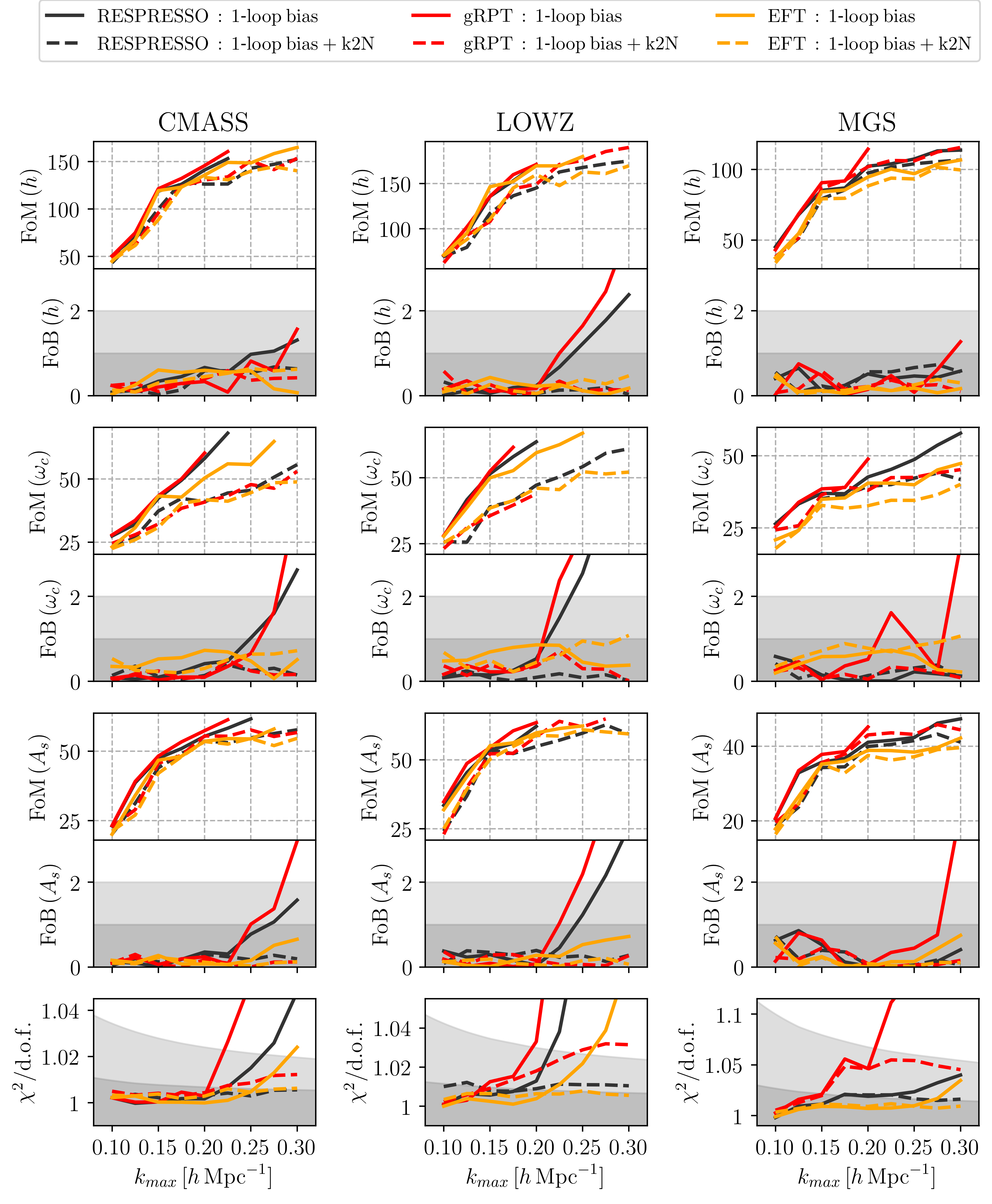}
  \caption{FoM, FoB and goodness-of-fit extracted from fits of $P_{\rm gg}+P_{\rm gm}$. Performance metrics are separated into each individual component, i.e. $h$, $\Omega_{\rm c} h^2$ and $A_{\rm s}$. Different dark matter models and different one-loop modeling assumptions are color- and style-coded as shown in the legend. The two shaded grey regions mark the $68$th and $95$th percentiles of the FoB and $\chi^2$ distribution.}
  \label{fig:fcf_Pgg+Pgm_all}
\end{figure*}

So far we have fixed the modeling of the non-linear matter power spectrum using a hybrid approach such as \texttt{RESPRESSO}. Another way of proceeding is to employ a perturbative description not only for one-loop bias but also for $P_{\rm mm}$. In this section, we analyse the impact of a different modeling of $P_{\rm mm}$ inside Equation (\ref{eq:Pgg_1-loop}) and (\ref{eq:Pgm_1-loop}), and for this purpose we make use of the two additional models described in Section \ref{sec:matter_models}.

We remind the reader that these models correspond to 1) an EFT-like model based on resummation of IR modes and non-trivial stress tensor that at leading order in real-space can be described by a single counterterm, and 2) a RPT-based approach with a tighter requirement of preserving the Galilean invariance of equal time correlators (gRPT). For the three different matter models, we show results for the STD model and for the extension including higher-order stochastic contributions, as in the previous sections we assessed how this extension is preferred from the three data samples we are considering. Differently from the other matter models, the EFT-based approach requires an additional free parameter, represented by the counterterm $c_1$, that is completely degenerate with leading order higher derivatives terms. For this reason, in this case the STD model already includes a higher-derivative correction. The full list of configurations, together with free model parameters, is given in Table \ref{tab:models}.

Figure \ref{fig:fcf_Pgg+Pgm_all} shows the performance metrics for the three different dark matter models when fitting the combination of $P_{\rm gg}$ and $P_{\rm gm}$. In this case, we show the individual contributions to FoM and FoB from the three cosmological parameters, $h$, $\Omega_{\rm c} h^2$ and $A_{\rm s}$. In addition, for the sake of simplicity, we identify the model breaking scale $k^\dag$ with the end of the corresponding line in the FoM panels. 

We first notice how the recovered goodness-of-fit is severely worsened when adopting gRPT to describe the non-linear matter power spectrum, particularly when considering the STD model. We separately tested the impact of extending this model to include higher derivatives, and while this extension slightly improves the range of validity of the gRPT-based model, it still fails in being accepted up to $k_{\rm max}=0.30\,h\,\mathrm{Mpc}^{-1}$. The corresponding k2N model definitely improves the recovered trends but the net result is a worse performance respect to \texttt{RESPRESSO} and EFT. The STD model of EFT breaks at higher values of $k_{\rm max}$, and this is somewhat expected given the presence of the additional free parameter $c_1$. Nevertheless in order to provide a $\chi^2$ consistent within $95\%$ confidence interval at all scales we still have to include additional stochastic terms. 

The FoB is overall consistent among the three different parameters, showing a strong bias on the marginalised posterior ($>2$-$\sigma$ at $k_{\rm max}>0.25\,h\,\mathrm{Mpc}^{-1}$) when adopting the STD for either gRPT or \texttt{RESPRESSO}. Similarly to the goodness-of-fit, the STD case combined with EFT provides better results (consistent within $1$-$\sigma$) since it already incorporates a $k^2P_{\rm L}(k)$ correction which is absent from the other matter models. The overall interpretation is that all three matter models strongly hint for the necessity of adding $k^2$-dependent stochastic contributions, both to produce an accurate description of the joint data vector $P_{\rm gg}+P_{\rm gm}$ and to provide unbiased measurements of the cosmological parameters.

As for the FoM, we notice how its dependency on $k_{\rm max}$ is significantly different between the $(h,\,A_{\rm s})$ pair and $\Omega_{\rm c} h^2$. For the former, the difference in performance between STD and extended models is tiny $(\sim10\%)$, while, for the latter, adding $k^2$-dependent stochasticity (or higher derivatives) results in the suppression of the FoM at a reference scale of $k_{\rm max}=0.2\,h\,\mathrm{Mpc}^{-1}$ by a factor of $\sim50\%$ and $\sim25\%$, respectively.

Focusing on the maximum FoM achieved by each model in the range of scales we consider, we obtain different results. The usage of gRPT typically leads to the breaking of the model at lower wave modes, $k_{\rm max}\sim0.2\,h\,\mathrm{Mpc}^{-1}$, when using the STD model (slightly larger with the HD extension). Substituting gRPT with \texttt{RESPRESSO} helps in extending the range of validity of the HD model, as shown in Section \ref{sec:consistency-ax}, while leaving almost untouched the model breaking scale of the STD case. However, this last configuration, i.e. \texttt{RESPRESSO} with the STD biasing scheme systematically results in one of the best performing models in terms of maximally achievable FoM. Adding $k^2$-dependent stochastic terms to both matter models extends the range of validity up to $k_{\rm max}=0.3\,h\,\mathrm{Mpc}^{-1}$, with a value of FoM which is typically consistent with the one of the STD case at the corresponding $k^\dag$. Finally, the range of validity of the EFT model is hitting our maximum value of $0.3\,h\,\mathrm{Mpc}^{-1}$ already with the STD model for both CMASS and MGS, while being limited to intermediate values $(k_{\rm max}\sim0.25\,h\,\mathrm{Mpc}^{-1})$ for LOWZ. The latter can be also be adjusted by considering $k^2$-dependent shot-noise parameters. 

\subsection{Impact of $P_{\rm gm}$ on the Figure of Merit}
\label{sec:Pgg_vs_Pgg+Pgm}

\begin{figure*}
\includegraphics[width=0.495\textwidth]{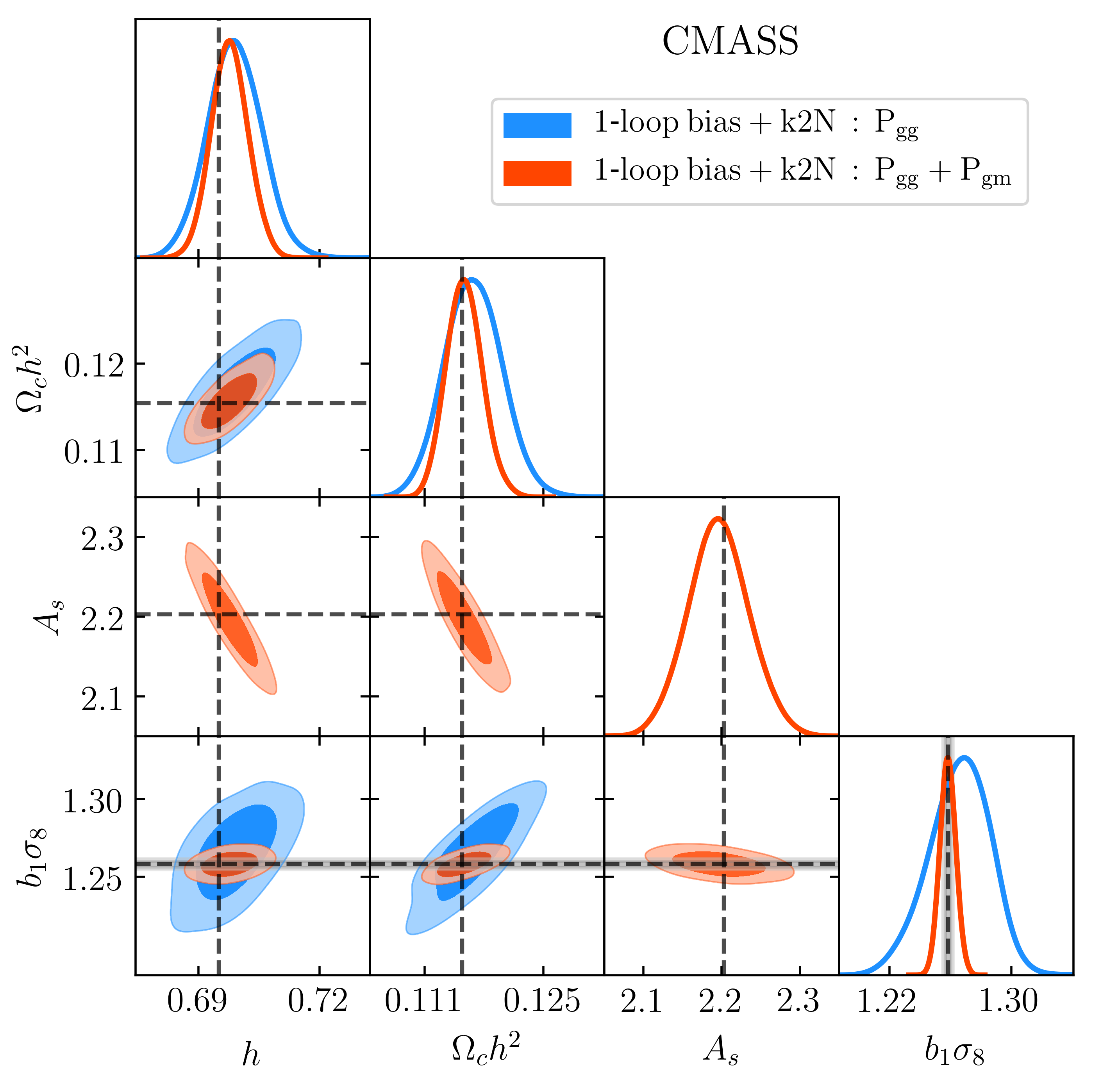}
\includegraphics[width=0.495\textwidth]{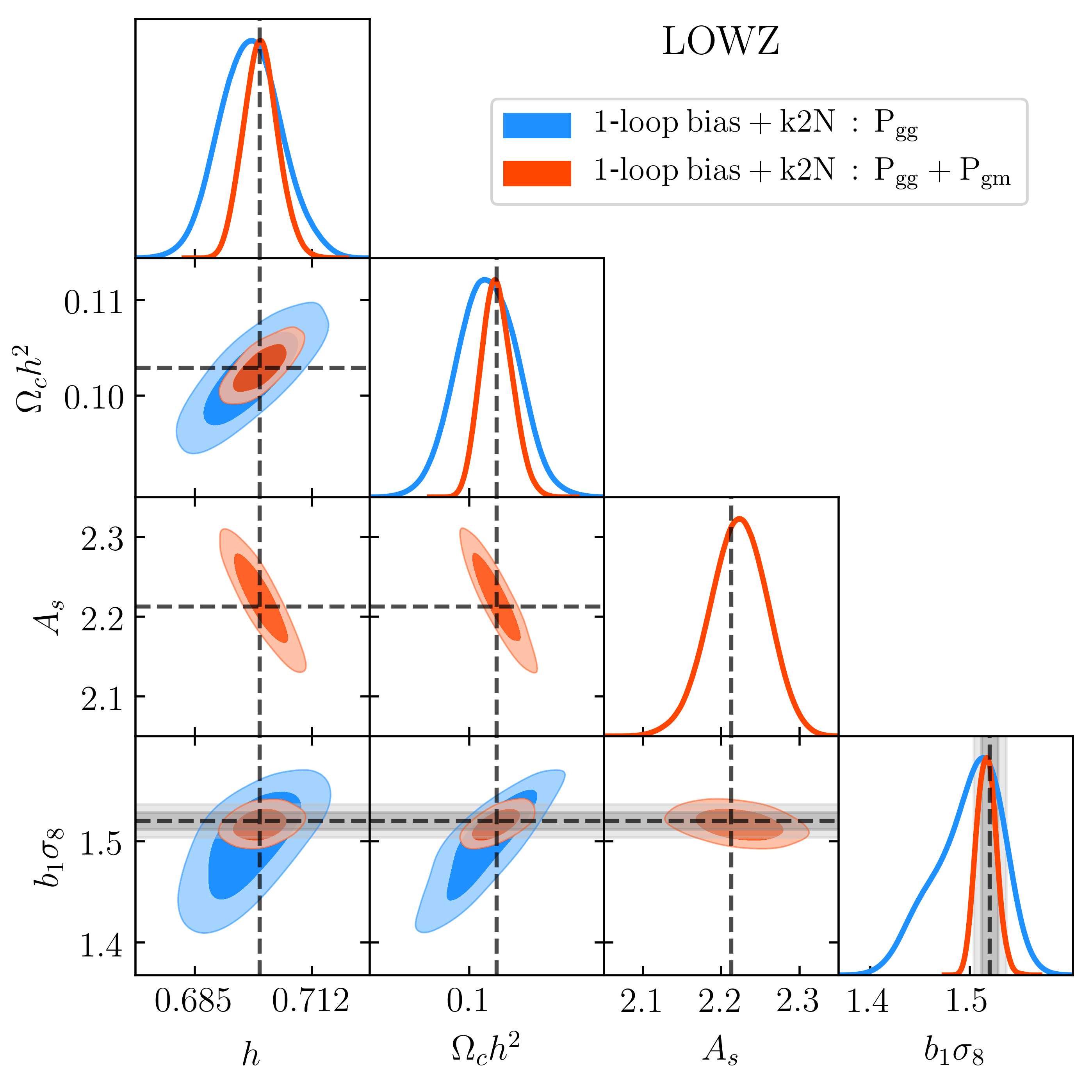}
\includegraphics[width=0.495\textwidth]{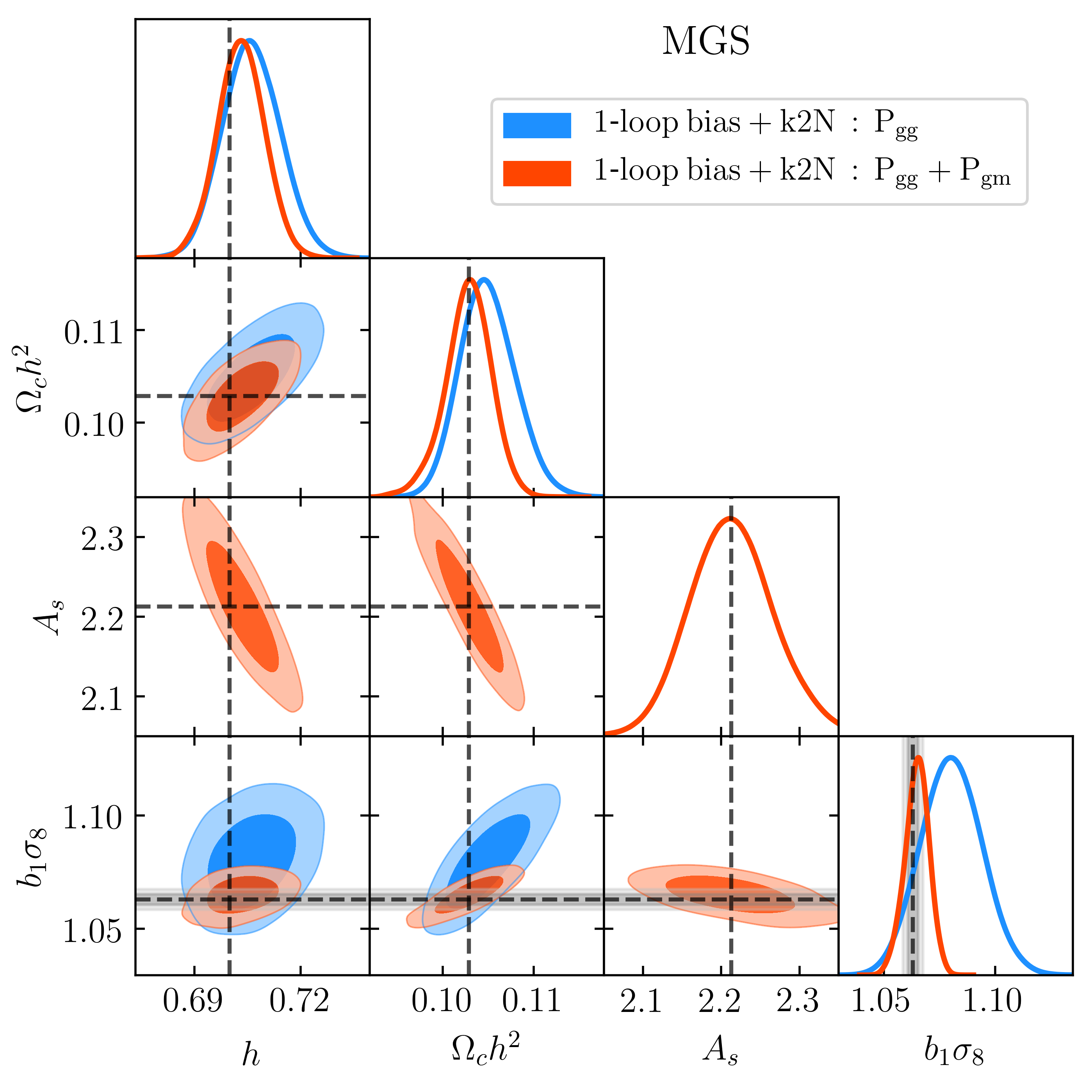}
\caption{Comparison between the posterior parameter distribution ($68\%$ and $95\%$ confidence intervals) obtained by fitting the galaxy auto power spectrum $P_{\rm gg}$ (blue) and the combination of galaxy-galaxy and galaxy matter power spectrum $P_{\rm gg}+P_{\rm gm}$ (red). We show constraints obtained at $k_{\rm max}=0.3\,h\,\mathrm{Mpc}^{-1}$ for the cosmological parameters $h$, $\Omega_{\rm c} h^2$, $A_{\rm s}$, and the linear bias $b_1\sigma_8$. In all cases we assume non-linear dark-matter evolution to be described by \texttt{RESPRESSO}. Dashed solid lines correspond to the fiducial values of the corresponding parameter. Grey solid bands mark the $1$-$\sigma$ and $2$-$\sigma$ error on the fiducial linear bias parameter.}
\label{fig:comp_post_Pgg_vs_Pgm}
\end{figure*}

After having assessed the impact of different modeling assumptions we can now check the statistical significance of adding the galaxy-matter cross power spectrum to a full shape measurement fit like the one we carried out in this analysis. Indeed, as we already pointed out in Section \ref{sec:consistency-ax} $3\times2$-point analysis will be the source of several cosmological constraints in next-generation galaxy surveys. Although in this analysis we are excluding redshift-space distortions, in order to separate the additional model systematics from the ones of one-loop galaxy bias, it is anyhow interesting to quantify the improvement in FoM when including constraints from $P_{\rm gm}$.

In Figure \ref{fig:comp_post_Pgg_vs_Pgm} we compare the marginalised posterior distribution of the cosmological parameters and linear bias obtained when fitting only $P_{gg}$ and the combination $P_{\rm gg}+P_{\rm gm}$. For this comparison we model $P_{\rm mm}$ with \texttt{RESPRESSO}, include $k^2$-dependent stochastic contributions, and include all scales up to $k_{\rm max}=0.3\,h\,\mathrm{Mpc}^{-1}$. We remind that the comparison is not properly fair, as we fix the scalar amplitude $A_{\rm s}$ in the fits of the auto power spectrum alone, while this is treated as a free parameter in the fits of the combined statistics.

The observed trend when including $P_{\rm gm}$ in the fit is a significant reduction of the statistical uncertainty of all parameters. In particular we measure a $1$-$\sigma$ standard deviation for $h$ and $\Omega_{\rm c} h^2$ that is $1.4$, $2$, and $1.2$ times smaller for CMASS, LOWZ and MGS, respectively. On the other hand, the linear bias absorbs most of the additional constraining power, leading to more precise measurements of $b_1\sigma_8$ by a factor $3$ for all three samples.

\section{Conclusions}
\label{sec:conclusions}

In this work we performed a full-shape analysis  of the galaxy power spectrum meant to assess the robustness of one-loop galaxy bias models, extending the investigation carried out in \cite{EggScoCro2011} to include also the sampling over cosmological parameters. Since we deal with galaxy clustering in real space, we decided to fix both the baryon density $\Omega_{\rm b}\,h^2$ and the spectral index $n_s$ while leaving as free parameters the cold dark matter density $\Omega_{\rm c} \,h^2$, the Hubble constant $h$, and the scalar amplitude $A_{\rm s}$. The latter is kept as a free parameter only when considering joint fits to the galaxy-galaxy and galaxy-matter power spectra.

We measured both observables from a set of three different synthetic galaxy samples, whose clustering properties are meant to reproduce the ones of three real data catalogues, i.e. CMASS and LOWZ from BOSS, and MGS from SDSS. We rescaled the statistical uncertainties of each of these samples to match an effective volume of $6\,\left(h^{-1}\,\mathrm{Gpc}\right)^3$ (i.e. the one of the LOWZ sample), which is representative of the volume of tomographic bins from next-generation galaxy surveys (e.g. Euclid \cite{EucFor}).

The analytical recipes we adopted to model these observables are based on a standard one-loop expansion of the galaxy density field on the matter density field, collecting terms related to both spherically-symmetric gravitational collapse and tidal fields up to third-order in perturbations. In the first part of the paper, we fixed the description of the non-linear matter power spectrum using a hybrid perturbative-simulated approach, i.e. \texttt{RESPRESSO}, which was already shown in \cite{EggScoCro2011} to provide the closest behaviour to the true measured matter power spectrum on the overall range of scales and redshifts that we consider. In a later section we explicitly tested the impact of different matter modeling assumptions, by employing an EFT-like model and a RPT-derived model (gRPT). In all the considered cases we fixed the quadratic tidal bias $\gamma_2$ using an excursion-set-derived relation, in order to break the strong degeneracy between the latter and the cubic non-local parameter $\gamma_{21}$.

Our main interest was in understanding the range of validity of the standard one-loop expansion, and if our synthetic galaxy sample would hint for the necessity of adding additional terms to the standard recipe. For this, we considered two extensions of the standard model that take into account either higher-derivatives of the gravitational potential or scale-dependent stochasticity, as expected from short-range non-locality due to galaxy formation and the halo-exclusion effect, respectively.

In order to quantify the goodness of each model we make use of a set of three different performance metrics, which are 1) the Figure of Bias, to assess the level of bias introduced in the recovery of cosmological parameters, 2) the goodness-of-fit, to quantify the likelihood between the input data vector and the best fit model, and 3) the Figure of Merit, to compare the constraining power of each model at any given $k_{\rm max}$. All of these quantities are easily obtained by post-processing the MCMC chains that we run for each combination of matter modeling, one-loop bias extensions, and maximum mode $k_{\rm max}$.

Our results can be summarised as follows:
\begin{enumerate}[label=(\roman*)]
\item When using the fiducial description of the non-linear matter power spectrum (i.e \texttt{RESPRESSO}), we find that a standard one-loop bias model (linear bias, quadratic bias, cubic non-local bias, constant shot-noise parameter) with fixed quadratic tidal bias can provide a good description of the galaxy power spectrum for all our samples up to $k_{\rm max}=0.3\,h\,\mathrm{Mpc}^{-1}$ while also returning unbiased value of the cosmological parameter set $\left(h,\,\Omega_{\rm c} h^2\right)$.
\item Similarly to the fixed cosmology analysis of \cite{EggScoCro2011} we notice that when we consider the additional information from the galaxy-matter cross power spectrum, the standard model is no longer capable of providing a good performance on the overall range of scales we consider for all the three samples. In particular, this model breaks at a scale of $k_{\rm max}=0.25\,h\,\mathrm{Mpc}^{-1}$ and $k_{\rm max}=0.2\,h\,\mathrm{Mpc}^{-1}$ for CMASS and LOWZ, respectively, while being accepted all the way up to $k_{\rm max}=0.3\,h\,\mathrm{Mpc}^{-1}$ for MGS. This might be representative of the level of non-linearities in the galaxy-matter relationship, which at first order can be captured by the parameter $b_1\sigma_8$ (that is $\sim1.52$ for LOWZ, $\sim1.26$ for CMASS and $\sim1.06$ for MGS). 
\item Extending the standard model to account also for the presence of either higher derivatives or scale-dependent stochasticity provides a better performance both in terms of Figure of Bias (in this case for the parameter set $(h,\,\Omega_{\rm c}h^2,\,A_{\rm s})$) and goodness-of-fit. Adding scale-dependent stochastic terms restores the range of validity of the model up to $k_{\rm max}=0.3\,h\,\mathrm{Mpc}^{-1}$ for both CMASS and LOWZ, while the importance of higher-derivatives is limited by a worse $\chi^2$ for $k_{\rm max}\gtrsim0.2\,h\,\mathrm{Mpc}^{-1}$, and slightly worse recovery of the linear bias parameter and the constant shot-noise correction. As an additional check, we verify that the marginalised posterior distribution of both $k^2$-dependent terms and higher derivatives parameters is stable as a function of $k_{\rm max}$. However we notice how the maximally achievable FoM of the extended models is consistent with the one of the standard configuration at its lower $k_{\rm max}$, leading to equivalent statistical constraints on the cosmological parameters.
\item Changing the description of the non-linear matter power spectrum to either EFT or gRPT induces modifications to the range of validity of one-loop galaxy bias models. Overall, we find that, as highlighted in \cite{EggScoCro2011}, \texttt{RESPRESSO} is the best performing model, followed by EFT and gRPT. The EFT-based model is penalised in terms of Figure of Merit because of the presence of an additional degree of freedom (two in case of the combined fits), but it compensates this with an extended range of validity. On the contrary, gRPT breaks sooner than the other models, at a typical scale of $k_{\rm max}\sim0.2\,h\,\mathrm{Mpc}^{-1}$, which is consistent with the range of validity observed in direct fits of the matter power spectrum. This indicates that this model would also benefit of the presence of an additional free parameter representing a non-zero stress tensor such as in the case of the EFT-based model, as expected. Although the gRPT-based model breaks earlier than the other two models, its maximally achievable FoM is consistent with the one recovered using \texttt{RESPRESSO}.
\item The additional constraints coming from the galaxy-matter cross power spectrum results in an improved statistical precision on measurements of cosmological parameters. Although we fix the scalar amplitude $A_{\rm s}$ when fitting only the galaxy auto power spectrum, we still find that in the combined case we achieve better constraints on both $h$ and $\Omega_{\rm c} h^2$ by a factor $1.4$, $2$ and $1.2$ for CMASS, LOWZ and MGS, respectively. At the same time we notice how a significant fraction of the additional constraining power can be absorbed by nuisance parameters, such as the linear bias, for which constraints are tighter by a factor of $3$ for all the three samples.
\end{enumerate}

In this work we analyse three samples with an effective volume which is rescaled to a value of $6\,h^{-3}\,\mathrm{Gpc}^3$. Although the results  we obtained are partially based on this choice, as shown in \cite{EggScoCro2011}, this volume can be well representative of individual tomographic redshift bins that will be adopted by next-generation galaxy surveys. In addition we explore the additional constraining power brought by adding the galaxy-mass cross power spectrum to galaxy power spectrum fits, which is a relevant study in the context of $3\times2$pt analyses of next-generation imaging surveys.  Therefore, these results can provide useful insights when adopting one-loop perturbation theory to describe the relationship between the galaxy and matter density fields in that context \cite{Pandey2020}. In order to completely concentrate on galaxy bias, we performed this analysis using real-space coordinates, avoiding a further modeling layer for redshift-space distortions. We leave to future works the exploration of this effect, both using the power spectrum alone, and combining it with the galaxy bispectrum.

\acknowledgements

AP and MC acknowledge support from the Spanish Ministry of Science and Innovation through grants PGC2018-102021-B-100 and ESP2017-89838-C3-1-R, and EU grants LACEGAL 734374 and EWC 776247 with ERDF funds. IEEC is funded by the CERCA program of the Generalitat de Catalunya.  AE acknowledges support from the European Research Council (grant num- ber ERC-StG-716532-PUNCA. AGS acknowledges support from the Excellence Cluster ORIGINS, which is funded by the Deutsche Forschungsgemeinschaft (DFG, German Research Foundation) under Germany’s Excellence Strategy - EXC-2094 - 390783311
AP and AGS would like to thank Daniel Farrow, Jiamin Hou, Martha Lippich and Agne Semenaite for their help and useful discussions.
The analysis presented here has been performed on the high-performance computing resources
of the Max Planck Computing and Data Facility (MPCDF) in Garching.
This research made use of matplotlib, a Python library for
publication quality graphics \cite{Hunter:2007}.

\bibliography{main}

\begin{thebibliography}{105}%
\makeatletter
\providecommand \@ifxundefined [1]{%
 \@ifx{#1\undefined}
}%
\providecommand \@ifnum [1]{%
 \ifnum #1\expandafter \@firstoftwo
 \else \expandafter \@secondoftwo
 \fi
}%
\providecommand \@ifx [1]{%
 \ifx #1\expandafter \@firstoftwo
 \else \expandafter \@secondoftwo
 \fi
}%
\providecommand \natexlab [1]{#1}%
\providecommand \enquote  [1]{``#1''}%
\providecommand \bibnamefont  [1]{#1}%
\providecommand \bibfnamefont [1]{#1}%
\providecommand \citenamefont [1]{#1}%
\providecommand \href@noop [0]{\@secondoftwo}%
\providecommand \href [0]{\begingroup \@sanitize@url \@href}%
\providecommand \@href[1]{\@@startlink{#1}\@@href}%
\providecommand \@@href[1]{\endgroup#1\@@endlink}%
\providecommand \@sanitize@url [0]{\catcode `\\12\catcode `\$12\catcode
  `\&12\catcode `\#12\catcode `\^12\catcode `\_12\catcode `\%12\relax}%
\providecommand \@@startlink[1]{}%
\providecommand \@@endlink[0]{}%
\providecommand \url  [0]{\begingroup\@sanitize@url \@url }%
\providecommand \@url [1]{\endgroup\@href {#1}{\urlprefix }}%
\providecommand \urlprefix  [0]{URL }%
\providecommand \Eprint [0]{\href }%
\providecommand \doibase [0]{http://dx.doi.org/}%
\providecommand \selectlanguage [0]{\@gobble}%
\providecommand \bibinfo  [0]{\@secondoftwo}%
\providecommand \bibfield  [0]{\@secondoftwo}%
\providecommand \translation [1]{[#1]}%
\providecommand \BibitemOpen [0]{}%
\providecommand \bibitemStop [0]{}%
\providecommand \bibitemNoStop [0]{.\EOS\space}%
\providecommand \EOS [0]{\spacefactor3000\relax}%
\providecommand \BibitemShut  [1]{\csname bibitem#1\endcsname}%
\let\auto@bib@innerbib\@empty
\bibitem [{\citenamefont {{Davis}}\ and\ \citenamefont
  {{Peebles}}(1983)}]{DavPee8304}%
  \BibitemOpen
  \bibfield  {author} {\bibinfo {author} {\bibfnamefont {M.}~\bibnamefont
  {{Davis}}}\ and\ \bibinfo {author} {\bibfnamefont {P.~J.~E.}\ \bibnamefont
  {{Peebles}}},\ }\href {\doibase 10.1086/160884} {\bibfield  {journal}
  {\bibinfo  {journal} {\apj}\ }\textbf {\bibinfo {volume} {267}},\ \bibinfo
  {pages} {465} (\bibinfo {year} {1983})}\BibitemShut {NoStop}%
\bibitem [{\citenamefont {{Maddox}}\ \emph {et~al.}(1990)\citenamefont
  {{Maddox}}, \citenamefont {{Efstathiou}}, \citenamefont {{Sutherland}},\ and\
  \citenamefont {{Loveday}}}]{MadEfsSut9001}%
  \BibitemOpen
  \bibfield  {author} {\bibinfo {author} {\bibfnamefont {S.~J.}\ \bibnamefont
  {{Maddox}}}, \bibinfo {author} {\bibfnamefont {G.}~\bibnamefont
  {{Efstathiou}}}, \bibinfo {author} {\bibfnamefont {W.~J.}\ \bibnamefont
  {{Sutherland}}}, \ and\ \bibinfo {author} {\bibfnamefont {J.}~\bibnamefont
  {{Loveday}}},\ }\href {\doibase 10.1093/mnras/242.1.43P} {\bibfield
  {journal} {\bibinfo  {journal} {\mnras}\ }\textbf {\bibinfo {volume} {242}},\
  \bibinfo {pages} {43} (\bibinfo {year} {1990})}\BibitemShut {NoStop}%
\bibitem [{\citenamefont {Tegmark}\ \emph {et~al.}(2004)\citenamefont
  {Tegmark}, \citenamefont {Blanton}, \citenamefont {Strauss}, \citenamefont
  {Hoyle}, \citenamefont {Schlegel}, \citenamefont {Scoccimarro}, \citenamefont
  {Vogeley}, \citenamefont {Weinberg}, \citenamefont {Zehavi}, \citenamefont
  {Berlind},\ and\ \citenamefont {et~al.}}]{Tegmark2004}%
  \BibitemOpen
  \bibfield  {author} {\bibinfo {author} {\bibfnamefont {M.}~\bibnamefont
  {Tegmark}}, \bibinfo {author} {\bibfnamefont {M.~R.}\ \bibnamefont
  {Blanton}}, \bibinfo {author} {\bibfnamefont {M.~A.}\ \bibnamefont
  {Strauss}}, \bibinfo {author} {\bibfnamefont {F.}~\bibnamefont {Hoyle}},
  \bibinfo {author} {\bibfnamefont {D.}~\bibnamefont {Schlegel}}, \bibinfo
  {author} {\bibfnamefont {R.}~\bibnamefont {Scoccimarro}}, \bibinfo {author}
  {\bibfnamefont {M.~S.}\ \bibnamefont {Vogeley}}, \bibinfo {author}
  {\bibfnamefont {D.~H.}\ \bibnamefont {Weinberg}}, \bibinfo {author}
  {\bibfnamefont {I.}~\bibnamefont {Zehavi}}, \bibinfo {author} {\bibfnamefont
  {A.}~\bibnamefont {Berlind}}, \ and\ \bibinfo {author} {\bibnamefont
  {et~al.}},\ }\href {\doibase 10.1086/382125} {\bibfield  {journal} {\bibinfo
  {journal} {The Astrophysical Journal}\ }\textbf {\bibinfo {volume} {606}},\
  \bibinfo {pages} {702–740} (\bibinfo {year} {2004})}\BibitemShut {NoStop}%
\bibitem [{\citenamefont {Cole}\ \emph {et~al.}(2005)\citenamefont {Cole},
  \citenamefont {Percival}, \citenamefont {Peacock}, \citenamefont {Norberg},
  \citenamefont {Baugh}, \citenamefont {Frenk}, \citenamefont {Baldry},
  \citenamefont {Bland-Hawthorn}, \citenamefont {Bridges}, \citenamefont
  {Cannon},\ and\ \citenamefont {et~al.}}]{Cole2005}%
  \BibitemOpen
  \bibfield  {author} {\bibinfo {author} {\bibfnamefont {S.}~\bibnamefont
  {Cole}}, \bibinfo {author} {\bibfnamefont {W.~J.}\ \bibnamefont {Percival}},
  \bibinfo {author} {\bibfnamefont {J.~A.}\ \bibnamefont {Peacock}}, \bibinfo
  {author} {\bibfnamefont {P.}~\bibnamefont {Norberg}}, \bibinfo {author}
  {\bibfnamefont {C.~M.}\ \bibnamefont {Baugh}}, \bibinfo {author}
  {\bibfnamefont {C.~S.}\ \bibnamefont {Frenk}}, \bibinfo {author}
  {\bibfnamefont {I.}~\bibnamefont {Baldry}}, \bibinfo {author} {\bibfnamefont
  {J.}~\bibnamefont {Bland-Hawthorn}}, \bibinfo {author} {\bibfnamefont
  {T.}~\bibnamefont {Bridges}}, \bibinfo {author} {\bibfnamefont
  {R.}~\bibnamefont {Cannon}}, \ and\ \bibinfo {author} {\bibnamefont
  {et~al.}},\ }\href {\doibase 10.1111/j.1365-2966.2005.09318.x} {\bibfield
  {journal} {\bibinfo  {journal} {Monthly Notices of the Royal Astronomical
  Society}\ }\textbf {\bibinfo {volume} {362}},\ \bibinfo {pages} {505–534}
  (\bibinfo {year} {2005})}\BibitemShut {NoStop}%
\bibitem [{\citenamefont {{Hauser}}\ and\ \citenamefont
  {{Peebles}}(1973)}]{HauPee7311}%
  \BibitemOpen
  \bibfield  {author} {\bibinfo {author} {\bibfnamefont {M.~G.}\ \bibnamefont
  {{Hauser}}}\ and\ \bibinfo {author} {\bibfnamefont {P.~J.~E.}\ \bibnamefont
  {{Peebles}}},\ }\href {\doibase 10.1086/152453} {\bibfield  {journal}
  {\bibinfo  {journal} {\apj}\ }\textbf {\bibinfo {volume} {185}},\ \bibinfo
  {pages} {757} (\bibinfo {year} {1973})}\BibitemShut {NoStop}%
\bibitem [{\citenamefont {{Kaiser}}(1984)}]{Kai8409}%
  \BibitemOpen
  \bibfield  {author} {\bibinfo {author} {\bibfnamefont {N.}~\bibnamefont
  {{Kaiser}}},\ }\href {\doibase 10.1086/184341} {\bibfield  {journal}
  {\bibinfo  {journal} {\apjl}\ }\textbf {\bibinfo {volume} {284}},\ \bibinfo
  {pages} {L9} (\bibinfo {year} {1984})}\BibitemShut {NoStop}%
\bibitem [{\citenamefont {{Davis}}\ \emph {et~al.}(1985)\citenamefont
  {{Davis}}, \citenamefont {{Efstathiou}}, \citenamefont {{Frenk}},\ and\
  \citenamefont {{White}}}]{DavEfsFre8505}%
  \BibitemOpen
  \bibfield  {author} {\bibinfo {author} {\bibfnamefont {M.}~\bibnamefont
  {{Davis}}}, \bibinfo {author} {\bibfnamefont {G.}~\bibnamefont
  {{Efstathiou}}}, \bibinfo {author} {\bibfnamefont {C.~S.}\ \bibnamefont
  {{Frenk}}}, \ and\ \bibinfo {author} {\bibfnamefont {S.~D.~M.}\ \bibnamefont
  {{White}}},\ }\href {\doibase 10.1086/163168} {\bibfield  {journal} {\bibinfo
   {journal} {\apj}\ }\textbf {\bibinfo {volume} {292}},\ \bibinfo {pages}
  {371} (\bibinfo {year} {1985})}\BibitemShut {NoStop}%
\bibitem [{\citenamefont {{Bardeen}}\ \emph {et~al.}(1986)\citenamefont
  {{Bardeen}}, \citenamefont {{Bond}}, \citenamefont {{Kaiser}},\ and\
  \citenamefont {{Szalay}}}]{BarBonKai8605}%
  \BibitemOpen
  \bibfield  {author} {\bibinfo {author} {\bibfnamefont {J.~M.}\ \bibnamefont
  {{Bardeen}}}, \bibinfo {author} {\bibfnamefont {J.~R.}\ \bibnamefont
  {{Bond}}}, \bibinfo {author} {\bibfnamefont {N.}~\bibnamefont {{Kaiser}}}, \
  and\ \bibinfo {author} {\bibfnamefont {A.~S.}\ \bibnamefont {{Szalay}}},\
  }\href {\doibase 10.1086/164143} {\bibfield  {journal} {\bibinfo  {journal}
  {\apj}\ }\textbf {\bibinfo {volume} {304}},\ \bibinfo {pages} {15} (\bibinfo
  {year} {1986})}\BibitemShut {NoStop}%
\bibitem [{\citenamefont {{Tegmark}}\ \emph {et~al.}(2004)\citenamefont
  {{Tegmark}}, \citenamefont {{Blanton}}, \citenamefont {{Strauss}},
  \citenamefont {{Hoyle}}, \citenamefont {{Schlegel}}, \citenamefont
  {{Scoccimarro}}, \citenamefont {{Vogeley}}, \citenamefont {{Weinberg}},
  \citenamefont {{Zehavi}}, \citenamefont {{Berlind}}, \citenamefont
  {{Budavari}}, \citenamefont {{Connolly}}, \citenamefont {{Eisenstein}},
  \citenamefont {{Finkbeiner}}, \citenamefont {{Frieman}}, \citenamefont
  {{Gunn}}, \citenamefont {{Hamilton}}, \citenamefont {{Hui}}, \citenamefont
  {{Jain}}, \citenamefont {{Johnston}}, \citenamefont {{Kent}}, \citenamefont
  {{Lin}}, \citenamefont {{Nakajima}}, \citenamefont {{Nichol}}, \citenamefont
  {{Ostriker}}, \citenamefont {{Pope}}, \citenamefont {{Scranton}},
  \citenamefont {{Seljak}}, \citenamefont {{Sheth}}, \citenamefont
  {{Stebbins}}, \citenamefont {{Szalay}}, \citenamefont {{Szapudi}},
  \citenamefont {{Verde}}, \citenamefont {{Xu}}, \citenamefont {{Annis}},
  \citenamefont {{Bahcall}}, \citenamefont {{Brinkmann}}, \citenamefont
  {{Burles}}, \citenamefont {{Castander}}, \citenamefont {{Csabai}},
  \citenamefont {{Loveday}}, \citenamefont {{Doi}}, \citenamefont {{Fukugita}},
  \citenamefont {{Gott}}, \citenamefont {{Hennessy}}, \citenamefont {{Hogg}},
  \citenamefont {{Ivezi{\'c}}}, \citenamefont {{Knapp}}, \citenamefont
  {{Lamb}}, \citenamefont {{Lee}}, \citenamefont {{Lupton}}, \citenamefont
  {{McKay}}, \citenamefont {{Kunszt}}, \citenamefont {{Munn}}, \citenamefont
  {{O'Connell}}, \citenamefont {{Peoples}}, \citenamefont {{Pier}},
  \citenamefont {{Richmond}}, \citenamefont {{Rockosi}}, \citenamefont
  {{Schneider}}, \citenamefont {{Stoughton}}, \citenamefont {{Tucker}},
  \citenamefont {{Vanden Berk}}, \citenamefont {{Yanny}},\ and\ \citenamefont
  {{York}}}]{TegBlaStr04}%
  \BibitemOpen
  \bibfield  {author} {\bibinfo {author} {\bibfnamefont {M.}~\bibnamefont
  {{Tegmark}}}, \bibinfo {author} {\bibfnamefont {M.}~\bibnamefont
  {{Blanton}}}, \bibinfo {author} {\bibfnamefont {M.}~\bibnamefont
  {{Strauss}}}, \bibinfo {author} {\bibfnamefont {F.}~\bibnamefont {{Hoyle}}},
  \bibinfo {author} {\bibfnamefont {D.}~\bibnamefont {{Schlegel}}}, \bibinfo
  {author} {\bibfnamefont {R.}~\bibnamefont {{Scoccimarro}}}, \bibinfo {author}
  {\bibfnamefont {M.}~\bibnamefont {{Vogeley}}}, \bibinfo {author}
  {\bibfnamefont {D.}~\bibnamefont {{Weinberg}}}, \bibinfo {author}
  {\bibfnamefont {I.}~\bibnamefont {{Zehavi}}}, \bibinfo {author}
  {\bibfnamefont {A.}~\bibnamefont {{Berlind}}}, \bibinfo {author}
  {\bibfnamefont {T.}~\bibnamefont {{Budavari}}}, \bibinfo {author}
  {\bibfnamefont {A.}~\bibnamefont {{Connolly}}}, \bibinfo {author}
  {\bibfnamefont {D.}~\bibnamefont {{Eisenstein}}}, \bibinfo {author}
  {\bibfnamefont {D.}~\bibnamefont {{Finkbeiner}}}, \bibinfo {author}
  {\bibfnamefont {J.}~\bibnamefont {{Frieman}}}, \bibinfo {author}
  {\bibfnamefont {J.}~\bibnamefont {{Gunn}}}, \bibinfo {author} {\bibfnamefont
  {A.}~\bibnamefont {{Hamilton}}}, \bibinfo {author} {\bibfnamefont
  {L.}~\bibnamefont {{Hui}}}, \bibinfo {author} {\bibfnamefont
  {B.}~\bibnamefont {{Jain}}}, \bibinfo {author} {\bibfnamefont
  {D.}~\bibnamefont {{Johnston}}}, \bibinfo {author} {\bibfnamefont
  {S.}~\bibnamefont {{Kent}}}, \bibinfo {author} {\bibfnamefont
  {H.}~\bibnamefont {{Lin}}}, \bibinfo {author} {\bibfnamefont
  {R.}~\bibnamefont {{Nakajima}}}, \bibinfo {author} {\bibfnamefont
  {R.}~\bibnamefont {{Nichol}}}, \bibinfo {author} {\bibfnamefont
  {J.}~\bibnamefont {{Ostriker}}}, \bibinfo {author} {\bibfnamefont
  {A.}~\bibnamefont {{Pope}}}, \bibinfo {author} {\bibfnamefont
  {R.}~\bibnamefont {{Scranton}}}, \bibinfo {author} {\bibfnamefont
  {U.}~\bibnamefont {{Seljak}}}, \bibinfo {author} {\bibfnamefont
  {R.}~\bibnamefont {{Sheth}}}, \bibinfo {author} {\bibfnamefont
  {A.}~\bibnamefont {{Stebbins}}}, \bibinfo {author} {\bibfnamefont
  {A.}~\bibnamefont {{Szalay}}}, \bibinfo {author} {\bibfnamefont
  {I.}~\bibnamefont {{Szapudi}}}, \bibinfo {author} {\bibfnamefont
  {L.}~\bibnamefont {{Verde}}}, \bibinfo {author} {\bibfnamefont
  {Y.}~\bibnamefont {{Xu}}}, \bibinfo {author} {\bibfnamefont {J.}~\bibnamefont
  {{Annis}}}, \bibinfo {author} {\bibfnamefont {N.}~\bibnamefont {{Bahcall}}},
  \bibinfo {author} {\bibfnamefont {J.}~\bibnamefont {{Brinkmann}}}, \bibinfo
  {author} {\bibfnamefont {S.}~\bibnamefont {{Burles}}}, \bibinfo {author}
  {\bibfnamefont {F.}~\bibnamefont {{Castander}}}, \bibinfo {author}
  {\bibfnamefont {I.}~\bibnamefont {{Csabai}}}, \bibinfo {author}
  {\bibfnamefont {J.}~\bibnamefont {{Loveday}}}, \bibinfo {author}
  {\bibfnamefont {M.}~\bibnamefont {{Doi}}}, \bibinfo {author} {\bibfnamefont
  {M.}~\bibnamefont {{Fukugita}}}, \bibinfo {author} {\bibfnamefont
  {J.}~\bibnamefont {{Gott}}}, \bibinfo {author} {\bibfnamefont
  {G.}~\bibnamefont {{Hennessy}}}, \bibinfo {author} {\bibfnamefont
  {D.}~\bibnamefont {{Hogg}}}, \bibinfo {author} {\bibfnamefont {{\v
  Z}.}~\bibnamefont {{Ivezi{\'c}}}}, \bibinfo {author} {\bibfnamefont
  {G.}~\bibnamefont {{Knapp}}}, \bibinfo {author} {\bibfnamefont
  {D.}~\bibnamefont {{Lamb}}}, \bibinfo {author} {\bibfnamefont
  {B.}~\bibnamefont {{Lee}}}, \bibinfo {author} {\bibfnamefont
  {R.}~\bibnamefont {{Lupton}}}, \bibinfo {author} {\bibfnamefont
  {T.}~\bibnamefont {{McKay}}}, \bibinfo {author} {\bibfnamefont
  {P.}~\bibnamefont {{Kunszt}}}, \bibinfo {author} {\bibfnamefont
  {J.}~\bibnamefont {{Munn}}}, \bibinfo {author} {\bibfnamefont
  {L.}~\bibnamefont {{O'Connell}}}, \bibinfo {author} {\bibfnamefont
  {J.}~\bibnamefont {{Peoples}}}, \bibinfo {author} {\bibfnamefont
  {J.}~\bibnamefont {{Pier}}}, \bibinfo {author} {\bibfnamefont
  {M.}~\bibnamefont {{Richmond}}}, \bibinfo {author} {\bibfnamefont
  {C.}~\bibnamefont {{Rockosi}}}, \bibinfo {author} {\bibfnamefont
  {D.}~\bibnamefont {{Schneider}}}, \bibinfo {author} {\bibfnamefont
  {C.}~\bibnamefont {{Stoughton}}}, \bibinfo {author} {\bibfnamefont
  {D.}~\bibnamefont {{Tucker}}}, \bibinfo {author} {\bibfnamefont
  {D.}~\bibnamefont {{Vanden Berk}}}, \bibinfo {author} {\bibfnamefont
  {B.}~\bibnamefont {{Yanny}}}, \ and\ \bibinfo {author} {\bibfnamefont
  {D.}~\bibnamefont {{York}}},\ }\href@noop {} {\bibfield  {journal} {\bibinfo
  {journal} {\apj}\ }\textbf {\bibinfo {volume} {606}},\ \bibinfo {pages} {702}
  (\bibinfo {year} {2004})}\BibitemShut {NoStop}%
\bibitem [{\citenamefont {{Zehavi}}\ \emph {et~al.}(2005)\citenamefont
  {{Zehavi}}, \citenamefont {{Zheng}}, \citenamefont {{Weinberg}},
  \citenamefont {{Frieman}}, \citenamefont {{Berlind}}, \citenamefont
  {{Blanton}}, \citenamefont {{Scoccimarro}}, \citenamefont {{Sheth}},
  \citenamefont {{Strauss}}, \citenamefont {{Kayo}}, \citenamefont {{Suto}},
  \citenamefont {{Fukugita}}, \citenamefont {{Nakamura}}, \citenamefont
  {{Bahcall}}, \citenamefont {{Brinkmann}}, \citenamefont {{Gunn}},
  \citenamefont {{Hennessy}}, \citenamefont {{Ivezi{\'c}}}, \citenamefont
  {{Knapp}}, \citenamefont {{Loveday}}, \citenamefont {{Meiksin}},
  \citenamefont {{Schlegel}}, \citenamefont {{Schneider}}, \citenamefont
  {{Szapudi}}, \citenamefont {{Tegmark}}, \citenamefont {{Vogeley}},\ and\
  \citenamefont {{York}}}]{ZehZheWei05}%
  \BibitemOpen
  \bibfield  {author} {\bibinfo {author} {\bibfnamefont {I.}~\bibnamefont
  {{Zehavi}}}, \bibinfo {author} {\bibfnamefont {Z.}~\bibnamefont {{Zheng}}},
  \bibinfo {author} {\bibfnamefont {D.~H.}\ \bibnamefont {{Weinberg}}},
  \bibinfo {author} {\bibfnamefont {J.~A.}\ \bibnamefont {{Frieman}}}, \bibinfo
  {author} {\bibfnamefont {A.~A.}\ \bibnamefont {{Berlind}}}, \bibinfo {author}
  {\bibfnamefont {M.~R.}\ \bibnamefont {{Blanton}}}, \bibinfo {author}
  {\bibfnamefont {R.}~\bibnamefont {{Scoccimarro}}}, \bibinfo {author}
  {\bibfnamefont {R.~K.}\ \bibnamefont {{Sheth}}}, \bibinfo {author}
  {\bibfnamefont {M.~A.}\ \bibnamefont {{Strauss}}}, \bibinfo {author}
  {\bibfnamefont {I.}~\bibnamefont {{Kayo}}}, \bibinfo {author} {\bibfnamefont
  {Y.}~\bibnamefont {{Suto}}}, \bibinfo {author} {\bibfnamefont
  {M.}~\bibnamefont {{Fukugita}}}, \bibinfo {author} {\bibfnamefont
  {O.}~\bibnamefont {{Nakamura}}}, \bibinfo {author} {\bibfnamefont {N.~A.}\
  \bibnamefont {{Bahcall}}}, \bibinfo {author} {\bibfnamefont {J.}~\bibnamefont
  {{Brinkmann}}}, \bibinfo {author} {\bibfnamefont {J.~E.}\ \bibnamefont
  {{Gunn}}}, \bibinfo {author} {\bibfnamefont {G.~S.}\ \bibnamefont
  {{Hennessy}}}, \bibinfo {author} {\bibfnamefont {{\v Z}.}~\bibnamefont
  {{Ivezi{\'c}}}}, \bibinfo {author} {\bibfnamefont {G.~R.}\ \bibnamefont
  {{Knapp}}}, \bibinfo {author} {\bibfnamefont {J.}~\bibnamefont {{Loveday}}},
  \bibinfo {author} {\bibfnamefont {A.}~\bibnamefont {{Meiksin}}}, \bibinfo
  {author} {\bibfnamefont {D.~J.}\ \bibnamefont {{Schlegel}}}, \bibinfo
  {author} {\bibfnamefont {D.~P.}\ \bibnamefont {{Schneider}}}, \bibinfo
  {author} {\bibfnamefont {I.}~\bibnamefont {{Szapudi}}}, \bibinfo {author}
  {\bibfnamefont {M.}~\bibnamefont {{Tegmark}}}, \bibinfo {author}
  {\bibfnamefont {M.~S.}\ \bibnamefont {{Vogeley}}}, \ and\ \bibinfo {author}
  {\bibfnamefont {D.~G.}\ \bibnamefont {{York}}},\ }\href {\doibase
  10.1086/431891} {\bibfield  {journal} {\bibinfo  {journal} {\apj}\ }\textbf
  {\bibinfo {volume} {630}},\ \bibinfo {pages} {1} (\bibinfo {year}
  {2005})}\BibitemShut {NoStop}%
\bibitem [{\citenamefont {{Desjacques}}\ \emph {et~al.}(2018)\citenamefont
  {{Desjacques}}, \citenamefont {{Jeong}},\ and\ \citenamefont
  {{Schmidt}}}]{Desjacques:2018}%
  \BibitemOpen
  \bibfield  {author} {\bibinfo {author} {\bibfnamefont {V.}~\bibnamefont
  {{Desjacques}}}, \bibinfo {author} {\bibfnamefont {D.}~\bibnamefont
  {{Jeong}}}, \ and\ \bibinfo {author} {\bibfnamefont {F.}~\bibnamefont
  {{Schmidt}}},\ }\href {\doibase 10.1016/j.physrep.2017.12.002} {\bibfield
  {journal} {\bibinfo  {journal} {\physrep}\ }\textbf {\bibinfo {volume}
  {733}},\ \bibinfo {pages} {1} (\bibinfo {year} {2018})}\BibitemShut {NoStop}%
\bibitem [{\citenamefont {{Mo}}\ and\ \citenamefont
  {{White}}(1996)}]{MoWhi9609}%
  \BibitemOpen
  \bibfield  {author} {\bibinfo {author} {\bibfnamefont {H.~J.}\ \bibnamefont
  {{Mo}}}\ and\ \bibinfo {author} {\bibfnamefont {S.~D.~M.}\ \bibnamefont
  {{White}}},\ }\href@noop {} {\bibfield  {journal} {\bibinfo  {journal}
  {\mnras}\ }\textbf {\bibinfo {volume} {282}},\ \bibinfo {pages} {347}
  (\bibinfo {year} {1996})},\ \Eprint
  {http://arxiv.org/abs/arXiv:astro-ph/9512127} {arXiv:astro-ph/9512127}
  \BibitemShut {NoStop}%
\bibitem [{\citenamefont {{Mo}}\ \emph {et~al.}(1997)\citenamefont {{Mo}},
  \citenamefont {{Jing}},\ and\ \citenamefont {{White}}}]{MoJinWhi97}%
  \BibitemOpen
  \bibfield  {author} {\bibinfo {author} {\bibfnamefont {H.~J.}\ \bibnamefont
  {{Mo}}}, \bibinfo {author} {\bibfnamefont {Y.~P.}\ \bibnamefont {{Jing}}}, \
  and\ \bibinfo {author} {\bibfnamefont {S.~D.~M.}\ \bibnamefont {{White}}},\
  }\href@noop {} {\bibfield  {journal} {\bibinfo  {journal} {\mnras}\ }\textbf
  {\bibinfo {volume} {284}},\ \bibinfo {pages} {189} (\bibinfo {year}
  {1997})}\BibitemShut {NoStop}%
\bibitem [{\citenamefont {{Frieman}}\ and\ \citenamefont
  {{Gaztanaga}}(1994)}]{FriGaz94}%
  \BibitemOpen
  \bibfield  {author} {\bibinfo {author} {\bibfnamefont {J.}~\bibnamefont
  {{Frieman}}}\ and\ \bibinfo {author} {\bibfnamefont {E.}~\bibnamefont
  {{Gaztanaga}}},\ }\href@noop {} {\bibfield  {journal} {\bibinfo  {journal}
  {\apj}\ }\textbf {\bibinfo {volume} {425}},\ \bibinfo {pages} {392} (\bibinfo
  {year} {1994})}\BibitemShut {NoStop}%
\bibitem [{\citenamefont {{Chan}}\ \emph {et~al.}(2012)\citenamefont {{Chan}},
  \citenamefont {{Scoccimarro}},\ and\ \citenamefont {{Sheth}}}]{Chan:2012}%
  \BibitemOpen
  \bibfield  {author} {\bibinfo {author} {\bibfnamefont {K.~C.}\ \bibnamefont
  {{Chan}}}, \bibinfo {author} {\bibfnamefont {R.}~\bibnamefont
  {{Scoccimarro}}}, \ and\ \bibinfo {author} {\bibfnamefont {R.~K.}\
  \bibnamefont {{Sheth}}},\ }\href {\doibase 10.1103/PhysRevD.85.083509}
  {\bibfield  {journal} {\bibinfo  {journal} {\prd}\ }\textbf {\bibinfo
  {volume} {85}},\ \bibinfo {eid} {083509} (\bibinfo {year} {2012})},\ \Eprint
  {http://arxiv.org/abs/1201.3614} {arXiv:1201.3614 [astro-ph.CO]} \BibitemShut
  {NoStop}%
\bibitem [{\citenamefont {{Baldauf}}\ \emph {et~al.}(2012)\citenamefont
  {{Baldauf}}, \citenamefont {{Seljak}}, \citenamefont {{Desjacques}},\ and\
  \citenamefont {{McDonald}}}]{Baldauf:2012}%
  \BibitemOpen
  \bibfield  {author} {\bibinfo {author} {\bibfnamefont {T.}~\bibnamefont
  {{Baldauf}}}, \bibinfo {author} {\bibfnamefont {U.}~\bibnamefont {{Seljak}}},
  \bibinfo {author} {\bibfnamefont {V.}~\bibnamefont {{Desjacques}}}, \ and\
  \bibinfo {author} {\bibfnamefont {P.}~\bibnamefont {{McDonald}}},\ }\href
  {\doibase 10.1103/PhysRevD.86.083540} {\bibfield  {journal} {\bibinfo
  {journal} {\prd}\ }\textbf {\bibinfo {volume} {86}},\ \bibinfo {eid} {083540}
  (\bibinfo {year} {2012})},\ \Eprint {http://arxiv.org/abs/1201.4827}
  {arXiv:1201.4827 [astro-ph.CO]} \BibitemShut {NoStop}%
\bibitem [{\citenamefont {{Manera}}\ and\ \citenamefont
  {{Gazta{\~n}aga}}(2011)}]{ManGaz1107}%
  \BibitemOpen
  \bibfield  {author} {\bibinfo {author} {\bibfnamefont {M.}~\bibnamefont
  {{Manera}}}\ and\ \bibinfo {author} {\bibfnamefont {E.}~\bibnamefont
  {{Gazta{\~n}aga}}},\ }\href {\doibase 10.1111/j.1365-2966.2011.18705.x}
  {\bibfield  {journal} {\bibinfo  {journal} {\mnras}\ }\textbf {\bibinfo
  {volume} {415}},\ \bibinfo {pages} {383} (\bibinfo {year} {2011})},\ \Eprint
  {http://arxiv.org/abs/0912.0446} {arXiv:0912.0446 [astro-ph.CO]} \BibitemShut
  {NoStop}%
\bibitem [{\citenamefont {{Roth}}\ and\ \citenamefont
  {{Porciani}}(2011)}]{RotPor1107}%
  \BibitemOpen
  \bibfield  {author} {\bibinfo {author} {\bibfnamefont {N.}~\bibnamefont
  {{Roth}}}\ and\ \bibinfo {author} {\bibfnamefont {C.}~\bibnamefont
  {{Porciani}}},\ }\href {\doibase 10.1111/j.1365-2966.2011.18768.x} {\bibfield
   {journal} {\bibinfo  {journal} {\mnras}\ }\textbf {\bibinfo {volume}
  {415}},\ \bibinfo {pages} {829} (\bibinfo {year} {2011})},\ \Eprint
  {http://arxiv.org/abs/1101.1520} {arXiv:1101.1520 [astro-ph.CO]} \BibitemShut
  {NoStop}%
\bibitem [{\citenamefont {{Pollack}}\ \emph {et~al.}(2012)\citenamefont
  {{Pollack}}, \citenamefont {{Smith}},\ and\ \citenamefont
  {{Porciani}}}]{Pollack:2012}%
  \BibitemOpen
  \bibfield  {author} {\bibinfo {author} {\bibfnamefont {J.~E.}\ \bibnamefont
  {{Pollack}}}, \bibinfo {author} {\bibfnamefont {R.~E.}\ \bibnamefont
  {{Smith}}}, \ and\ \bibinfo {author} {\bibfnamefont {C.}~\bibnamefont
  {{Porciani}}},\ }\href {\doibase 10.1111/j.1365-2966.2011.20279.x} {\bibfield
   {journal} {\bibinfo  {journal} {\mnras}\ }\textbf {\bibinfo {volume}
  {420}},\ \bibinfo {pages} {3469} (\bibinfo {year} {2012})},\ \Eprint
  {http://arxiv.org/abs/1109.3458} {arXiv:1109.3458} \BibitemShut {NoStop}%
\bibitem [{\citenamefont {{Pollack}}\ \emph {et~al.}(2014)\citenamefont
  {{Pollack}}, \citenamefont {{Smith}},\ and\ \citenamefont
  {{Porciani}}}]{PolSmiPor1405}%
  \BibitemOpen
  \bibfield  {author} {\bibinfo {author} {\bibfnamefont {J.~E.}\ \bibnamefont
  {{Pollack}}}, \bibinfo {author} {\bibfnamefont {R.~E.}\ \bibnamefont
  {{Smith}}}, \ and\ \bibinfo {author} {\bibfnamefont {C.}~\bibnamefont
  {{Porciani}}},\ }\href {\doibase 10.1093/mnras/stu322} {\bibfield  {journal}
  {\bibinfo  {journal} {\mnras}\ }\textbf {\bibinfo {volume} {440}},\ \bibinfo
  {pages} {555} (\bibinfo {year} {2014})},\ \Eprint
  {http://arxiv.org/abs/1309.0504} {arXiv:1309.0504 [astro-ph.CO]} \BibitemShut
  {NoStop}%
\bibitem [{\citenamefont {{McDonald}}\ and\ \citenamefont
  {{Roy}}(2009)}]{McDonald:2009}%
  \BibitemOpen
  \bibfield  {author} {\bibinfo {author} {\bibfnamefont {P.}~\bibnamefont
  {{McDonald}}}\ and\ \bibinfo {author} {\bibfnamefont {A.}~\bibnamefont
  {{Roy}}},\ }\href {\doibase 10.1088/1475-7516/2009/08/020} {\bibfield
  {journal} {\bibinfo  {journal} {\jcap}\ }\textbf {\bibinfo {volume} {8}},\
  \bibinfo {eid} {020} (\bibinfo {year} {2009})},\ \Eprint
  {http://arxiv.org/abs/0902.0991} {arXiv:0902.0991 [astro-ph.CO]} \BibitemShut
  {NoStop}%
\bibitem [{\citenamefont {{Desjacques}}(2008)}]{Des0811}%
  \BibitemOpen
  \bibfield  {author} {\bibinfo {author} {\bibfnamefont {V.}~\bibnamefont
  {{Desjacques}}},\ }\href {\doibase 10.1103/PhysRevD.78.103503} {\bibfield
  {journal} {\bibinfo  {journal} {\prd}\ }\textbf {\bibinfo {volume} {78}},\
  \bibinfo {eid} {103503} (\bibinfo {year} {2008})},\ \Eprint
  {http://arxiv.org/abs/0806.0007} {arXiv:0806.0007 [astro-ph]} \BibitemShut
  {NoStop}%
\bibitem [{\citenamefont {{Desjacques}}\ \emph {et~al.}(2010)\citenamefont
  {{Desjacques}}, \citenamefont {{Crocce}}, \citenamefont {{Scoccimarro}},\
  and\ \citenamefont {{Sheth}}}]{DesCroSco1011}%
  \BibitemOpen
  \bibfield  {author} {\bibinfo {author} {\bibfnamefont {V.}~\bibnamefont
  {{Desjacques}}}, \bibinfo {author} {\bibfnamefont {M.}~\bibnamefont
  {{Crocce}}}, \bibinfo {author} {\bibfnamefont {R.}~\bibnamefont
  {{Scoccimarro}}}, \ and\ \bibinfo {author} {\bibfnamefont {R.~K.}\
  \bibnamefont {{Sheth}}},\ }\href {\doibase 10.1103/PhysRevD.82.103529}
  {\bibfield  {journal} {\bibinfo  {journal} {\prd}\ }\textbf {\bibinfo
  {volume} {82}},\ \bibinfo {pages} {103529} (\bibinfo {year} {2010})},\
  \Eprint {http://arxiv.org/abs/1009.3449} {arXiv:1009.3449 [astro-ph.CO]}
  \BibitemShut {NoStop}%
\bibitem [{\citenamefont {{Scherrer}}\ and\ \citenamefont
  {{Weinberg}}(1998)}]{Scherrer:1998}%
  \BibitemOpen
  \bibfield  {author} {\bibinfo {author} {\bibfnamefont {R.~J.}\ \bibnamefont
  {{Scherrer}}}\ and\ \bibinfo {author} {\bibfnamefont {D.~H.}\ \bibnamefont
  {{Weinberg}}},\ }\href {\doibase 10.1086/306113} {\bibfield  {journal}
  {\bibinfo  {journal} {\apj}\ }\textbf {\bibinfo {volume} {504}},\ \bibinfo
  {pages} {607} (\bibinfo {year} {1998})},\ \Eprint
  {http://arxiv.org/abs/astro-ph/9712192} {arXiv:astro-ph/9712192 [astro-ph]}
  \BibitemShut {NoStop}%
\bibitem [{\citenamefont {{Dekel}}\ and\ \citenamefont
  {{Lahav}}(1999)}]{Dekel:1999}%
  \BibitemOpen
  \bibfield  {author} {\bibinfo {author} {\bibfnamefont {A.}~\bibnamefont
  {{Dekel}}}\ and\ \bibinfo {author} {\bibfnamefont {O.}~\bibnamefont
  {{Lahav}}},\ }\href {\doibase 10.1086/307428} {\bibfield  {journal} {\bibinfo
   {journal} {\apj}\ }\textbf {\bibinfo {volume} {520}},\ \bibinfo {pages} {24}
  (\bibinfo {year} {1999})},\ \Eprint {http://arxiv.org/abs/astro-ph/9806193}
  {arXiv:astro-ph/9806193 [astro-ph]} \BibitemShut {NoStop}%
\bibitem [{\citenamefont {{Taruya}}\ and\ \citenamefont
  {{Soda}}(1999)}]{TarSod9909}%
  \BibitemOpen
  \bibfield  {author} {\bibinfo {author} {\bibfnamefont {A.}~\bibnamefont
  {{Taruya}}}\ and\ \bibinfo {author} {\bibfnamefont {J.}~\bibnamefont
  {{Soda}}},\ }\href {\doibase 10.1086/307612} {\bibfield  {journal} {\bibinfo
  {journal} {\apj}\ }\textbf {\bibinfo {volume} {522}},\ \bibinfo {pages} {46}
  (\bibinfo {year} {1999})},\ \Eprint {http://arxiv.org/abs/astro-ph/9809204}
  {arXiv:astro-ph/9809204 [astro-ph]} \BibitemShut {NoStop}%
\bibitem [{\citenamefont {{Matsubara}}(1999)}]{Mat9911}%
  \BibitemOpen
  \bibfield  {author} {\bibinfo {author} {\bibfnamefont {T.}~\bibnamefont
  {{Matsubara}}},\ }\href {\doibase 10.1086/307931} {\bibfield  {journal}
  {\bibinfo  {journal} {\apj}\ }\textbf {\bibinfo {volume} {525}},\ \bibinfo
  {pages} {543} (\bibinfo {year} {1999})},\ \Eprint
  {http://arxiv.org/abs/astro-ph/9906029} {arXiv:astro-ph/9906029 [astro-ph]}
  \BibitemShut {NoStop}%
\bibitem [{\citenamefont {{Sheth}}\ and\ \citenamefont
  {{Lemson}}(1999)}]{SheLem99}%
  \BibitemOpen
  \bibfield  {author} {\bibinfo {author} {\bibfnamefont {R.}~\bibnamefont
  {{Sheth}}}\ and\ \bibinfo {author} {\bibfnamefont {G.}~\bibnamefont
  {{Lemson}}},\ }\href@noop {} {\bibfield  {journal} {\bibinfo  {journal}
  {\mnras}\ }\textbf {\bibinfo {volume} {304}},\ \bibinfo {pages} {767}
  (\bibinfo {year} {1999})}\BibitemShut {NoStop}%
\bibitem [{\citenamefont {{Smith}}\ \emph {et~al.}(2007)\citenamefont
  {{Smith}}, \citenamefont {{Scoccimarro}},\ and\ \citenamefont
  {{Sheth}}}]{SmiScoShe0703}%
  \BibitemOpen
  \bibfield  {author} {\bibinfo {author} {\bibfnamefont {R.~E.}\ \bibnamefont
  {{Smith}}}, \bibinfo {author} {\bibfnamefont {R.}~\bibnamefont
  {{Scoccimarro}}}, \ and\ \bibinfo {author} {\bibfnamefont {R.~K.}\
  \bibnamefont {{Sheth}}},\ }\href {\doibase 10.1103/PhysRevD.75.063512}
  {\bibfield  {journal} {\bibinfo  {journal} {\prd}\ }\textbf {\bibinfo
  {volume} {75}},\ \bibinfo {pages} {063512} (\bibinfo {year} {2007})},\
  \Eprint {http://arxiv.org/abs/arXiv:astro-ph/0609547}
  {arXiv:astro-ph/0609547} \BibitemShut {NoStop}%
\bibitem [{\citenamefont {{Baldauf}}\ \emph {et~al.}(2013)\citenamefont
  {{Baldauf}}, \citenamefont {{Seljak}}, \citenamefont {{Smith}}, \citenamefont
  {{Hamaus}},\ and\ \citenamefont {{Desjacques}}}]{BalSelSmi1310}%
  \BibitemOpen
  \bibfield  {author} {\bibinfo {author} {\bibfnamefont {T.}~\bibnamefont
  {{Baldauf}}}, \bibinfo {author} {\bibfnamefont {U.}~\bibnamefont {{Seljak}}},
  \bibinfo {author} {\bibfnamefont {R.~E.}\ \bibnamefont {{Smith}}}, \bibinfo
  {author} {\bibfnamefont {N.}~\bibnamefont {{Hamaus}}}, \ and\ \bibinfo
  {author} {\bibfnamefont {V.}~\bibnamefont {{Desjacques}}},\ }\href {\doibase
  10.1103/PhysRevD.88.083507} {\bibfield  {journal} {\bibinfo  {journal}
  {\prd}\ }\textbf {\bibinfo {volume} {88}},\ \bibinfo {eid} {083507} (\bibinfo
  {year} {2013})},\ \Eprint {http://arxiv.org/abs/1305.2917} {arXiv:1305.2917
  [astro-ph.CO]} \BibitemShut {NoStop}%
\bibitem [{\citenamefont {Eggemeier}\ \emph {et~al.}(2020)\citenamefont
  {Eggemeier}, \citenamefont {Scoccimarro}, \citenamefont {Crocce},
  \citenamefont {Pezzotta},\ and\ \citenamefont {S\'anchez}}]{EggScoCro2011}%
  \BibitemOpen
  \bibfield  {author} {\bibinfo {author} {\bibfnamefont {A.}~\bibnamefont
  {Eggemeier}}, \bibinfo {author} {\bibfnamefont {R.}~\bibnamefont
  {Scoccimarro}}, \bibinfo {author} {\bibfnamefont {M.}~\bibnamefont {Crocce}},
  \bibinfo {author} {\bibfnamefont {A.}~\bibnamefont {Pezzotta}}, \ and\
  \bibinfo {author} {\bibfnamefont {A.~G.}\ \bibnamefont {S\'anchez}},\ }\href
  {\doibase 10.1103/PhysRevD.102.103530} {\bibfield  {journal} {\bibinfo
  {journal} {Phys. Rev. D}\ }\textbf {\bibinfo {volume} {102}},\ \bibinfo
  {pages} {103530} (\bibinfo {year} {2020})}\BibitemShut {NoStop}%
\bibitem [{\citenamefont {{S{\'a}nchez}}\ \emph {et~al.}(2017)\citenamefont
  {{S{\'a}nchez}}, \citenamefont {{Scoccimarro}}, \citenamefont {{Crocce}},
  \citenamefont {{Grieb}}, \citenamefont {{Salazar-Albornoz}}, \citenamefont
  {{Dalla Vecchia}}, \citenamefont {{Lippich}}, \citenamefont {{Beutler}},
  \citenamefont {{Brownstein}}, \citenamefont {{Chuang}}, \citenamefont
  {{Eisenstein}}, \citenamefont {{Kitaura}}, \citenamefont {{Olmstead}},
  \citenamefont {{Percival}}, \citenamefont {{Prada}}, \citenamefont
  {{Rodr{\'{\i}}guez-Torres}}, \citenamefont {{Ross}}, \citenamefont
  {{Samushia}}, \citenamefont {{Seo}}, \citenamefont {{Tinker}}, \citenamefont
  {{Tojeiro}}, \citenamefont {{Vargas-Maga{\~n}a}}, \citenamefont {{Wang}},\
  and\ \citenamefont {{Zhao}}}]{SanScoCro1701}%
  \BibitemOpen
  \bibfield  {author} {\bibinfo {author} {\bibfnamefont {A.~G.}\ \bibnamefont
  {{S{\'a}nchez}}}, \bibinfo {author} {\bibfnamefont {R.}~\bibnamefont
  {{Scoccimarro}}}, \bibinfo {author} {\bibfnamefont {M.}~\bibnamefont
  {{Crocce}}}, \bibinfo {author} {\bibfnamefont {J.~N.}\ \bibnamefont
  {{Grieb}}}, \bibinfo {author} {\bibfnamefont {S.}~\bibnamefont
  {{Salazar-Albornoz}}}, \bibinfo {author} {\bibfnamefont {C.}~\bibnamefont
  {{Dalla Vecchia}}}, \bibinfo {author} {\bibfnamefont {M.}~\bibnamefont
  {{Lippich}}}, \bibinfo {author} {\bibfnamefont {F.}~\bibnamefont
  {{Beutler}}}, \bibinfo {author} {\bibfnamefont {J.~R.}\ \bibnamefont
  {{Brownstein}}}, \bibinfo {author} {\bibfnamefont {C.-H.}\ \bibnamefont
  {{Chuang}}}, \bibinfo {author} {\bibfnamefont {D.~J.}\ \bibnamefont
  {{Eisenstein}}}, \bibinfo {author} {\bibfnamefont {F.-S.}\ \bibnamefont
  {{Kitaura}}}, \bibinfo {author} {\bibfnamefont {M.~D.}\ \bibnamefont
  {{Olmstead}}}, \bibinfo {author} {\bibfnamefont {W.~J.}\ \bibnamefont
  {{Percival}}}, \bibinfo {author} {\bibfnamefont {F.}~\bibnamefont {{Prada}}},
  \bibinfo {author} {\bibfnamefont {S.}~\bibnamefont
  {{Rodr{\'{\i}}guez-Torres}}}, \bibinfo {author} {\bibfnamefont {A.~J.}\
  \bibnamefont {{Ross}}}, \bibinfo {author} {\bibfnamefont {L.}~\bibnamefont
  {{Samushia}}}, \bibinfo {author} {\bibfnamefont {H.-J.}\ \bibnamefont
  {{Seo}}}, \bibinfo {author} {\bibfnamefont {J.}~\bibnamefont {{Tinker}}},
  \bibinfo {author} {\bibfnamefont {R.}~\bibnamefont {{Tojeiro}}}, \bibinfo
  {author} {\bibfnamefont {M.}~\bibnamefont {{Vargas-Maga{\~n}a}}}, \bibinfo
  {author} {\bibfnamefont {Y.}~\bibnamefont {{Wang}}}, \ and\ \bibinfo {author}
  {\bibfnamefont {G.-B.}\ \bibnamefont {{Zhao}}},\ }\href {\doibase
  10.1093/mnras/stw2443} {\bibfield  {journal} {\bibinfo  {journal} {\mnras}\
  }\textbf {\bibinfo {volume} {464}},\ \bibinfo {pages} {1640} (\bibinfo {year}
  {2017})},\ \Eprint {http://arxiv.org/abs/1607.03147} {arXiv:1607.03147}
  \BibitemShut {NoStop}%
\bibitem [{\citenamefont {{Gil-Mar{\'\i}n}}\ \emph {et~al.}(2017)\citenamefont
  {{Gil-Mar{\'\i}n}}, \citenamefont {{Percival}}, \citenamefont {{Verde}},
  \citenamefont {{Brownstein}}, \citenamefont {{Chuang}}, \citenamefont
  {{Kitaura}}, \citenamefont {{Rodr{\'\i}guez-Torres}},\ and\ \citenamefont
  {{Olmstead}}}]{GilPerVer1702}%
  \BibitemOpen
  \bibfield  {author} {\bibinfo {author} {\bibfnamefont {H.}~\bibnamefont
  {{Gil-Mar{\'\i}n}}}, \bibinfo {author} {\bibfnamefont {W.~J.}\ \bibnamefont
  {{Percival}}}, \bibinfo {author} {\bibfnamefont {L.}~\bibnamefont {{Verde}}},
  \bibinfo {author} {\bibfnamefont {J.~R.}\ \bibnamefont {{Brownstein}}},
  \bibinfo {author} {\bibfnamefont {C.-H.}\ \bibnamefont {{Chuang}}}, \bibinfo
  {author} {\bibfnamefont {F.-S.}\ \bibnamefont {{Kitaura}}}, \bibinfo {author}
  {\bibfnamefont {S.~A.}\ \bibnamefont {{Rodr{\'\i}guez-Torres}}}, \ and\
  \bibinfo {author} {\bibfnamefont {M.~D.}\ \bibnamefont {{Olmstead}}},\ }\href
  {\doibase 10.1093/mnras/stw2679} {\bibfield  {journal} {\bibinfo  {journal}
  {\mnras}\ }\textbf {\bibinfo {volume} {465}},\ \bibinfo {pages} {1757}
  (\bibinfo {year} {2017})},\ \Eprint {http://arxiv.org/abs/1606.00439}
  {arXiv:1606.00439 [astro-ph.CO]} \BibitemShut {NoStop}%
\bibitem [{\citenamefont {{Beutler}}\ \emph {et~al.}(2017)\citenamefont
  {{Beutler}}, \citenamefont {{Seo}}, \citenamefont {{Saito}}, \citenamefont
  {{Chuang}}, \citenamefont {{Cuesta}}, \citenamefont {{Eisenstein}},
  \citenamefont {{Gil-Mar{\'\i}n}}, \citenamefont {{Grieb}}, \citenamefont
  {{Hand}}, \citenamefont {{Kitaura}}, \citenamefont {{Modi}}, \citenamefont
  {{Nichol}}, \citenamefont {{Olmstead}}, \citenamefont {{Percival}},
  \citenamefont {{Prada}}, \citenamefont {{S{\'a}nchez}}, \citenamefont
  {{Rodriguez-Torres}}, \citenamefont {{Ross}}, \citenamefont {{Ross}},
  \citenamefont {{Schneider}}, \citenamefont {{Tinker}}, \citenamefont
  {{Tojeiro}},\ and\ \citenamefont {{Vargas-Maga{\~n}a}}}]{BeuSeoSai1704}%
  \BibitemOpen
  \bibfield  {author} {\bibinfo {author} {\bibfnamefont {F.}~\bibnamefont
  {{Beutler}}}, \bibinfo {author} {\bibfnamefont {H.-J.}\ \bibnamefont
  {{Seo}}}, \bibinfo {author} {\bibfnamefont {S.}~\bibnamefont {{Saito}}},
  \bibinfo {author} {\bibfnamefont {C.-H.}\ \bibnamefont {{Chuang}}}, \bibinfo
  {author} {\bibfnamefont {A.~J.}\ \bibnamefont {{Cuesta}}}, \bibinfo {author}
  {\bibfnamefont {D.~J.}\ \bibnamefont {{Eisenstein}}}, \bibinfo {author}
  {\bibfnamefont {H.}~\bibnamefont {{Gil-Mar{\'\i}n}}}, \bibinfo {author}
  {\bibfnamefont {J.~N.}\ \bibnamefont {{Grieb}}}, \bibinfo {author}
  {\bibfnamefont {N.}~\bibnamefont {{Hand}}}, \bibinfo {author} {\bibfnamefont
  {F.-S.}\ \bibnamefont {{Kitaura}}}, \bibinfo {author} {\bibfnamefont
  {C.}~\bibnamefont {{Modi}}}, \bibinfo {author} {\bibfnamefont {R.~C.}\
  \bibnamefont {{Nichol}}}, \bibinfo {author} {\bibfnamefont {M.~D.}\
  \bibnamefont {{Olmstead}}}, \bibinfo {author} {\bibfnamefont {W.~J.}\
  \bibnamefont {{Percival}}}, \bibinfo {author} {\bibfnamefont
  {F.}~\bibnamefont {{Prada}}}, \bibinfo {author} {\bibfnamefont {A.~G.}\
  \bibnamefont {{S{\'a}nchez}}}, \bibinfo {author} {\bibfnamefont
  {S.}~\bibnamefont {{Rodriguez-Torres}}}, \bibinfo {author} {\bibfnamefont
  {A.~J.}\ \bibnamefont {{Ross}}}, \bibinfo {author} {\bibfnamefont {N.~P.}\
  \bibnamefont {{Ross}}}, \bibinfo {author} {\bibfnamefont {D.~P.}\
  \bibnamefont {{Schneider}}}, \bibinfo {author} {\bibfnamefont
  {J.}~\bibnamefont {{Tinker}}}, \bibinfo {author} {\bibfnamefont
  {R.}~\bibnamefont {{Tojeiro}}}, \ and\ \bibinfo {author} {\bibfnamefont
  {M.}~\bibnamefont {{Vargas-Maga{\~n}a}}},\ }\href {\doibase
  10.1093/mnras/stw3298} {\bibfield  {journal} {\bibinfo  {journal} {\mnras}\
  }\textbf {\bibinfo {volume} {466}},\ \bibinfo {pages} {2242} (\bibinfo {year}
  {2017})},\ \Eprint {http://arxiv.org/abs/1607.03150} {arXiv:1607.03150
  [astro-ph.CO]} \BibitemShut {NoStop}%
\bibitem [{\citenamefont {{Grieb}}\ \emph {et~al.}(2017)\citenamefont
  {{Grieb}}, \citenamefont {{S{\'a}nchez}}, \citenamefont {{Salazar-Albornoz}},
  \citenamefont {{Scoccimarro}}, \citenamefont {{Crocce}}, \citenamefont
  {{Dalla Vecchia}}, \citenamefont {{Montesano}}, \citenamefont
  {{Gil-Mar{\'{\i}}n}}, \citenamefont {{Ross}}, \citenamefont {{Beutler}},
  \citenamefont {{Rodr{\'{\i}}guez-Torres}}, \citenamefont {{Chuang}},
  \citenamefont {{Prada}}, \citenamefont {{Kitaura}}, \citenamefont {{Cuesta}},
  \citenamefont {{Eisenstein}}, \citenamefont {{Percival}}, \citenamefont
  {{Vargas-Maga{\~n}a}}, \citenamefont {{Tinker}}, \citenamefont {{Tojeiro}},
  \citenamefont {{Brownstein}}, \citenamefont {{Maraston}}, \citenamefont
  {{Nichol}}, \citenamefont {{Olmstead}}, \citenamefont {{Samushia}},
  \citenamefont {{Seo}}, \citenamefont {{Streblyanska}},\ and\ \citenamefont
  {{Zhao}}}]{GriSanSal1705}%
  \BibitemOpen
  \bibfield  {author} {\bibinfo {author} {\bibfnamefont {J.~N.}\ \bibnamefont
  {{Grieb}}}, \bibinfo {author} {\bibfnamefont {A.~G.}\ \bibnamefont
  {{S{\'a}nchez}}}, \bibinfo {author} {\bibfnamefont {S.}~\bibnamefont
  {{Salazar-Albornoz}}}, \bibinfo {author} {\bibfnamefont {R.}~\bibnamefont
  {{Scoccimarro}}}, \bibinfo {author} {\bibfnamefont {M.}~\bibnamefont
  {{Crocce}}}, \bibinfo {author} {\bibfnamefont {C.}~\bibnamefont {{Dalla
  Vecchia}}}, \bibinfo {author} {\bibfnamefont {F.}~\bibnamefont
  {{Montesano}}}, \bibinfo {author} {\bibfnamefont {H.}~\bibnamefont
  {{Gil-Mar{\'{\i}}n}}}, \bibinfo {author} {\bibfnamefont {A.~J.}\ \bibnamefont
  {{Ross}}}, \bibinfo {author} {\bibfnamefont {F.}~\bibnamefont {{Beutler}}},
  \bibinfo {author} {\bibfnamefont {S.}~\bibnamefont
  {{Rodr{\'{\i}}guez-Torres}}}, \bibinfo {author} {\bibfnamefont {C.-H.}\
  \bibnamefont {{Chuang}}}, \bibinfo {author} {\bibfnamefont {F.}~\bibnamefont
  {{Prada}}}, \bibinfo {author} {\bibfnamefont {F.-S.}\ \bibnamefont
  {{Kitaura}}}, \bibinfo {author} {\bibfnamefont {A.~J.}\ \bibnamefont
  {{Cuesta}}}, \bibinfo {author} {\bibfnamefont {D.~J.}\ \bibnamefont
  {{Eisenstein}}}, \bibinfo {author} {\bibfnamefont {W.~J.}\ \bibnamefont
  {{Percival}}}, \bibinfo {author} {\bibfnamefont {M.}~\bibnamefont
  {{Vargas-Maga{\~n}a}}}, \bibinfo {author} {\bibfnamefont {J.~L.}\
  \bibnamefont {{Tinker}}}, \bibinfo {author} {\bibfnamefont {R.}~\bibnamefont
  {{Tojeiro}}}, \bibinfo {author} {\bibfnamefont {J.~R.}\ \bibnamefont
  {{Brownstein}}}, \bibinfo {author} {\bibfnamefont {C.}~\bibnamefont
  {{Maraston}}}, \bibinfo {author} {\bibfnamefont {R.~C.}\ \bibnamefont
  {{Nichol}}}, \bibinfo {author} {\bibfnamefont {M.~D.}\ \bibnamefont
  {{Olmstead}}}, \bibinfo {author} {\bibfnamefont {L.}~\bibnamefont
  {{Samushia}}}, \bibinfo {author} {\bibfnamefont {H.-J.}\ \bibnamefont
  {{Seo}}}, \bibinfo {author} {\bibfnamefont {A.}~\bibnamefont
  {{Streblyanska}}}, \ and\ \bibinfo {author} {\bibfnamefont {G.-b.}\
  \bibnamefont {{Zhao}}},\ }\href {\doibase 10.1093/mnras/stw3384} {\bibfield
  {journal} {\bibinfo  {journal} {\mnras}\ }\textbf {\bibinfo {volume} {467}},\
  \bibinfo {pages} {2085} (\bibinfo {year} {2017})},\ \Eprint
  {http://arxiv.org/abs/1607.03143} {arXiv:1607.03143} \BibitemShut {NoStop}%
\bibitem [{\citenamefont {{Ivanov}}\ \emph {et~al.}(2019)\citenamefont
  {{Ivanov}}, \citenamefont {{Simonovi{\'c}}},\ and\ \citenamefont
  {{Zaldarriaga}}}]{IvaSimZal1909}%
  \BibitemOpen
  \bibfield  {author} {\bibinfo {author} {\bibfnamefont {M.~M.}\ \bibnamefont
  {{Ivanov}}}, \bibinfo {author} {\bibfnamefont {M.}~\bibnamefont
  {{Simonovi{\'c}}}}, \ and\ \bibinfo {author} {\bibfnamefont {M.}~\bibnamefont
  {{Zaldarriaga}}},\ }\href@noop {} {\bibfield  {journal} {\bibinfo  {journal}
  {arXiv e-prints}\ ,\ \bibinfo {eid} {arXiv:1909.05277}} (\bibinfo {year}
  {2019})},\ \Eprint {http://arxiv.org/abs/1909.05277} {arXiv:1909.05277
  [astro-ph.CO]} \BibitemShut {NoStop}%
\bibitem [{\citenamefont {{D'Amico}}\ \emph {et~al.}(2019)\citenamefont
  {{D'Amico}}, \citenamefont {{Gleyzes}}, \citenamefont {{Kokron}},
  \citenamefont {{Markovic}}, \citenamefont {{Senatore}}, \citenamefont
  {{Zhang}}, \citenamefont {{Beutler}},\ and\ \citenamefont
  {{Gil-Mar{\'\i}n}}}]{AmiGleKok1909}%
  \BibitemOpen
  \bibfield  {author} {\bibinfo {author} {\bibfnamefont {G.}~\bibnamefont
  {{D'Amico}}}, \bibinfo {author} {\bibfnamefont {J.}~\bibnamefont
  {{Gleyzes}}}, \bibinfo {author} {\bibfnamefont {N.}~\bibnamefont {{Kokron}}},
  \bibinfo {author} {\bibfnamefont {D.}~\bibnamefont {{Markovic}}}, \bibinfo
  {author} {\bibfnamefont {L.}~\bibnamefont {{Senatore}}}, \bibinfo {author}
  {\bibfnamefont {P.}~\bibnamefont {{Zhang}}}, \bibinfo {author} {\bibfnamefont
  {F.}~\bibnamefont {{Beutler}}}, \ and\ \bibinfo {author} {\bibfnamefont
  {H.}~\bibnamefont {{Gil-Mar{\'\i}n}}},\ }\href@noop {} {\bibfield  {journal}
  {\bibinfo  {journal} {arXiv e-prints}\ ,\ \bibinfo {eid} {arXiv:1909.05271}}
  (\bibinfo {year} {2019})},\ \Eprint {http://arxiv.org/abs/1909.05271}
  {arXiv:1909.05271 [astro-ph.CO]} \BibitemShut {NoStop}%
\bibitem [{\citenamefont {{Tr{\"o}ster}}\ \emph {et~al.}(2020)\citenamefont
  {{Tr{\"o}ster}}, \citenamefont {{S{\'a}nchez}}, \citenamefont {{Asgari}},
  \citenamefont {{Blake}}, \citenamefont {{Crocce}}, \citenamefont {{Heymans}},
  \citenamefont {{Hildebrandt}}, \citenamefont {{Joachimi}}, \citenamefont
  {{Joudaki}}, \citenamefont {{Kannawadi}}, \citenamefont {{Lin}},\ and\
  \citenamefont {{Wright}}}]{TroSanAsg2001}%
  \BibitemOpen
  \bibfield  {author} {\bibinfo {author} {\bibfnamefont {T.}~\bibnamefont
  {{Tr{\"o}ster}}}, \bibinfo {author} {\bibfnamefont {A.~G.}\ \bibnamefont
  {{S{\'a}nchez}}}, \bibinfo {author} {\bibfnamefont {M.}~\bibnamefont
  {{Asgari}}}, \bibinfo {author} {\bibfnamefont {C.}~\bibnamefont {{Blake}}},
  \bibinfo {author} {\bibfnamefont {M.}~\bibnamefont {{Crocce}}}, \bibinfo
  {author} {\bibfnamefont {C.}~\bibnamefont {{Heymans}}}, \bibinfo {author}
  {\bibfnamefont {H.}~\bibnamefont {{Hildebrandt}}}, \bibinfo {author}
  {\bibfnamefont {B.}~\bibnamefont {{Joachimi}}}, \bibinfo {author}
  {\bibfnamefont {S.}~\bibnamefont {{Joudaki}}}, \bibinfo {author}
  {\bibfnamefont {A.}~\bibnamefont {{Kannawadi}}}, \bibinfo {author}
  {\bibfnamefont {C.-A.}\ \bibnamefont {{Lin}}}, \ and\ \bibinfo {author}
  {\bibfnamefont {A.}~\bibnamefont {{Wright}}},\ }\href {\doibase
  10.1051/0004-6361/201936772} {\bibfield  {journal} {\bibinfo  {journal}
  {\aap}\ }\textbf {\bibinfo {volume} {633}},\ \bibinfo {eid} {L10} (\bibinfo
  {year} {2020})},\ \Eprint {http://arxiv.org/abs/1909.11006} {arXiv:1909.11006
  [astro-ph.CO]} \BibitemShut {NoStop}%
\bibitem [{\citenamefont {Eggemeier}\ \emph {et~al.}(2021)\citenamefont
  {Eggemeier}, \citenamefont {Scoccimarro}, \citenamefont {Smith},
  \citenamefont {Crocce}, \citenamefont {Pezzotta},\ and\ \citenamefont
  {Sánchez}}]{EggScoSmi2021}%
  \BibitemOpen
  \bibfield  {author} {\bibinfo {author} {\bibfnamefont {A.}~\bibnamefont
  {Eggemeier}}, \bibinfo {author} {\bibfnamefont {R.}~\bibnamefont
  {Scoccimarro}}, \bibinfo {author} {\bibfnamefont {R.~E.}\ \bibnamefont
  {Smith}}, \bibinfo {author} {\bibfnamefont {M.}~\bibnamefont {Crocce}},
  \bibinfo {author} {\bibfnamefont {A.}~\bibnamefont {Pezzotta}}, \ and\
  \bibinfo {author} {\bibfnamefont {A.~G.}\ \bibnamefont {Sánchez}},\
  }\href@noop {} {\enquote {\bibinfo {title} {Testing one-loop galaxy bias:
  joint analysis of power spectrum and bispectrum},}\ } (\bibinfo {year}
  {2021}),\ \Eprint {http://arxiv.org/abs/2102.06902} {arXiv:2102.06902
  [astro-ph.CO]} \BibitemShut {NoStop}%
\bibitem [{\citenamefont {Eisenstein}\ \emph {et~al.}(2011)\citenamefont
  {Eisenstein}, \citenamefont {Weinberg}, \citenamefont {Agol}, \citenamefont
  {Aihara}, \citenamefont {Allende~Prieto}, \citenamefont {Anderson},
  \citenamefont {Arns}, \citenamefont {Aubourg}, \citenamefont {Bailey},
  \citenamefont {Balbinot},\ and\ \citenamefont {et~al.}}]{Eisenstein2011}%
  \BibitemOpen
  \bibfield  {author} {\bibinfo {author} {\bibfnamefont {D.~J.}\ \bibnamefont
  {Eisenstein}}, \bibinfo {author} {\bibfnamefont {D.~H.}\ \bibnamefont
  {Weinberg}}, \bibinfo {author} {\bibfnamefont {E.}~\bibnamefont {Agol}},
  \bibinfo {author} {\bibfnamefont {H.}~\bibnamefont {Aihara}}, \bibinfo
  {author} {\bibfnamefont {C.}~\bibnamefont {Allende~Prieto}}, \bibinfo
  {author} {\bibfnamefont {S.~F.}\ \bibnamefont {Anderson}}, \bibinfo {author}
  {\bibfnamefont {J.~A.}\ \bibnamefont {Arns}}, \bibinfo {author}
  {\bibfnamefont {E.}~\bibnamefont {Aubourg}}, \bibinfo {author} {\bibfnamefont
  {S.}~\bibnamefont {Bailey}}, \bibinfo {author} {\bibfnamefont
  {E.}~\bibnamefont {Balbinot}}, \ and\ \bibinfo {author} {\bibnamefont
  {et~al.}},\ }\href {\doibase 10.1088/0004-6256/142/3/72} {\bibfield
  {journal} {\bibinfo  {journal} {The Astronomical Journal}\ }\textbf {\bibinfo
  {volume} {142}},\ \bibinfo {pages} {72} (\bibinfo {year} {2011})}\BibitemShut
  {NoStop}%
\bibitem [{\citenamefont {{Dawson}}\ \emph {et~al.}(2013)\citenamefont
  {{Dawson}}, \citenamefont {{Schlegel}},\ and\ \citenamefont
  {et~al.}}]{Dawson2012}%
  \BibitemOpen
  \bibfield  {author} {\bibinfo {author} {\bibfnamefont {K.~S.}\ \bibnamefont
  {{Dawson}}}, \bibinfo {author} {\bibfnamefont {D.~J.}\ \bibnamefont
  {{Schlegel}}}, \ and\ \bibinfo {author} {\bibnamefont {et~al.}},\ }\href
  {\doibase 10.1088/0004-6256/145/1/10} {\bibfield  {journal} {\bibinfo
  {journal} {\aj}\ }\textbf {\bibinfo {volume} {145}},\ \bibinfo {eid} {10}
  (\bibinfo {year} {2013})},\ \Eprint {http://arxiv.org/abs/1208.0022}
  {arXiv:1208.0022 [astro-ph.CO]} \BibitemShut {NoStop}%
\bibitem [{\citenamefont {Reid}\ \emph {et~al.}(2015)\citenamefont {Reid},
  \citenamefont {Ho}, \citenamefont {Padmanabhan}, \citenamefont {Percival},
  \citenamefont {Tinker}, \citenamefont {Tojeiro}, \citenamefont {White},
  \citenamefont {Eisenstein}, \citenamefont {Maraston},\ and\ \citenamefont
  {et~al.}}]{reid2015}%
  \BibitemOpen
  \bibfield  {author} {\bibinfo {author} {\bibfnamefont {B.}~\bibnamefont
  {Reid}}, \bibinfo {author} {\bibfnamefont {S.}~\bibnamefont {Ho}}, \bibinfo
  {author} {\bibfnamefont {N.}~\bibnamefont {Padmanabhan}}, \bibinfo {author}
  {\bibfnamefont {W.~J.}\ \bibnamefont {Percival}}, \bibinfo {author}
  {\bibfnamefont {J.}~\bibnamefont {Tinker}}, \bibinfo {author} {\bibfnamefont
  {R.}~\bibnamefont {Tojeiro}}, \bibinfo {author} {\bibfnamefont
  {M.}~\bibnamefont {White}}, \bibinfo {author} {\bibfnamefont {D.~J.}\
  \bibnamefont {Eisenstein}}, \bibinfo {author} {\bibfnamefont
  {C.}~\bibnamefont {Maraston}}, \ and\ \bibinfo {author} {\bibnamefont
  {et~al.}},\ }\href@noop {} {\  (\bibinfo {year} {2015})},\ \Eprint
  {http://arxiv.org/abs/1509.06529} {arXiv:1509.06529 [astro-ph.CO]}
  \BibitemShut {NoStop}%
\bibitem [{\citenamefont {{Strauss}}\ \emph {et~al.}(2002)\citenamefont
  {{Strauss}}, \citenamefont {{Weinberg}}, \citenamefont {{Lupton}},
  \citenamefont {{Narayanan}},\ and\ \citenamefont {et~al.}}]{Strauss2002}%
  \BibitemOpen
  \bibfield  {author} {\bibinfo {author} {\bibfnamefont {M.~A.}\ \bibnamefont
  {{Strauss}}}, \bibinfo {author} {\bibfnamefont {D.~H.}\ \bibnamefont
  {{Weinberg}}}, \bibinfo {author} {\bibfnamefont {R.~H.}\ \bibnamefont
  {{Lupton}}}, \bibinfo {author} {\bibfnamefont {V.~K.}\ \bibnamefont
  {{Narayanan}}}, \ and\ \bibinfo {author} {\bibnamefont {et~al.}},\ }\href
  {\doibase 10.1086/342343} {\bibfield  {journal} {\bibinfo  {journal} {\aj}\
  }\textbf {\bibinfo {volume} {124}},\ \bibinfo {pages} {1810} (\bibinfo {year}
  {2002})},\ \Eprint {http://arxiv.org/abs/astro-ph/0206225}
  {arXiv:astro-ph/0206225 [astro-ph]} \BibitemShut {NoStop}%
\bibitem [{\citenamefont {{Eggemeier}}\ \emph {et~al.}(2019)\citenamefont
  {{Eggemeier}}, \citenamefont {{Scoccimarro}},\ and\ \citenamefont
  {{Smith}}}]{EggScoSmi1906}%
  \BibitemOpen
  \bibfield  {author} {\bibinfo {author} {\bibfnamefont {A.}~\bibnamefont
  {{Eggemeier}}}, \bibinfo {author} {\bibfnamefont {R.}~\bibnamefont
  {{Scoccimarro}}}, \ and\ \bibinfo {author} {\bibfnamefont {R.~E.}\
  \bibnamefont {{Smith}}},\ }\href {\doibase 10.1103/PhysRevD.99.123514}
  {\bibfield  {journal} {\bibinfo  {journal} {\prd}\ }\textbf {\bibinfo
  {volume} {99}},\ \bibinfo {eid} {123514} (\bibinfo {year} {2019})},\ \Eprint
  {http://arxiv.org/abs/1812.03208} {arXiv:1812.03208 [astro-ph.CO]}
  \BibitemShut {NoStop}%
\bibitem [{\citenamefont {{Chan}}\ \emph {et~al.}(2017)\citenamefont {{Chan}},
  \citenamefont {{Sheth}},\ and\ \citenamefont
  {{Scoccimarro}}}]{ChaSheSco1711}%
  \BibitemOpen
  \bibfield  {author} {\bibinfo {author} {\bibfnamefont {K.~C.}\ \bibnamefont
  {{Chan}}}, \bibinfo {author} {\bibfnamefont {R.~K.}\ \bibnamefont {{Sheth}}},
  \ and\ \bibinfo {author} {\bibfnamefont {R.}~\bibnamefont {{Scoccimarro}}},\
  }\href {\doibase 10.1103/PhysRevD.96.103543} {\bibfield  {journal} {\bibinfo
  {journal} {\prd}\ }\textbf {\bibinfo {volume} {96}},\ \bibinfo {eid} {103543}
  (\bibinfo {year} {2017})},\ \Eprint {http://arxiv.org/abs/1511.01909}
  {arXiv:1511.01909} \BibitemShut {NoStop}%
\bibitem [{\citenamefont {{Musso}}\ and\ \citenamefont
  {{Sheth}}(2012)}]{MusShe1206}%
  \BibitemOpen
  \bibfield  {author} {\bibinfo {author} {\bibfnamefont {M.}~\bibnamefont
  {{Musso}}}\ and\ \bibinfo {author} {\bibfnamefont {R.~K.}\ \bibnamefont
  {{Sheth}}},\ }\href {\doibase 10.1111/j.1745-3933.2012.01266.x} {\bibfield
  {journal} {\bibinfo  {journal} {\mnras}\ }\textbf {\bibinfo {volume} {423}},\
  \bibinfo {pages} {L102} (\bibinfo {year} {2012})},\ \Eprint
  {http://arxiv.org/abs/1201.3876} {arXiv:1201.3876 [astro-ph.CO]} \BibitemShut
  {NoStop}%
\bibitem [{\citenamefont {{Lazeyras}}\ and\ \citenamefont
  {{Schmidt}}(2019)}]{LazSch1911}%
  \BibitemOpen
  \bibfield  {author} {\bibinfo {author} {\bibfnamefont {T.}~\bibnamefont
  {{Lazeyras}}}\ and\ \bibinfo {author} {\bibfnamefont {F.}~\bibnamefont
  {{Schmidt}}},\ }\href {\doibase 10.1088/1475-7516/2019/11/041} {\bibfield
  {journal} {\bibinfo  {journal} {\jcap}\ }\textbf {\bibinfo {volume} {2019}},\
  \bibinfo {eid} {041} (\bibinfo {year} {2019})},\ \Eprint
  {http://arxiv.org/abs/1904.11294} {arXiv:1904.11294 [astro-ph.CO]}
  \BibitemShut {NoStop}%
\bibitem [{\citenamefont {{McDonald}}(2006)}]{McDonald:2006}%
  \BibitemOpen
  \bibfield  {author} {\bibinfo {author} {\bibfnamefont {P.}~\bibnamefont
  {{McDonald}}},\ }\href {\doibase 10.1103/PhysRevD.74.103512} {\bibfield
  {journal} {\bibinfo  {journal} {\prd}\ }\textbf {\bibinfo {volume} {74}},\
  \bibinfo {eid} {103512} (\bibinfo {year} {2006})},\ \Eprint
  {http://arxiv.org/abs/astro-ph/0609413} {astro-ph/0609413} \BibitemShut
  {NoStop}%
\bibitem [{\citenamefont {{Assassi}}\ \emph {et~al.}(2014)\citenamefont
  {{Assassi}}, \citenamefont {{Baumann}}, \citenamefont {{Green}},\ and\
  \citenamefont {{Zaldarriaga}}}]{Assassi:2014}%
  \BibitemOpen
  \bibfield  {author} {\bibinfo {author} {\bibfnamefont {V.}~\bibnamefont
  {{Assassi}}}, \bibinfo {author} {\bibfnamefont {D.}~\bibnamefont
  {{Baumann}}}, \bibinfo {author} {\bibfnamefont {D.}~\bibnamefont {{Green}}},
  \ and\ \bibinfo {author} {\bibfnamefont {M.}~\bibnamefont {{Zaldarriaga}}},\
  }\href {\doibase 10.1088/1475-7516/2014/08/056} {\bibfield  {journal}
  {\bibinfo  {journal} {\jcap}\ }\textbf {\bibinfo {volume} {8}},\ \bibinfo
  {eid} {056} (\bibinfo {year} {2014})},\ \Eprint
  {http://arxiv.org/abs/1402.5916} {arXiv:1402.5916} \BibitemShut {NoStop}%
\bibitem [{\citenamefont {{Schmidt}}\ \emph {et~al.}(2013)\citenamefont
  {{Schmidt}}, \citenamefont {{Jeong}},\ and\ \citenamefont
  {{Desjacques}}}]{SchJeoDes1307}%
  \BibitemOpen
  \bibfield  {author} {\bibinfo {author} {\bibfnamefont {F.}~\bibnamefont
  {{Schmidt}}}, \bibinfo {author} {\bibfnamefont {D.}~\bibnamefont {{Jeong}}},
  \ and\ \bibinfo {author} {\bibfnamefont {V.}~\bibnamefont {{Desjacques}}},\
  }\href {\doibase 10.1103/PhysRevD.88.023515} {\bibfield  {journal} {\bibinfo
  {journal} {\prd}\ }\textbf {\bibinfo {volume} {88}},\ \bibinfo {eid} {023515}
  (\bibinfo {year} {2013})},\ \Eprint {http://arxiv.org/abs/1212.0868}
  {arXiv:1212.0868 [astro-ph.CO]} \BibitemShut {NoStop}%
\bibitem [{\citenamefont {{Crocce}}\ and\ \citenamefont
  {{Scoccimarro}}(2006{\natexlab{a}})}]{CroSco0603a}%
  \BibitemOpen
  \bibfield  {author} {\bibinfo {author} {\bibfnamefont {M.}~\bibnamefont
  {{Crocce}}}\ and\ \bibinfo {author} {\bibfnamefont {R.}~\bibnamefont
  {{Scoccimarro}}},\ }\href {\doibase 10.1103/PhysRevD.73.063519} {\bibfield
  {journal} {\bibinfo  {journal} {\prd}\ }\textbf {\bibinfo {volume} {73}},\
  \bibinfo {pages} {063519} (\bibinfo {year} {2006}{\natexlab{a}})},\ \Eprint
  {http://arxiv.org/abs/arXiv:astro-ph/0509418} {arXiv:astro-ph/0509418}
  \BibitemShut {NoStop}%
\bibitem [{\citenamefont {{Bernardeau}}\ \emph {et~al.}(2002)\citenamefont
  {{Bernardeau}}, \citenamefont {{Colombi}}, \citenamefont {{Gazta{\~n}aga}},\
  and\ \citenamefont {{Scoccimarro}}}]{Bernardeau:2002}%
  \BibitemOpen
  \bibfield  {author} {\bibinfo {author} {\bibfnamefont {F.}~\bibnamefont
  {{Bernardeau}}}, \bibinfo {author} {\bibfnamefont {S.}~\bibnamefont
  {{Colombi}}}, \bibinfo {author} {\bibfnamefont {E.}~\bibnamefont
  {{Gazta{\~n}aga}}}, \ and\ \bibinfo {author} {\bibfnamefont {R.}~\bibnamefont
  {{Scoccimarro}}},\ }\href {\doibase 10.1016/S0370-1573(02)00135-7} {\bibfield
   {journal} {\bibinfo  {journal} {\physrep}\ }\textbf {\bibinfo {volume}
  {367}},\ \bibinfo {pages} {1} (\bibinfo {year} {2002})},\ \Eprint
  {http://arxiv.org/abs/astro-ph/0112551} {astro-ph/0112551} \BibitemShut
  {NoStop}%
\bibitem [{\citenamefont {{Grieb}}\ \emph {et~al.}(2016)\citenamefont
  {{Grieb}}, \citenamefont {{S{\'a}nchez}}, \citenamefont
  {{Salazar-Albornoz}},\ and\ \citenamefont {{Dalla Vecchia}}}]{GriSanSal1604}%
  \BibitemOpen
  \bibfield  {author} {\bibinfo {author} {\bibfnamefont {J.~N.}\ \bibnamefont
  {{Grieb}}}, \bibinfo {author} {\bibfnamefont {A.~G.}\ \bibnamefont
  {{S{\'a}nchez}}}, \bibinfo {author} {\bibfnamefont {S.}~\bibnamefont
  {{Salazar-Albornoz}}}, \ and\ \bibinfo {author} {\bibfnamefont
  {C.}~\bibnamefont {{Dalla Vecchia}}},\ }\href {\doibase 10.1093/mnras/stw065}
  {\bibfield  {journal} {\bibinfo  {journal} {\mnras}\ }\textbf {\bibinfo
  {volume} {457}},\ \bibinfo {pages} {1577} (\bibinfo {year} {2016})},\ \Eprint
  {http://arxiv.org/abs/1509.04293} {arXiv:1509.04293 [astro-ph.CO]}
  \BibitemShut {NoStop}%
\bibitem [{\citenamefont {{Blas}}\ \emph {et~al.}(2013)\citenamefont {{Blas}},
  \citenamefont {{Garny}},\ and\ \citenamefont {{Konstandin}}}]{BlaGarKon1309}%
  \BibitemOpen
  \bibfield  {author} {\bibinfo {author} {\bibfnamefont {D.}~\bibnamefont
  {{Blas}}}, \bibinfo {author} {\bibfnamefont {M.}~\bibnamefont {{Garny}}}, \
  and\ \bibinfo {author} {\bibfnamefont {T.}~\bibnamefont {{Konstandin}}},\
  }\href@noop {} {\bibfield  {journal} {\bibinfo  {journal} {ArXiv e-prints}\ }
  (\bibinfo {year} {2013})},\ \Eprint {http://arxiv.org/abs/1309.3308}
  {arXiv:1309.3308 [astro-ph.CO]} \BibitemShut {NoStop}%
\bibitem [{\citenamefont {{Crocce}}\ and\ \citenamefont
  {{Scoccimarro}}(2006{\natexlab{b}})}]{CroSco0603b}%
  \BibitemOpen
  \bibfield  {author} {\bibinfo {author} {\bibfnamefont {M.}~\bibnamefont
  {{Crocce}}}\ and\ \bibinfo {author} {\bibfnamefont {R.}~\bibnamefont
  {{Scoccimarro}}},\ }\href {\doibase 10.1103/PhysRevD.73.063520} {\bibfield
  {journal} {\bibinfo  {journal} {\prd}\ }\textbf {\bibinfo {volume} {73}},\
  \bibinfo {pages} {063520} (\bibinfo {year} {2006}{\natexlab{b}})},\ \Eprint
  {http://arxiv.org/abs/arXiv:astro-ph/0509419} {arXiv:astro-ph/0509419}
  \BibitemShut {NoStop}%
\bibitem [{\citenamefont {{Eisenstein}}\ \emph
  {et~al.}(2007{\natexlab{a}})\citenamefont {{Eisenstein}}, \citenamefont
  {{Seo}},\ and\ \citenamefont {{White}}}]{EisSeoWhi0708}%
  \BibitemOpen
  \bibfield  {author} {\bibinfo {author} {\bibfnamefont {D.~J.}\ \bibnamefont
  {{Eisenstein}}}, \bibinfo {author} {\bibfnamefont {H.-J.}\ \bibnamefont
  {{Seo}}}, \ and\ \bibinfo {author} {\bibfnamefont {M.}~\bibnamefont
  {{White}}},\ }\href {\doibase 10.1086/518755} {\bibfield  {journal} {\bibinfo
   {journal} {\apj}\ }\textbf {\bibinfo {volume} {664}},\ \bibinfo {pages}
  {660} (\bibinfo {year} {2007}{\natexlab{a}})},\ \Eprint
  {http://arxiv.org/abs/arXiv:astro-ph/0604361} {arXiv:astro-ph/0604361}
  \BibitemShut {NoStop}%
\bibitem [{\citenamefont {{Eisenstein}}\ \emph
  {et~al.}(2007{\natexlab{b}})\citenamefont {{Eisenstein}}, \citenamefont
  {{Seo}}, \citenamefont {{Sirko}},\ and\ \citenamefont
  {{Spergel}}}]{EisSeoSir0708}%
  \BibitemOpen
  \bibfield  {author} {\bibinfo {author} {\bibfnamefont {D.~J.}\ \bibnamefont
  {{Eisenstein}}}, \bibinfo {author} {\bibfnamefont {H.-J.}\ \bibnamefont
  {{Seo}}}, \bibinfo {author} {\bibfnamefont {E.}~\bibnamefont {{Sirko}}}, \
  and\ \bibinfo {author} {\bibfnamefont {D.~N.}\ \bibnamefont {{Spergel}}},\
  }\href {\doibase 10.1086/518712} {\bibfield  {journal} {\bibinfo  {journal}
  {\apj}\ }\textbf {\bibinfo {volume} {664}},\ \bibinfo {pages} {675} (\bibinfo
  {year} {2007}{\natexlab{b}})},\ \Eprint
  {http://arxiv.org/abs/astro-ph/0604362} {arXiv:astro-ph/0604362 [astro-ph]}
  \BibitemShut {NoStop}%
\bibitem [{\citenamefont {{Matarrese}}\ and\ \citenamefont
  {{Pietroni}}(2008)}]{MatPie08}%
  \BibitemOpen
  \bibfield  {author} {\bibinfo {author} {\bibfnamefont {S.}~\bibnamefont
  {{Matarrese}}}\ and\ \bibinfo {author} {\bibfnamefont {M.}~\bibnamefont
  {{Pietroni}}},\ }\href {\doibase 10.1142/S0217732308026182} {\bibfield
  {journal} {\bibinfo  {journal} {Modern Physics Letters A}\ }\textbf {\bibinfo
  {volume} {23}},\ \bibinfo {pages} {25} (\bibinfo {year} {2008})},\ \Eprint
  {http://arxiv.org/abs/arXiv:astro-ph/0702653} {arXiv:astro-ph/0702653}
  \BibitemShut {NoStop}%
\bibitem [{\citenamefont {{Crocce}}\ and\ \citenamefont
  {{Scoccimarro}}(2008)}]{CroSco0801}%
  \BibitemOpen
  \bibfield  {author} {\bibinfo {author} {\bibfnamefont {M.}~\bibnamefont
  {{Crocce}}}\ and\ \bibinfo {author} {\bibfnamefont {R.}~\bibnamefont
  {{Scoccimarro}}},\ }\href {\doibase 10.1103/PhysRevD.77.023533} {\bibfield
  {journal} {\bibinfo  {journal} {\prd}\ }\textbf {\bibinfo {volume} {77}},\
  \bibinfo {eid} {023533} (\bibinfo {year} {2008})},\ \Eprint
  {http://arxiv.org/abs/0704.2783} {arXiv:0704.2783 [astro-ph]} \BibitemShut
  {NoStop}%
\bibitem [{\citenamefont {{Seo}}\ \emph {et~al.}(2008)\citenamefont {{Seo}},
  \citenamefont {{Siegel}}, \citenamefont {{Eisenstein}},\ and\ \citenamefont
  {{White}}}]{SeoSieEis0810}%
  \BibitemOpen
  \bibfield  {author} {\bibinfo {author} {\bibfnamefont {H.-J.}\ \bibnamefont
  {{Seo}}}, \bibinfo {author} {\bibfnamefont {E.~R.}\ \bibnamefont {{Siegel}}},
  \bibinfo {author} {\bibfnamefont {D.~J.}\ \bibnamefont {{Eisenstein}}}, \
  and\ \bibinfo {author} {\bibfnamefont {M.}~\bibnamefont {{White}}},\ }\href
  {\doibase 10.1086/589921} {\bibfield  {journal} {\bibinfo  {journal} {\apj}\
  }\textbf {\bibinfo {volume} {686}},\ \bibinfo {pages} {13} (\bibinfo {year}
  {2008})},\ \Eprint {http://arxiv.org/abs/0805.0117} {arXiv:0805.0117}
  \BibitemShut {NoStop}%
\bibitem [{\citenamefont {{Vlah}}\ \emph {et~al.}(2016)\citenamefont {{Vlah}},
  \citenamefont {{Seljak}}, \citenamefont {{Yat Chu}},\ and\ \citenamefont
  {{Feng}}}]{VlaSelChu1603}%
  \BibitemOpen
  \bibfield  {author} {\bibinfo {author} {\bibfnamefont {Z.}~\bibnamefont
  {{Vlah}}}, \bibinfo {author} {\bibfnamefont {U.}~\bibnamefont {{Seljak}}},
  \bibinfo {author} {\bibfnamefont {M.}~\bibnamefont {{Yat Chu}}}, \ and\
  \bibinfo {author} {\bibfnamefont {Y.}~\bibnamefont {{Feng}}},\ }\href
  {\doibase 10.1088/1475-7516/2016/03/057} {\bibfield  {journal} {\bibinfo
  {journal} {\jcap}\ }\textbf {\bibinfo {volume} {2016}},\ \bibinfo {eid} {057}
  (\bibinfo {year} {2016})},\ \Eprint {http://arxiv.org/abs/1509.02120}
  {arXiv:1509.02120 [astro-ph.CO]} \BibitemShut {NoStop}%
\bibitem [{\citenamefont {{Osato}}\ \emph {et~al.}(2019)\citenamefont
  {{Osato}}, \citenamefont {{Nishimichi}}, \citenamefont {{Bernardeau}},\ and\
  \citenamefont {{Taruya}}}]{OsaNisBer1903}%
  \BibitemOpen
  \bibfield  {author} {\bibinfo {author} {\bibfnamefont {K.}~\bibnamefont
  {{Osato}}}, \bibinfo {author} {\bibfnamefont {T.}~\bibnamefont
  {{Nishimichi}}}, \bibinfo {author} {\bibfnamefont {F.}~\bibnamefont
  {{Bernardeau}}}, \ and\ \bibinfo {author} {\bibfnamefont {A.}~\bibnamefont
  {{Taruya}}},\ }\href {\doibase 10.1103/PhysRevD.99.063530} {\bibfield
  {journal} {\bibinfo  {journal} {\prd}\ }\textbf {\bibinfo {volume} {99}},\
  \bibinfo {eid} {063530} (\bibinfo {year} {2019})},\ \Eprint
  {http://arxiv.org/abs/1810.10104} {arXiv:1810.10104 [astro-ph.CO]}
  \BibitemShut {NoStop}%
\bibitem [{\citenamefont {Eisenstein}\ and\ \citenamefont
  {Hu}(1999)}]{EisHu1901}%
  \BibitemOpen
  \bibfield  {author} {\bibinfo {author} {\bibfnamefont {D.~J.}\ \bibnamefont
  {Eisenstein}}\ and\ \bibinfo {author} {\bibfnamefont {W.}~\bibnamefont
  {Hu}},\ }\href {\doibase 10.1086/306640} {\bibfield  {journal} {\bibinfo
  {journal} {The Astrophysical Journal}\ }\textbf {\bibinfo {volume} {511}},\
  \bibinfo {pages} {5} (\bibinfo {year} {1999})}\BibitemShut {NoStop}%
\bibitem [{\citenamefont {{Blas}}\ \emph {et~al.}(2016)\citenamefont {{Blas}},
  \citenamefont {{Garny}}, \citenamefont {{Ivanov}},\ and\ \citenamefont
  {{Sibiryakov}}}]{BlaGarIva1607}%
  \BibitemOpen
  \bibfield  {author} {\bibinfo {author} {\bibfnamefont {D.}~\bibnamefont
  {{Blas}}}, \bibinfo {author} {\bibfnamefont {M.}~\bibnamefont {{Garny}}},
  \bibinfo {author} {\bibfnamefont {M.~M.}\ \bibnamefont {{Ivanov}}}, \ and\
  \bibinfo {author} {\bibfnamefont {S.}~\bibnamefont {{Sibiryakov}}},\ }\href
  {\doibase 10.1088/1475-7516/2016/07/028} {\bibfield  {journal} {\bibinfo
  {journal} {\jcap}\ }\textbf {\bibinfo {volume} {2016}},\ \bibinfo {eid} {028}
  (\bibinfo {year} {2016})},\ \Eprint {http://arxiv.org/abs/1605.02149}
  {arXiv:1605.02149 [astro-ph.CO]} \BibitemShut {NoStop}%
\bibitem [{\citenamefont {{Pueblas}}\ and\ \citenamefont
  {{Scoccimarro}}(2009)}]{PueSco0908}%
  \BibitemOpen
  \bibfield  {author} {\bibinfo {author} {\bibfnamefont {S.}~\bibnamefont
  {{Pueblas}}}\ and\ \bibinfo {author} {\bibfnamefont {R.}~\bibnamefont
  {{Scoccimarro}}},\ }\href {\doibase 10.1103/PhysRevD.80.043504} {\bibfield
  {journal} {\bibinfo  {journal} {\prd}\ }\textbf {\bibinfo {volume} {80}},\
  \bibinfo {pages} {043504} (\bibinfo {year} {2009})},\ \Eprint
  {http://arxiv.org/abs/0809.4606} {arXiv:0809.4606} \BibitemShut {NoStop}%
\bibitem [{\citenamefont {{Carrasco}}\ \emph {et~al.}(2012)\citenamefont
  {{Carrasco}}, \citenamefont {{Hertzberg}},\ and\ \citenamefont
  {{Senatore}}}]{CarHerSen1206}%
  \BibitemOpen
  \bibfield  {author} {\bibinfo {author} {\bibfnamefont {J.~J.~M.}\
  \bibnamefont {{Carrasco}}}, \bibinfo {author} {\bibfnamefont {M.~P.}\
  \bibnamefont {{Hertzberg}}}, \ and\ \bibinfo {author} {\bibfnamefont
  {L.}~\bibnamefont {{Senatore}}},\ }\href@noop {} {\bibfield  {journal}
  {\bibinfo  {journal} {ArXiv e-prints}\ } (\bibinfo {year} {2012})},\ \Eprint
  {http://arxiv.org/abs/1206.2926} {arXiv:1206.2926 [astro-ph.CO]} \BibitemShut
  {NoStop}%
\bibitem [{\citenamefont {{Baumann}}\ \emph {et~al.}(2012)\citenamefont
  {{Baumann}}, \citenamefont {{Nicolis}}, \citenamefont {{Senatore}},\ and\
  \citenamefont {{Zaldarriaga}}}]{BauNicSen1207}%
  \BibitemOpen
  \bibfield  {author} {\bibinfo {author} {\bibfnamefont {D.}~\bibnamefont
  {{Baumann}}}, \bibinfo {author} {\bibfnamefont {A.}~\bibnamefont
  {{Nicolis}}}, \bibinfo {author} {\bibfnamefont {L.}~\bibnamefont
  {{Senatore}}}, \ and\ \bibinfo {author} {\bibfnamefont {M.}~\bibnamefont
  {{Zaldarriaga}}},\ }\href {\doibase 10.1088/1475-7516/2012/07/051} {\bibfield
   {journal} {\bibinfo  {journal} {\jcap}\ }\textbf {\bibinfo {volume} {7}},\
  \bibinfo {eid} {051} (\bibinfo {year} {2012})},\ \Eprint
  {http://arxiv.org/abs/1004.2488} {arXiv:1004.2488 [astro-ph.CO]} \BibitemShut
  {NoStop}%
\bibitem [{\citenamefont {{Carrasco}}\ \emph {et~al.}(2013)\citenamefont
  {{Carrasco}}, \citenamefont {{Foreman}}, \citenamefont {{Green}},\ and\
  \citenamefont {{Senatore}}}]{CarForGre1310}%
  \BibitemOpen
  \bibfield  {author} {\bibinfo {author} {\bibfnamefont {J.~J.~M.}\
  \bibnamefont {{Carrasco}}}, \bibinfo {author} {\bibfnamefont
  {S.}~\bibnamefont {{Foreman}}}, \bibinfo {author} {\bibfnamefont
  {D.}~\bibnamefont {{Green}}}, \ and\ \bibinfo {author} {\bibfnamefont
  {L.}~\bibnamefont {{Senatore}}},\ }\href@noop {} {\bibfield  {journal}
  {\bibinfo  {journal} {ArXiv e-prints}\ } (\bibinfo {year} {2013})},\ \Eprint
  {http://arxiv.org/abs/1310.0464} {arXiv:1310.0464 [astro-ph.CO]} \BibitemShut
  {NoStop}%
\bibitem [{\citenamefont {{Baldauf}}\ \emph {et~al.}(2015)\citenamefont
  {{Baldauf}}, \citenamefont {{Mercolli}},\ and\ \citenamefont
  {{Zaldarriaga}}}]{BalMerZal1512}%
  \BibitemOpen
  \bibfield  {author} {\bibinfo {author} {\bibfnamefont {T.}~\bibnamefont
  {{Baldauf}}}, \bibinfo {author} {\bibfnamefont {L.}~\bibnamefont
  {{Mercolli}}}, \ and\ \bibinfo {author} {\bibfnamefont {M.}~\bibnamefont
  {{Zaldarriaga}}},\ }\href {\doibase 10.1103/PhysRevD.92.123007} {\bibfield
  {journal} {\bibinfo  {journal} {\prd}\ }\textbf {\bibinfo {volume} {92}},\
  \bibinfo {eid} {123007} (\bibinfo {year} {2015})},\ \Eprint
  {http://arxiv.org/abs/1507.02256} {arXiv:1507.02256 [astro-ph.CO]}
  \BibitemShut {NoStop}%
\bibitem [{\citenamefont {{Bernardeau}}\ \emph {et~al.}(2014)\citenamefont
  {{Bernardeau}}, \citenamefont {{Taruya}},\ and\ \citenamefont
  {{Nishimichi}}}]{BerTarNis1401}%
  \BibitemOpen
  \bibfield  {author} {\bibinfo {author} {\bibfnamefont {F.}~\bibnamefont
  {{Bernardeau}}}, \bibinfo {author} {\bibfnamefont {A.}~\bibnamefont
  {{Taruya}}}, \ and\ \bibinfo {author} {\bibfnamefont {T.}~\bibnamefont
  {{Nishimichi}}},\ }\href {\doibase 10.1103/PhysRevD.89.023502} {\bibfield
  {journal} {\bibinfo  {journal} {\prd}\ }\textbf {\bibinfo {volume} {89}},\
  \bibinfo {eid} {023502} (\bibinfo {year} {2014})}\BibitemShut {NoStop}%
\bibitem [{\citenamefont {{Nishimichi}}\ \emph {et~al.}(2016)\citenamefont
  {{Nishimichi}}, \citenamefont {{Bernardeau}},\ and\ \citenamefont
  {{Taruya}}}]{NisBerTar1611}%
  \BibitemOpen
  \bibfield  {author} {\bibinfo {author} {\bibfnamefont {T.}~\bibnamefont
  {{Nishimichi}}}, \bibinfo {author} {\bibfnamefont {F.}~\bibnamefont
  {{Bernardeau}}}, \ and\ \bibinfo {author} {\bibfnamefont {A.}~\bibnamefont
  {{Taruya}}},\ }\href {\doibase 10.1016/j.physletb.2016.09.035} {\bibfield
  {journal} {\bibinfo  {journal} {Physics Letters B}\ }\textbf {\bibinfo
  {volume} {762}},\ \bibinfo {pages} {247} (\bibinfo {year} {2016})},\ \Eprint
  {http://arxiv.org/abs/1411.2970} {arXiv:1411.2970 [astro-ph.CO]} \BibitemShut
  {NoStop}%
\bibitem [{\citenamefont {{Nishimichi}}\ \emph {et~al.}(2017)\citenamefont
  {{Nishimichi}}, \citenamefont {{Bernardeau}},\ and\ \citenamefont
  {{Taruya}}}]{NisBerTar1712}%
  \BibitemOpen
  \bibfield  {author} {\bibinfo {author} {\bibfnamefont {T.}~\bibnamefont
  {{Nishimichi}}}, \bibinfo {author} {\bibfnamefont {F.}~\bibnamefont
  {{Bernardeau}}}, \ and\ \bibinfo {author} {\bibfnamefont {A.}~\bibnamefont
  {{Taruya}}},\ }\href {\doibase 10.1103/PhysRevD.96.123515} {\bibfield
  {journal} {\bibinfo  {journal} {\prd}\ }\textbf {\bibinfo {volume} {96}},\
  \bibinfo {eid} {123515} (\bibinfo {year} {2017})},\ \Eprint
  {http://arxiv.org/abs/1708.08946} {arXiv:1708.08946 [astro-ph.CO]}
  \BibitemShut {NoStop}%
\bibitem [{\citenamefont {{Planck Collaboration}}(2016)}]{Planck1609}%
  \BibitemOpen
  \bibfield  {author} {\bibinfo {author} {\bibnamefont {{Planck
  Collaboration}}},\ }\href {\doibase 10.1051/0004-6361/201525830} {\bibfield
  {journal} {\bibinfo  {journal} {\aap}\ }\textbf {\bibinfo {volume} {594}},\
  \bibinfo {eid} {A13} (\bibinfo {year} {2016})},\ \Eprint
  {http://arxiv.org/abs/1502.01589} {arXiv:1502.01589 [astro-ph.CO]}
  \BibitemShut {NoStop}%
\bibitem [{\citenamefont {{Angulo}}\ \emph {et~al.}(2015)\citenamefont
  {{Angulo}}, \citenamefont {{Fasiello}}, \citenamefont {{Senatore}},\ and\
  \citenamefont {{Vlah}}}]{AngFasSen1503}%
  \BibitemOpen
  \bibfield  {author} {\bibinfo {author} {\bibfnamefont {R.}~\bibnamefont
  {{Angulo}}}, \bibinfo {author} {\bibfnamefont {M.}~\bibnamefont
  {{Fasiello}}}, \bibinfo {author} {\bibfnamefont {L.}~\bibnamefont
  {{Senatore}}}, \ and\ \bibinfo {author} {\bibfnamefont {Z.}~\bibnamefont
  {{Vlah}}},\ }\href {\doibase 10.1088/1475-7516/2015/09/029} {\bibfield
  {journal} {\bibinfo  {journal} {Journal of Cosmology and Astro-Particle
  Physics}\ }\textbf {\bibinfo {volume} {2015}},\ \bibinfo {eid} {029}
  (\bibinfo {year} {2015})},\ \Eprint {http://arxiv.org/abs/1503.08826}
  {arXiv:1503.08826 [astro-ph.CO]} \BibitemShut {NoStop}%
\bibitem [{\citenamefont {{Bernardeau}}\ \emph {et~al.}(2008)\citenamefont
  {{Bernardeau}}, \citenamefont {{Crocce}},\ and\ \citenamefont
  {{Scoccimarro}}}]{BerCroSco0811}%
  \BibitemOpen
  \bibfield  {author} {\bibinfo {author} {\bibfnamefont {F.}~\bibnamefont
  {{Bernardeau}}}, \bibinfo {author} {\bibfnamefont {M.}~\bibnamefont
  {{Crocce}}}, \ and\ \bibinfo {author} {\bibfnamefont {R.}~\bibnamefont
  {{Scoccimarro}}},\ }\href {\doibase 10.1103/PhysRevD.78.103521} {\bibfield
  {journal} {\bibinfo  {journal} {\prd}\ }\textbf {\bibinfo {volume} {78}},\
  \bibinfo {pages} {103521} (\bibinfo {year} {2008})},\ \Eprint
  {http://arxiv.org/abs/0806.2334} {arXiv:0806.2334} \BibitemShut {NoStop}%
\bibitem [{\citenamefont {{Taruya}}\ \emph {et~al.}(2013)\citenamefont
  {{Taruya}}, \citenamefont {{Nishimichi}},\ and\ \citenamefont
  {{Bernardeau}}}]{TarNisBer1304}%
  \BibitemOpen
  \bibfield  {author} {\bibinfo {author} {\bibfnamefont {A.}~\bibnamefont
  {{Taruya}}}, \bibinfo {author} {\bibfnamefont {T.}~\bibnamefont
  {{Nishimichi}}}, \ and\ \bibinfo {author} {\bibfnamefont {F.}~\bibnamefont
  {{Bernardeau}}},\ }\href {\doibase 10.1103/PhysRevD.87.083509} {\bibfield
  {journal} {\bibinfo  {journal} {\prd}\ }\textbf {\bibinfo {volume} {87}},\
  \bibinfo {eid} {083509} (\bibinfo {year} {2013})},\ \Eprint
  {http://arxiv.org/abs/1301.3624} {arXiv:1301.3624 [astro-ph.CO]} \BibitemShut
  {NoStop}%
\bibitem [{\citenamefont {{Scoccimarro}}\ and\ \citenamefont
  {{Frieman}}(1996)}]{ScoFri9607}%
  \BibitemOpen
  \bibfield  {author} {\bibinfo {author} {\bibfnamefont {R.}~\bibnamefont
  {{Scoccimarro}}}\ and\ \bibinfo {author} {\bibfnamefont {J.}~\bibnamefont
  {{Frieman}}},\ }\href {\doibase 10.1086/192306} {\bibfield  {journal}
  {\bibinfo  {journal} {\apjs}\ }\textbf {\bibinfo {volume} {105}},\ \bibinfo
  {pages} {37} (\bibinfo {year} {1996})},\ \Eprint
  {http://arxiv.org/abs/arXiv:astro-ph/9509047} {arXiv:astro-ph/9509047}
  \BibitemShut {NoStop}%
\bibitem [{\citenamefont {{Casas-Miranda}}\ \emph {et~al.}(2002)\citenamefont
  {{Casas-Miranda}}, \citenamefont {{Mo}}, \citenamefont {{Sheth}},\ and\
  \citenamefont {{Boerner}}}]{CasMoShe0207}%
  \BibitemOpen
  \bibfield  {author} {\bibinfo {author} {\bibfnamefont {R.}~\bibnamefont
  {{Casas-Miranda}}}, \bibinfo {author} {\bibfnamefont {H.~J.}\ \bibnamefont
  {{Mo}}}, \bibinfo {author} {\bibfnamefont {R.~K.}\ \bibnamefont {{Sheth}}}, \
  and\ \bibinfo {author} {\bibfnamefont {G.}~\bibnamefont {{Boerner}}},\ }\href
  {\doibase 10.1046/j.1365-8711.2002.05378.x} {\bibfield  {journal} {\bibinfo
  {journal} {\mnras}\ }\textbf {\bibinfo {volume} {333}},\ \bibinfo {pages}
  {730} (\bibinfo {year} {2002})},\ \Eprint
  {http://arxiv.org/abs/astro-ph/0105008} {arXiv:astro-ph/0105008 [astro-ph]}
  \BibitemShut {NoStop}%
\bibitem [{\citenamefont {{Akbar Abolhasani}}\ \emph
  {et~al.}(2016)\citenamefont {{Akbar Abolhasani}}, \citenamefont
  {{Mirbabayi}},\ and\ \citenamefont {{Pajer}}}]{AboMirPaj1605}%
  \BibitemOpen
  \bibfield  {author} {\bibinfo {author} {\bibfnamefont {A.}~\bibnamefont
  {{Akbar Abolhasani}}}, \bibinfo {author} {\bibfnamefont {M.}~\bibnamefont
  {{Mirbabayi}}}, \ and\ \bibinfo {author} {\bibfnamefont {E.}~\bibnamefont
  {{Pajer}}},\ }\href {\doibase 10.1088/1475-7516/2016/05/063} {\bibfield
  {journal} {\bibinfo  {journal} {\jcap}\ }\textbf {\bibinfo {volume} {2016}},\
  \bibinfo {eid} {063} (\bibinfo {year} {2016})},\ \Eprint
  {http://arxiv.org/abs/1509.07886} {arXiv:1509.07886 [hep-th]} \BibitemShut
  {NoStop}%
\bibitem [{\citenamefont {{Catelan}}\ \emph {et~al.}(1998)\citenamefont
  {{Catelan}}, \citenamefont {{Lucchin}}, \citenamefont {{Matarrese}},\ and\
  \citenamefont {{Porciani}}}]{CatLucMat9807}%
  \BibitemOpen
  \bibfield  {author} {\bibinfo {author} {\bibfnamefont {P.}~\bibnamefont
  {{Catelan}}}, \bibinfo {author} {\bibfnamefont {F.}~\bibnamefont
  {{Lucchin}}}, \bibinfo {author} {\bibfnamefont {S.}~\bibnamefont
  {{Matarrese}}}, \ and\ \bibinfo {author} {\bibfnamefont {C.}~\bibnamefont
  {{Porciani}}},\ }\href {\doibase 10.1046/j.1365-8711.1998.01455.x} {\bibfield
   {journal} {\bibinfo  {journal} {\mnras}\ }\textbf {\bibinfo {volume}
  {297}},\ \bibinfo {pages} {692} (\bibinfo {year} {1998})},\ \Eprint
  {http://arxiv.org/abs/arXiv:astro-ph/9708067} {arXiv:astro-ph/9708067}
  \BibitemShut {NoStop}%
\bibitem [{\citenamefont {{Catelan}}\ \emph {et~al.}(2000)\citenamefont
  {{Catelan}}, \citenamefont {{Porciani}},\ and\ \citenamefont
  {{Kamionkowski}}}]{Catelan:2000}%
  \BibitemOpen
  \bibfield  {author} {\bibinfo {author} {\bibfnamefont {P.}~\bibnamefont
  {{Catelan}}}, \bibinfo {author} {\bibfnamefont {C.}~\bibnamefont
  {{Porciani}}}, \ and\ \bibinfo {author} {\bibfnamefont {M.}~\bibnamefont
  {{Kamionkowski}}},\ }\href {\doibase 10.1046/j.1365-8711.2000.04023.x}
  {\bibfield  {journal} {\bibinfo  {journal} {\mnras}\ }\textbf {\bibinfo
  {volume} {318}},\ \bibinfo {pages} {L39} (\bibinfo {year} {2000})},\ \Eprint
  {http://arxiv.org/abs/astro-ph/0005544} {astro-ph/0005544} \BibitemShut
  {NoStop}%
\bibitem [{\citenamefont {{Matsubara}}(2011)}]{Matsubara:2011}%
  \BibitemOpen
  \bibfield  {author} {\bibinfo {author} {\bibfnamefont {T.}~\bibnamefont
  {{Matsubara}}},\ }\href {\doibase 10.1103/PhysRevD.83.083518} {\bibfield
  {journal} {\bibinfo  {journal} {\prd}\ }\textbf {\bibinfo {volume} {83}},\
  \bibinfo {eid} {083518} (\bibinfo {year} {2011})},\ \Eprint
  {http://arxiv.org/abs/1102.4619} {arXiv:1102.4619 [astro-ph.CO]} \BibitemShut
  {NoStop}%
\bibitem [{\citenamefont {{Lazeyras}}\ and\ \citenamefont
  {{Schmidt}}(2018)}]{LazSch1712}%
  \BibitemOpen
  \bibfield  {author} {\bibinfo {author} {\bibfnamefont {T.}~\bibnamefont
  {{Lazeyras}}}\ and\ \bibinfo {author} {\bibfnamefont {F.}~\bibnamefont
  {{Schmidt}}},\ }\href {\doibase 10.1088/1475-7516/2018/09/008} {\bibfield
  {journal} {\bibinfo  {journal} {Journal of Cosmology and Astro-Particle
  Physics}\ }\textbf {\bibinfo {volume} {2018}},\ \bibinfo {eid} {008}
  (\bibinfo {year} {2018})},\ \Eprint {http://arxiv.org/abs/1712.07531}
  {arXiv:1712.07531 [astro-ph.CO]} \BibitemShut {NoStop}%
\bibitem [{\citenamefont {{Abidi}}\ and\ \citenamefont
  {{Baldauf}}(2018)}]{AbiBal1807}%
  \BibitemOpen
  \bibfield  {author} {\bibinfo {author} {\bibfnamefont {M.~M.}\ \bibnamefont
  {{Abidi}}}\ and\ \bibinfo {author} {\bibfnamefont {T.}~\bibnamefont
  {{Baldauf}}},\ }\href {\doibase 10.1088/1475-7516/2018/07/029} {\bibfield
  {journal} {\bibinfo  {journal} {\jcap}\ }\textbf {\bibinfo {volume} {7}},\
  \bibinfo {eid} {029} (\bibinfo {year} {2018})},\ \Eprint
  {http://arxiv.org/abs/1802.07622} {arXiv:1802.07622} \BibitemShut {NoStop}%
\bibitem [{\citenamefont {{Sheth}}\ \emph {et~al.}(2013)\citenamefont
  {{Sheth}}, \citenamefont {{Chan}},\ and\ \citenamefont
  {{Scoccimarro}}}]{SheChaSco1304}%
  \BibitemOpen
  \bibfield  {author} {\bibinfo {author} {\bibfnamefont {R.~K.}\ \bibnamefont
  {{Sheth}}}, \bibinfo {author} {\bibfnamefont {K.~C.}\ \bibnamefont {{Chan}}},
  \ and\ \bibinfo {author} {\bibfnamefont {R.}~\bibnamefont {{Scoccimarro}}},\
  }\href {\doibase 10.1103/PhysRevD.87.083002} {\bibfield  {journal} {\bibinfo
  {journal} {\prd}\ }\textbf {\bibinfo {volume} {87}},\ \bibinfo {eid} {083002}
  (\bibinfo {year} {2013})},\ \Eprint {http://arxiv.org/abs/1207.7117}
  {arXiv:1207.7117 [astro-ph.CO]} \BibitemShut {NoStop}%
\bibitem [{\citenamefont {Springel}\ \emph {et~al.}(2001)\citenamefont
  {Springel}, \citenamefont {Yoshida},\ and\ \citenamefont
  {White}}]{Springel2001}%
  \BibitemOpen
  \bibfield  {author} {\bibinfo {author} {\bibfnamefont {V.}~\bibnamefont
  {Springel}}, \bibinfo {author} {\bibfnamefont {N.}~\bibnamefont {Yoshida}}, \
  and\ \bibinfo {author} {\bibfnamefont {S.~D.}\ \bibnamefont {White}},\ }\href
  {\doibase 10.1016/s1384-1076(01)00042-2} {\bibfield  {journal} {\bibinfo
  {journal} {New Astronomy}\ }\textbf {\bibinfo {volume} {6}},\ \bibinfo
  {pages} {79–117} (\bibinfo {year} {2001})}\BibitemShut {NoStop}%
\bibitem [{\citenamefont {Springel}(2005)}]{Springel2005}%
  \BibitemOpen
  \bibfield  {author} {\bibinfo {author} {\bibfnamefont {V.}~\bibnamefont
  {Springel}},\ }\href {\doibase 10.1111/j.1365-2966.2005.09655.x} {\bibfield
  {journal} {\bibinfo  {journal} {Monthly Notices of the Royal Astronomical
  Society}\ }\textbf {\bibinfo {volume} {364}},\ \bibinfo {pages} {1105–1134}
  (\bibinfo {year} {2005})}\BibitemShut {NoStop}%
\bibitem [{\citenamefont {Lewis}\ and\ \citenamefont
  {Bridle}(2002)}]{Lewis2002}%
  \BibitemOpen
  \bibfield  {author} {\bibinfo {author} {\bibfnamefont {A.}~\bibnamefont
  {Lewis}}\ and\ \bibinfo {author} {\bibfnamefont {S.}~\bibnamefont {Bridle}},\
  }\href {\doibase 10.1103/PhysRevD.66.103511} {\bibfield  {journal} {\bibinfo
  {journal} {\prd}\ }\textbf {\bibinfo {volume} {66}},\ \bibinfo {pages}
  {103511} (\bibinfo {year} {2002})},\ \Eprint
  {http://arxiv.org/abs/astro-ph/0205436} {arXiv:astro-ph/0205436 [astro-ph]}
  \BibitemShut {NoStop}%
\bibitem [{\citenamefont {{Crocce}}\ \emph {et~al.}(2006)\citenamefont
  {{Crocce}}, \citenamefont {{Pueblas}},\ and\ \citenamefont
  {{Scoccimarro}}}]{CroPueSco0611}%
  \BibitemOpen
  \bibfield  {author} {\bibinfo {author} {\bibfnamefont {M.}~\bibnamefont
  {{Crocce}}}, \bibinfo {author} {\bibfnamefont {S.}~\bibnamefont {{Pueblas}}},
  \ and\ \bibinfo {author} {\bibfnamefont {R.}~\bibnamefont {{Scoccimarro}}},\
  }\href {\doibase 10.1111/j.1365-2966.2006.11040.x} {\bibfield  {journal}
  {\bibinfo  {journal} {\mnras}\ }\textbf {\bibinfo {volume} {373}},\ \bibinfo
  {pages} {369} (\bibinfo {year} {2006})},\ \Eprint
  {http://arxiv.org/abs/arXiv:astro-ph/0606505} {arXiv:astro-ph/0606505}
  \BibitemShut {NoStop}%
\bibitem [{\citenamefont {{Alam}}\ \emph {et~al.}(2017)\citenamefont {{Alam}},
  \citenamefont {{Ata}}, \citenamefont {{Bailey}}, \citenamefont {{Beutler}},
  \citenamefont {{Bizyaev}}, \citenamefont {{Blazek}}, \citenamefont
  {{Bolton}}, \citenamefont {{Brownstein}}, \citenamefont {{Burden}},
  \citenamefont {{Chuang}}, \citenamefont {{Comparat}}, \citenamefont
  {{Cuesta}}, \citenamefont {{Dawson}}, \citenamefont {{Eisenstein}},
  \citenamefont {{Escoffier}}, \citenamefont {{Gil-Mar{\'\i}n}}, \citenamefont
  {{Grieb}}, \citenamefont {{Hand}}, \citenamefont {{Ho}}, \citenamefont
  {{Kinemuchi}}, \citenamefont {{Kirkby}}, \citenamefont {{Kitaura}},
  \citenamefont {{Malanushenko}}, \citenamefont {{Malanushenko}}, \citenamefont
  {{Maraston}}, \citenamefont {{McBride}}, \citenamefont {{Nichol}},
  \citenamefont {{Olmstead}}, \citenamefont {{Oravetz}}, \citenamefont
  {{Padmanabhan}}, \citenamefont {{Palanque-Delabrouille}}, \citenamefont
  {{Pan}}, \citenamefont {{Pellejero-Ibanez}}, \citenamefont {{Percival}},
  \citenamefont {{Petitjean}}, \citenamefont {{Prada}}, \citenamefont
  {{Price-Whelan}}, \citenamefont {{Reid}}, \citenamefont
  {{Rodr{\'\i}guez-Torres}}, \citenamefont {{Roe}}, \citenamefont {{Ross}},
  \citenamefont {{Ross}}, \citenamefont {{Rossi}}, \citenamefont
  {{Rubi{\~n}o-Mart{\'\i}n}}, \citenamefont {{Saito}}, \citenamefont
  {{Salazar-Albornoz}}, \citenamefont {{Samushia}}, \citenamefont
  {{S{\'a}nchez}}, \citenamefont {{Satpathy}}, \citenamefont {{Schlegel}},
  \citenamefont {{Schneider}}, \citenamefont {{Sc{\'o}ccola}}, \citenamefont
  {{Seo}}, \citenamefont {{Sheldon}}, \citenamefont {{Simmons}}, \citenamefont
  {{Slosar}}, \citenamefont {{Strauss}}, \citenamefont {{Swanson}},
  \citenamefont {{Thomas}}, \citenamefont {{Tinker}}, \citenamefont
  {{Tojeiro}}, \citenamefont {{Maga{\~n}a}}, \citenamefont {{Vazquez}},
  \citenamefont {{Verde}}, \citenamefont {{Wake}}, \citenamefont {{Wang}},
  \citenamefont {{Weinberg}}, \citenamefont {{White}}, \citenamefont
  {{Wood-Vasey}}, \citenamefont {{Y{\`e}che}}, \citenamefont {{Zehavi}},
  \citenamefont {{Zhai}},\ and\ \citenamefont {{Zhao}}}]{AlaAtaBai1709}%
  \BibitemOpen
  \bibfield  {author} {\bibinfo {author} {\bibfnamefont {S.}~\bibnamefont
  {{Alam}}}, \bibinfo {author} {\bibfnamefont {M.}~\bibnamefont {{Ata}}},
  \bibinfo {author} {\bibfnamefont {S.}~\bibnamefont {{Bailey}}}, \bibinfo
  {author} {\bibfnamefont {F.}~\bibnamefont {{Beutler}}}, \bibinfo {author}
  {\bibfnamefont {D.}~\bibnamefont {{Bizyaev}}}, \bibinfo {author}
  {\bibfnamefont {J.~A.}\ \bibnamefont {{Blazek}}}, \bibinfo {author}
  {\bibfnamefont {A.~S.}\ \bibnamefont {{Bolton}}}, \bibinfo {author}
  {\bibfnamefont {J.~R.}\ \bibnamefont {{Brownstein}}}, \bibinfo {author}
  {\bibfnamefont {A.}~\bibnamefont {{Burden}}}, \bibinfo {author}
  {\bibfnamefont {C.-H.}\ \bibnamefont {{Chuang}}}, \bibinfo {author}
  {\bibfnamefont {J.}~\bibnamefont {{Comparat}}}, \bibinfo {author}
  {\bibfnamefont {A.~J.}\ \bibnamefont {{Cuesta}}}, \bibinfo {author}
  {\bibfnamefont {K.~S.}\ \bibnamefont {{Dawson}}}, \bibinfo {author}
  {\bibfnamefont {D.~J.}\ \bibnamefont {{Eisenstein}}}, \bibinfo {author}
  {\bibfnamefont {S.}~\bibnamefont {{Escoffier}}}, \bibinfo {author}
  {\bibfnamefont {H.}~\bibnamefont {{Gil-Mar{\'\i}n}}}, \bibinfo {author}
  {\bibfnamefont {J.~N.}\ \bibnamefont {{Grieb}}}, \bibinfo {author}
  {\bibfnamefont {N.}~\bibnamefont {{Hand}}}, \bibinfo {author} {\bibfnamefont
  {S.}~\bibnamefont {{Ho}}}, \bibinfo {author} {\bibfnamefont {K.}~\bibnamefont
  {{Kinemuchi}}}, \bibinfo {author} {\bibfnamefont {D.}~\bibnamefont
  {{Kirkby}}}, \bibinfo {author} {\bibfnamefont {F.}~\bibnamefont {{Kitaura}}},
  \bibinfo {author} {\bibfnamefont {E.}~\bibnamefont {{Malanushenko}}},
  \bibinfo {author} {\bibfnamefont {V.}~\bibnamefont {{Malanushenko}}},
  \bibinfo {author} {\bibfnamefont {C.}~\bibnamefont {{Maraston}}}, \bibinfo
  {author} {\bibfnamefont {C.~K.}\ \bibnamefont {{McBride}}}, \bibinfo {author}
  {\bibfnamefont {R.~C.}\ \bibnamefont {{Nichol}}}, \bibinfo {author}
  {\bibfnamefont {M.~D.}\ \bibnamefont {{Olmstead}}}, \bibinfo {author}
  {\bibfnamefont {D.}~\bibnamefont {{Oravetz}}}, \bibinfo {author}
  {\bibfnamefont {N.}~\bibnamefont {{Padmanabhan}}}, \bibinfo {author}
  {\bibfnamefont {N.}~\bibnamefont {{Palanque-Delabrouille}}}, \bibinfo
  {author} {\bibfnamefont {K.}~\bibnamefont {{Pan}}}, \bibinfo {author}
  {\bibfnamefont {M.}~\bibnamefont {{Pellejero-Ibanez}}}, \bibinfo {author}
  {\bibfnamefont {W.~J.}\ \bibnamefont {{Percival}}}, \bibinfo {author}
  {\bibfnamefont {P.}~\bibnamefont {{Petitjean}}}, \bibinfo {author}
  {\bibfnamefont {F.}~\bibnamefont {{Prada}}}, \bibinfo {author} {\bibfnamefont
  {A.~M.}\ \bibnamefont {{Price-Whelan}}}, \bibinfo {author} {\bibfnamefont
  {B.~A.}\ \bibnamefont {{Reid}}}, \bibinfo {author} {\bibfnamefont {S.~A.}\
  \bibnamefont {{Rodr{\'\i}guez-Torres}}}, \bibinfo {author} {\bibfnamefont
  {N.~A.}\ \bibnamefont {{Roe}}}, \bibinfo {author} {\bibfnamefont {A.~J.}\
  \bibnamefont {{Ross}}}, \bibinfo {author} {\bibfnamefont {N.~P.}\
  \bibnamefont {{Ross}}}, \bibinfo {author} {\bibfnamefont {G.}~\bibnamefont
  {{Rossi}}}, \bibinfo {author} {\bibfnamefont {J.~A.}\ \bibnamefont
  {{Rubi{\~n}o-Mart{\'\i}n}}}, \bibinfo {author} {\bibfnamefont
  {S.}~\bibnamefont {{Saito}}}, \bibinfo {author} {\bibfnamefont
  {S.}~\bibnamefont {{Salazar-Albornoz}}}, \bibinfo {author} {\bibfnamefont
  {L.}~\bibnamefont {{Samushia}}}, \bibinfo {author} {\bibfnamefont {A.~G.}\
  \bibnamefont {{S{\'a}nchez}}}, \bibinfo {author} {\bibfnamefont
  {S.}~\bibnamefont {{Satpathy}}}, \bibinfo {author} {\bibfnamefont {D.~J.}\
  \bibnamefont {{Schlegel}}}, \bibinfo {author} {\bibfnamefont {D.~P.}\
  \bibnamefont {{Schneider}}}, \bibinfo {author} {\bibfnamefont {C.~G.}\
  \bibnamefont {{Sc{\'o}ccola}}}, \bibinfo {author} {\bibfnamefont {H.-J.}\
  \bibnamefont {{Seo}}}, \bibinfo {author} {\bibfnamefont {E.~S.}\ \bibnamefont
  {{Sheldon}}}, \bibinfo {author} {\bibfnamefont {A.}~\bibnamefont
  {{Simmons}}}, \bibinfo {author} {\bibfnamefont {A.}~\bibnamefont {{Slosar}}},
  \bibinfo {author} {\bibfnamefont {M.~A.}\ \bibnamefont {{Strauss}}}, \bibinfo
  {author} {\bibfnamefont {M.~E.~C.}\ \bibnamefont {{Swanson}}}, \bibinfo
  {author} {\bibfnamefont {D.}~\bibnamefont {{Thomas}}}, \bibinfo {author}
  {\bibfnamefont {J.~L.}\ \bibnamefont {{Tinker}}}, \bibinfo {author}
  {\bibfnamefont {R.}~\bibnamefont {{Tojeiro}}}, \bibinfo {author}
  {\bibfnamefont {M.~V.}\ \bibnamefont {{Maga{\~n}a}}}, \bibinfo {author}
  {\bibfnamefont {J.~A.}\ \bibnamefont {{Vazquez}}}, \bibinfo {author}
  {\bibfnamefont {L.}~\bibnamefont {{Verde}}}, \bibinfo {author} {\bibfnamefont
  {D.~A.}\ \bibnamefont {{Wake}}}, \bibinfo {author} {\bibfnamefont
  {Y.}~\bibnamefont {{Wang}}}, \bibinfo {author} {\bibfnamefont {D.~H.}\
  \bibnamefont {{Weinberg}}}, \bibinfo {author} {\bibfnamefont
  {M.}~\bibnamefont {{White}}}, \bibinfo {author} {\bibfnamefont {W.~M.}\
  \bibnamefont {{Wood-Vasey}}}, \bibinfo {author} {\bibfnamefont
  {C.}~\bibnamefont {{Y{\`e}che}}}, \bibinfo {author} {\bibfnamefont
  {I.}~\bibnamefont {{Zehavi}}}, \bibinfo {author} {\bibfnamefont
  {Z.}~\bibnamefont {{Zhai}}}, \ and\ \bibinfo {author} {\bibfnamefont {G.-B.}\
  \bibnamefont {{Zhao}}},\ }\href {\doibase 10.1093/mnras/stx721} {\bibfield
  {journal} {\bibinfo  {journal} {\mnras}\ }\textbf {\bibinfo {volume} {470}},\
  \bibinfo {pages} {2617} (\bibinfo {year} {2017})},\ \Eprint
  {http://arxiv.org/abs/1607.03155} {arXiv:1607.03155 [astro-ph.CO]}
  \BibitemShut {NoStop}%
\bibitem [{\citenamefont {{McBride}}\ \emph {et~al.}(2009)\citenamefont
  {{McBride}}, \citenamefont {{Berlind}}, \citenamefont {{Scoccimarro}},
  \citenamefont {{Wechsler}}, \citenamefont {{Busha}}, \citenamefont
  {{Gardner}},\ and\ \citenamefont {{van den Bosch}}}]{McBBerSco0901}%
  \BibitemOpen
  \bibfield  {author} {\bibinfo {author} {\bibfnamefont {C.}~\bibnamefont
  {{McBride}}}, \bibinfo {author} {\bibfnamefont {A.}~\bibnamefont
  {{Berlind}}}, \bibinfo {author} {\bibfnamefont {R.}~\bibnamefont
  {{Scoccimarro}}}, \bibinfo {author} {\bibfnamefont {R.}~\bibnamefont
  {{Wechsler}}}, \bibinfo {author} {\bibfnamefont {M.}~\bibnamefont {{Busha}}},
  \bibinfo {author} {\bibfnamefont {J.}~\bibnamefont {{Gardner}}}, \ and\
  \bibinfo {author} {\bibfnamefont {F.}~\bibnamefont {{van den Bosch}}},\ }in\
  \href@noop {} {\emph {\bibinfo {booktitle} {American Astronomical Society
  Meeting Abstracts \#213}}},\ \bibinfo {series} {American Astronomical Society
  Meeting Abstracts}, Vol.\ \bibinfo {volume} {213}\ (\bibinfo {year} {2009})\
  p.\ \bibinfo {pages} {425.06}\BibitemShut {NoStop}%
\bibitem [{\citenamefont {{Sinha}}\ \emph {et~al.}(2018)\citenamefont
  {{Sinha}}, \citenamefont {{Berlind}}, \citenamefont {{McBride}},
  \citenamefont {{Scoccimarro}}, \citenamefont {{Piscionere}},\ and\
  \citenamefont {{Wibking}}}]{SinBerMcB1807}%
  \BibitemOpen
  \bibfield  {author} {\bibinfo {author} {\bibfnamefont {M.}~\bibnamefont
  {{Sinha}}}, \bibinfo {author} {\bibfnamefont {A.~A.}\ \bibnamefont
  {{Berlind}}}, \bibinfo {author} {\bibfnamefont {C.~K.}\ \bibnamefont
  {{McBride}}}, \bibinfo {author} {\bibfnamefont {R.}~\bibnamefont
  {{Scoccimarro}}}, \bibinfo {author} {\bibfnamefont {J.~A.}\ \bibnamefont
  {{Piscionere}}}, \ and\ \bibinfo {author} {\bibfnamefont {B.~D.}\
  \bibnamefont {{Wibking}}},\ }\href {\doibase 10.1093/mnras/sty967} {\bibfield
   {journal} {\bibinfo  {journal} {\mnras}\ }\textbf {\bibinfo {volume}
  {478}},\ \bibinfo {pages} {1042} (\bibinfo {year} {2018})},\ \Eprint
  {http://arxiv.org/abs/1708.04892} {arXiv:1708.04892 [astro-ph.CO]}
  \BibitemShut {NoStop}%
\bibitem [{\citenamefont {{Seljak}}\ and\ \citenamefont
  {{Zaldarriaga}}(1996)}]{SelZal96}%
  \BibitemOpen
  \bibfield  {author} {\bibinfo {author} {\bibfnamefont {U.}~\bibnamefont
  {{Seljak}}}\ and\ \bibinfo {author} {\bibfnamefont {M.}~\bibnamefont
  {{Zaldarriaga}}},\ }\href {\doibase 10.1086/177793} {\bibfield  {journal}
  {\bibinfo  {journal} {\apj}\ }\textbf {\bibinfo {volume} {469}},\ \bibinfo
  {pages} {437} (\bibinfo {year} {1996})}\BibitemShut {NoStop}%
\bibitem [{\citenamefont {{Sefusatti}}\ \emph {et~al.}(2016)\citenamefont
  {{Sefusatti}}, \citenamefont {{Crocce}}, \citenamefont {{Scoccimarro}},\ and\
  \citenamefont {{Couchman}}}]{SefCroSco1512}%
  \BibitemOpen
  \bibfield  {author} {\bibinfo {author} {\bibfnamefont {E.}~\bibnamefont
  {{Sefusatti}}}, \bibinfo {author} {\bibfnamefont {M.}~\bibnamefont
  {{Crocce}}}, \bibinfo {author} {\bibfnamefont {R.}~\bibnamefont
  {{Scoccimarro}}}, \ and\ \bibinfo {author} {\bibfnamefont {H.~M.~P.}\
  \bibnamefont {{Couchman}}},\ }\href {\doibase 10.1093/mnras/stw1229}
  {\bibfield  {journal} {\bibinfo  {journal} {\mnras}\ }\textbf {\bibinfo
  {volume} {460}},\ \bibinfo {pages} {3624} (\bibinfo {year} {2016})},\ \Eprint
  {http://arxiv.org/abs/1512.07295} {arXiv:1512.07295 [astro-ph.CO]}
  \BibitemShut {NoStop}%
\bibitem [{\citenamefont {{Feldman}}\ \emph {et~al.}(1994)\citenamefont
  {{Feldman}}, \citenamefont {{Kaiser}},\ and\ \citenamefont
  {{Peacock}}}]{FelKaiPea9405}%
  \BibitemOpen
  \bibfield  {author} {\bibinfo {author} {\bibfnamefont {H.~A.}\ \bibnamefont
  {{Feldman}}}, \bibinfo {author} {\bibfnamefont {N.}~\bibnamefont {{Kaiser}}},
  \ and\ \bibinfo {author} {\bibfnamefont {J.~A.}\ \bibnamefont {{Peacock}}},\
  }\href {\doibase 10.1086/174036} {\bibfield  {journal} {\bibinfo  {journal}
  {\apj}\ }\textbf {\bibinfo {volume} {426}},\ \bibinfo {pages} {23} (\bibinfo
  {year} {1994})},\ \Eprint {http://arxiv.org/abs/arXiv:astro-ph/9304022}
  {arXiv:astro-ph/9304022} \BibitemShut {NoStop}%
\bibitem [{\citenamefont {{Tegmark}}(1997)}]{Teg9711}%
  \BibitemOpen
  \bibfield  {author} {\bibinfo {author} {\bibfnamefont {M.}~\bibnamefont
  {{Tegmark}}},\ }\href {\doibase 10.1103/PhysRevLett.79.3806} {\bibfield
  {journal} {\bibinfo  {journal} {\prl}\ }\textbf {\bibinfo {volume} {79}},\
  \bibinfo {pages} {3806} (\bibinfo {year} {1997})},\ \Eprint
  {http://arxiv.org/abs/astro-ph/9706198} {arXiv:astro-ph/9706198 [astro-ph]}
  \BibitemShut {NoStop}%
\bibitem [{\citenamefont {{Oddo}}\ \emph {et~al.}(2020)\citenamefont {{Oddo}},
  \citenamefont {{Sefusatti}}, \citenamefont {{Porciani}}, \citenamefont
  {{Monaco}},\ and\ \citenamefont {{S{\'a}nchez}}}]{OddSefPor2003}%
  \BibitemOpen
  \bibfield  {author} {\bibinfo {author} {\bibfnamefont {A.}~\bibnamefont
  {{Oddo}}}, \bibinfo {author} {\bibfnamefont {E.}~\bibnamefont {{Sefusatti}}},
  \bibinfo {author} {\bibfnamefont {C.}~\bibnamefont {{Porciani}}}, \bibinfo
  {author} {\bibfnamefont {P.}~\bibnamefont {{Monaco}}}, \ and\ \bibinfo
  {author} {\bibfnamefont {A.~G.}\ \bibnamefont {{S{\'a}nchez}}},\ }\href
  {\doibase 10.1088/1475-7516/2020/03/056} {\bibfield  {journal} {\bibinfo
  {journal} {\jcap}\ }\textbf {\bibinfo {volume} {2020}},\ \bibinfo {eid} {056}
  (\bibinfo {year} {2020})},\ \Eprint {http://arxiv.org/abs/1908.01774}
  {arXiv:1908.01774 [astro-ph.CO]} \BibitemShut {NoStop}%
\bibitem [{\citenamefont {{Gelman}}\ and\ \citenamefont
  {{Rubin}}(1992)}]{GelRub9201}%
  \BibitemOpen
  \bibfield  {author} {\bibinfo {author} {\bibfnamefont {A.}~\bibnamefont
  {{Gelman}}}\ and\ \bibinfo {author} {\bibfnamefont {D.~B.}\ \bibnamefont
  {{Rubin}}},\ }\href {\doibase 10.1214/ss/1177011136} {\bibfield  {journal}
  {\bibinfo  {journal} {Statistical Science}\ }\textbf {\bibinfo {volume}
  {7}},\ \bibinfo {pages} {457} (\bibinfo {year} {1992})}\BibitemShut {NoStop}%
\bibitem [{\citenamefont {Lewis}(2019)}]{Lew2019}%
  \BibitemOpen
  \bibfield  {author} {\bibinfo {author} {\bibfnamefont {A.}~\bibnamefont
  {Lewis}},\ }\href {https://getdist.readthedocs.io} {\  (\bibinfo {year}
  {2019})},\ \Eprint {http://arxiv.org/abs/1910.13970} {arXiv:1910.13970
  [astro-ph.IM]} \BibitemShut {NoStop}%
\bibitem [{\citenamefont {S\'anchez}(2020)}]{Sanchez2012}%
  \BibitemOpen
  \bibfield  {author} {\bibinfo {author} {\bibfnamefont {A.~G.}\ \bibnamefont
  {S\'anchez}},\ }\href {\doibase 10.1103/PhysRevD.102.123511} {\bibfield
  {journal} {\bibinfo  {journal} {Phys. Rev. D}\ }\textbf {\bibinfo {volume}
  {102}},\ \bibinfo {pages} {123511} (\bibinfo {year} {2020})}\BibitemShut
  {NoStop}%
\bibitem [{\citenamefont {Abbott}\ \emph {et~al.}(2018)\citenamefont {Abbott},
  \citenamefont {Abdalla}, \citenamefont {Alarcon}, \citenamefont {Aleksić},
  \citenamefont {Allam}, \citenamefont {Allen}, \citenamefont {Amara},
  \citenamefont {Annis}, \citenamefont {Asorey}, \citenamefont {Avila},\ and\
  \citenamefont {et~al.}}]{Abbott2018}%
  \BibitemOpen
  \bibfield  {author} {\bibinfo {author} {\bibfnamefont {T.}~\bibnamefont
  {Abbott}}, \bibinfo {author} {\bibfnamefont {F.}~\bibnamefont {Abdalla}},
  \bibinfo {author} {\bibfnamefont {A.}~\bibnamefont {Alarcon}}, \bibinfo
  {author} {\bibfnamefont {J.}~\bibnamefont {Aleksić}}, \bibinfo {author}
  {\bibfnamefont {S.}~\bibnamefont {Allam}}, \bibinfo {author} {\bibfnamefont
  {S.}~\bibnamefont {Allen}}, \bibinfo {author} {\bibfnamefont
  {A.}~\bibnamefont {Amara}}, \bibinfo {author} {\bibfnamefont
  {J.}~\bibnamefont {Annis}}, \bibinfo {author} {\bibfnamefont
  {J.}~\bibnamefont {Asorey}}, \bibinfo {author} {\bibfnamefont
  {S.}~\bibnamefont {Avila}}, \ and\ \bibinfo {author} {\bibnamefont
  {et~al.}},\ }\href {\doibase 10.1103/physrevd.98.043526} {\bibfield
  {journal} {\bibinfo  {journal} {Physical Review D}\ }\textbf {\bibinfo
  {volume} {98}} (\bibinfo {year} {2018}),\
  10.1103/physrevd.98.043526}\BibitemShut {NoStop}%
\bibitem [{\citenamefont {Heymans}\ \emph {et~al.}(2020)\citenamefont
  {Heymans}, \citenamefont {Tröster}, \citenamefont {Asgari}, \citenamefont
  {Blake}, \citenamefont {Hildebrandt}, \citenamefont {Joachimi}, \citenamefont
  {Kuijken}, \citenamefont {Lin}, \citenamefont {Sánchez}, \citenamefont
  {van~den Busch}, \citenamefont {Wright}, \citenamefont {Amon}, \citenamefont
  {Bilicki}, \citenamefont {de~Jong}, \citenamefont {Crocce}, \citenamefont
  {Dvornik}, \citenamefont {Erben}, \citenamefont {Fortuna}, \citenamefont
  {Getman}, \citenamefont {Giblin}, \citenamefont {Glazebrook}, \citenamefont
  {Hoekstra}, \citenamefont {Joudaki}, \citenamefont {Kannawadi}, \citenamefont
  {Köhlinger}, \citenamefont {Lidman}, \citenamefont {Miller}, \citenamefont
  {Napolitano}, \citenamefont {Parkinson}, \citenamefont {Schneider},
  \citenamefont {Shan}, \citenamefont {Valentijn}, \citenamefont {Kleijn},\
  and\ \citenamefont {Wolf}}]{heymans2020}%
  \BibitemOpen
  \bibfield  {author} {\bibinfo {author} {\bibfnamefont {C.}~\bibnamefont
  {Heymans}}, \bibinfo {author} {\bibfnamefont {T.}~\bibnamefont {Tröster}},
  \bibinfo {author} {\bibfnamefont {M.}~\bibnamefont {Asgari}}, \bibinfo
  {author} {\bibfnamefont {C.}~\bibnamefont {Blake}}, \bibinfo {author}
  {\bibfnamefont {H.}~\bibnamefont {Hildebrandt}}, \bibinfo {author}
  {\bibfnamefont {B.}~\bibnamefont {Joachimi}}, \bibinfo {author}
  {\bibfnamefont {K.}~\bibnamefont {Kuijken}}, \bibinfo {author} {\bibfnamefont
  {C.-A.}\ \bibnamefont {Lin}}, \bibinfo {author} {\bibfnamefont {A.~G.}\
  \bibnamefont {Sánchez}}, \bibinfo {author} {\bibfnamefont {J.~L.}\
  \bibnamefont {van~den Busch}}, \bibinfo {author} {\bibfnamefont {A.~H.}\
  \bibnamefont {Wright}}, \bibinfo {author} {\bibfnamefont {A.}~\bibnamefont
  {Amon}}, \bibinfo {author} {\bibfnamefont {M.}~\bibnamefont {Bilicki}},
  \bibinfo {author} {\bibfnamefont {J.}~\bibnamefont {de~Jong}}, \bibinfo
  {author} {\bibfnamefont {M.}~\bibnamefont {Crocce}}, \bibinfo {author}
  {\bibfnamefont {A.}~\bibnamefont {Dvornik}}, \bibinfo {author} {\bibfnamefont
  {T.}~\bibnamefont {Erben}}, \bibinfo {author} {\bibfnamefont {M.~C.}\
  \bibnamefont {Fortuna}}, \bibinfo {author} {\bibfnamefont {F.}~\bibnamefont
  {Getman}}, \bibinfo {author} {\bibfnamefont {B.}~\bibnamefont {Giblin}},
  \bibinfo {author} {\bibfnamefont {K.}~\bibnamefont {Glazebrook}}, \bibinfo
  {author} {\bibfnamefont {H.}~\bibnamefont {Hoekstra}}, \bibinfo {author}
  {\bibfnamefont {S.}~\bibnamefont {Joudaki}}, \bibinfo {author} {\bibfnamefont
  {A.}~\bibnamefont {Kannawadi}}, \bibinfo {author} {\bibfnamefont
  {F.}~\bibnamefont {Köhlinger}}, \bibinfo {author} {\bibfnamefont
  {C.}~\bibnamefont {Lidman}}, \bibinfo {author} {\bibfnamefont
  {L.}~\bibnamefont {Miller}}, \bibinfo {author} {\bibfnamefont {N.~R.}\
  \bibnamefont {Napolitano}}, \bibinfo {author} {\bibfnamefont
  {D.}~\bibnamefont {Parkinson}}, \bibinfo {author} {\bibfnamefont
  {P.}~\bibnamefont {Schneider}}, \bibinfo {author} {\bibfnamefont
  {H.}~\bibnamefont {Shan}}, \bibinfo {author} {\bibfnamefont {E.}~\bibnamefont
  {Valentijn}}, \bibinfo {author} {\bibfnamefont {G.~V.}\ \bibnamefont
  {Kleijn}}, \ and\ \bibinfo {author} {\bibfnamefont {C.}~\bibnamefont
  {Wolf}},\ }\href@noop {} {\  (\bibinfo {year} {2020})},\ \Eprint
  {http://arxiv.org/abs/2007.15632} {arXiv:2007.15632 [astro-ph.CO]}
  \BibitemShut {NoStop}%
\bibitem [{\citenamefont {Pandey}\ \emph {et~al.}(2020)\citenamefont {Pandey},
  \citenamefont {Krause}, \citenamefont {Jain}, \citenamefont {MacCrann},
  \citenamefont {Blazek}, \citenamefont {Crocce}, \citenamefont {DeRose},
  \citenamefont {Fang}, \citenamefont {Ferrero}, \citenamefont {Friedrich},
  \citenamefont {Aguena}, \citenamefont {Allam}, \citenamefont {Annis},
  \citenamefont {Avila}, \citenamefont {Bernstein}, \citenamefont {Brooks},
  \citenamefont {Burke}, \citenamefont {Carnero~Rosell}, \citenamefont
  {Carrasco~Kind}, \citenamefont {Carretero}, \citenamefont {Costanzi},
  \citenamefont {da~Costa}, \citenamefont {De~Vicente}, \citenamefont {Desai},
  \citenamefont {Elvin-Poole}, \citenamefont {Everett}, \citenamefont
  {Fosalba}, \citenamefont {Frieman}, \citenamefont {Garc\'{\i}a-Bellido},
  \citenamefont {Gruen}, \citenamefont {Gruendl}, \citenamefont {Gschwend},
  \citenamefont {Gutierrez}, \citenamefont {Honscheid}, \citenamefont {Kuehn},
  \citenamefont {Kuropatkin}, \citenamefont {Maia}, \citenamefont {Marshall},
  \citenamefont {Menanteau}, \citenamefont {Miquel}, \citenamefont {Palmese},
  \citenamefont {Paz-Chinch\'on}, \citenamefont {Plazas}, \citenamefont
  {Roodman}, \citenamefont {Sanchez}, \citenamefont {Scarpine}, \citenamefont
  {Schubnell}, \citenamefont {Serrano}, \citenamefont {Sevilla-Noarbe},
  \citenamefont {Smith}, \citenamefont {Soares-Santos}, \citenamefont
  {Suchyta}, \citenamefont {Swanson}, \citenamefont {Tarle},\ and\
  \citenamefont {Weller}}]{Pandey2020}%
  \BibitemOpen
  \bibfield  {author} {\bibinfo {author} {\bibfnamefont {S.}~\bibnamefont
  {Pandey}}, \bibinfo {author} {\bibfnamefont {E.}~\bibnamefont {Krause}},
  \bibinfo {author} {\bibfnamefont {B.}~\bibnamefont {Jain}}, \bibinfo {author}
  {\bibfnamefont {N.}~\bibnamefont {MacCrann}}, \bibinfo {author}
  {\bibfnamefont {J.}~\bibnamefont {Blazek}}, \bibinfo {author} {\bibfnamefont
  {M.}~\bibnamefont {Crocce}}, \bibinfo {author} {\bibfnamefont
  {J.}~\bibnamefont {DeRose}}, \bibinfo {author} {\bibfnamefont
  {X.}~\bibnamefont {Fang}}, \bibinfo {author} {\bibfnamefont {I.}~\bibnamefont
  {Ferrero}}, \bibinfo {author} {\bibfnamefont {O.}~\bibnamefont {Friedrich}},
  \bibinfo {author} {\bibfnamefont {M.}~\bibnamefont {Aguena}}, \bibinfo
  {author} {\bibfnamefont {S.}~\bibnamefont {Allam}}, \bibinfo {author}
  {\bibfnamefont {J.}~\bibnamefont {Annis}}, \bibinfo {author} {\bibfnamefont
  {S.}~\bibnamefont {Avila}}, \bibinfo {author} {\bibfnamefont {G.~M.}\
  \bibnamefont {Bernstein}}, \bibinfo {author} {\bibfnamefont {D.}~\bibnamefont
  {Brooks}}, \bibinfo {author} {\bibfnamefont {D.~L.}\ \bibnamefont {Burke}},
  \bibinfo {author} {\bibfnamefont {A.}~\bibnamefont {Carnero~Rosell}},
  \bibinfo {author} {\bibfnamefont {M.}~\bibnamefont {Carrasco~Kind}}, \bibinfo
  {author} {\bibfnamefont {J.}~\bibnamefont {Carretero}}, \bibinfo {author}
  {\bibfnamefont {M.}~\bibnamefont {Costanzi}}, \bibinfo {author}
  {\bibfnamefont {L.~N.}\ \bibnamefont {da~Costa}}, \bibinfo {author}
  {\bibfnamefont {J.}~\bibnamefont {De~Vicente}}, \bibinfo {author}
  {\bibfnamefont {S.}~\bibnamefont {Desai}}, \bibinfo {author} {\bibfnamefont
  {J.}~\bibnamefont {Elvin-Poole}}, \bibinfo {author} {\bibfnamefont
  {S.}~\bibnamefont {Everett}}, \bibinfo {author} {\bibfnamefont
  {P.}~\bibnamefont {Fosalba}}, \bibinfo {author} {\bibfnamefont
  {J.}~\bibnamefont {Frieman}}, \bibinfo {author} {\bibfnamefont
  {J.}~\bibnamefont {Garc\'{\i}a-Bellido}}, \bibinfo {author} {\bibfnamefont
  {D.}~\bibnamefont {Gruen}}, \bibinfo {author} {\bibfnamefont {R.~A.}\
  \bibnamefont {Gruendl}}, \bibinfo {author} {\bibfnamefont {J.}~\bibnamefont
  {Gschwend}}, \bibinfo {author} {\bibfnamefont {G.}~\bibnamefont {Gutierrez}},
  \bibinfo {author} {\bibfnamefont {K.}~\bibnamefont {Honscheid}}, \bibinfo
  {author} {\bibfnamefont {K.}~\bibnamefont {Kuehn}}, \bibinfo {author}
  {\bibfnamefont {N.}~\bibnamefont {Kuropatkin}}, \bibinfo {author}
  {\bibfnamefont {M.~A.~G.}\ \bibnamefont {Maia}}, \bibinfo {author}
  {\bibfnamefont {J.~L.}\ \bibnamefont {Marshall}}, \bibinfo {author}
  {\bibfnamefont {F.}~\bibnamefont {Menanteau}}, \bibinfo {author}
  {\bibfnamefont {R.}~\bibnamefont {Miquel}}, \bibinfo {author} {\bibfnamefont
  {A.}~\bibnamefont {Palmese}}, \bibinfo {author} {\bibfnamefont
  {F.}~\bibnamefont {Paz-Chinch\'on}}, \bibinfo {author} {\bibfnamefont
  {A.~A.}\ \bibnamefont {Plazas}}, \bibinfo {author} {\bibfnamefont
  {A.}~\bibnamefont {Roodman}}, \bibinfo {author} {\bibfnamefont
  {E.}~\bibnamefont {Sanchez}}, \bibinfo {author} {\bibfnamefont
  {V.}~\bibnamefont {Scarpine}}, \bibinfo {author} {\bibfnamefont
  {M.}~\bibnamefont {Schubnell}}, \bibinfo {author} {\bibfnamefont
  {S.}~\bibnamefont {Serrano}}, \bibinfo {author} {\bibfnamefont
  {I.}~\bibnamefont {Sevilla-Noarbe}}, \bibinfo {author} {\bibfnamefont
  {M.}~\bibnamefont {Smith}}, \bibinfo {author} {\bibfnamefont
  {M.}~\bibnamefont {Soares-Santos}}, \bibinfo {author} {\bibfnamefont
  {E.}~\bibnamefont {Suchyta}}, \bibinfo {author} {\bibfnamefont {M.~E.~C.}\
  \bibnamefont {Swanson}}, \bibinfo {author} {\bibfnamefont {G.}~\bibnamefont
  {Tarle}}, \ and\ \bibinfo {author} {\bibfnamefont {J.}~\bibnamefont {Weller}}
  (\bibinfo {collaboration} {DES Collaboration}),\ }\href {\doibase
  10.1103/PhysRevD.102.123522} {\bibfield  {journal} {\bibinfo  {journal}
  {Phys. Rev. D}\ }\textbf {\bibinfo {volume} {102}},\ \bibinfo {pages}
  {123522} (\bibinfo {year} {2020})}\BibitemShut {NoStop}%
\bibitem [{\citenamefont {Blanchard}\ \emph {et~al.}(2020)\citenamefont
  {Blanchard}, \citenamefont {Camera}, \citenamefont {Carbone}, \citenamefont
  {Cardone}, \citenamefont {Casas}, \citenamefont {Clesse}, \citenamefont
  {Ilić}, \citenamefont {Kilbinger}, \citenamefont {Kitching},\ and\
  \citenamefont {et~al.}}]{EucFor}%
  \BibitemOpen
  \bibfield  {author} {\bibinfo {author} {\bibfnamefont {A.}~\bibnamefont
  {Blanchard}}, \bibinfo {author} {\bibfnamefont {S.}~\bibnamefont {Camera}},
  \bibinfo {author} {\bibfnamefont {C.}~\bibnamefont {Carbone}}, \bibinfo
  {author} {\bibfnamefont {V.~F.}\ \bibnamefont {Cardone}}, \bibinfo {author}
  {\bibfnamefont {S.}~\bibnamefont {Casas}}, \bibinfo {author} {\bibfnamefont
  {S.}~\bibnamefont {Clesse}}, \bibinfo {author} {\bibfnamefont
  {S.}~\bibnamefont {Ilić}}, \bibinfo {author} {\bibfnamefont
  {M.}~\bibnamefont {Kilbinger}}, \bibinfo {author} {\bibfnamefont
  {T.}~\bibnamefont {Kitching}}, \ and\ \bibinfo {author} {\bibnamefont
  {et~al.}},\ }\href {\doibase 10.1051/0004-6361/202038071} {\bibfield
  {journal} {\bibinfo  {journal} {\aap}\ }\textbf {\bibinfo {volume} {642}},\
  \bibinfo {pages} {A191} (\bibinfo {year} {2020})}\BibitemShut {NoStop}%
\bibitem [{\citenamefont {Hunter}(2007)}]{Hunter:2007}%
  \BibitemOpen
  \bibfield  {author} {\bibinfo {author} {\bibfnamefont {J.~D.}\ \bibnamefont
  {Hunter}},\ }\href {\doibase 10.1109/MCSE.2007.55} {\bibfield  {journal}
  {\bibinfo  {journal} {Computing in Science \& Engineering}\ }\textbf
  {\bibinfo {volume} {9}},\ \bibinfo {pages} {90} (\bibinfo {year}
  {2007})}\BibitemShut {NoStop}%
\end{thebibliography}%

\end{document}